\documentclass[english]{article}

\usepackage[latin9]{inputenc}
\usepackage[T1]{fontenc}
\usepackage[english]{babel}
\usepackage{csquotes}

\usepackage{geometry}
\usepackage{float}
\usepackage{mathrsfs}
\usepackage{mathtools}
\usepackage{amsmath}
\usepackage{amsthm}
\usepackage{amssymb}
\usepackage{graphicx}
\usepackage{titling}
\usepackage[colorlinks=true, allcolors=blue]{hyperref}
\usepackage[toc,page]{appendix}

\usepackage{colonequals}

\makeatletter

\theoremstyle{remark}
\newtheorem{rem}{\protect\remarkname}
\theoremstyle{definition}
\newtheorem{defn}{\protect\definitionname}
\theoremstyle{definition}
\newtheorem{example}{\protect\examplename}
\theoremstyle{plain}
\newtheorem{thm}{\protect\theoremname}
\theoremstyle{plain}

\theoremstyle{plain}

\allowdisplaybreaks
\usepackage{ dsfont }
\usepackage[all]{xy}
\usepackage{cite}
\usepackage[skins,theorems]{tcolorbox}
\tcbset{highlight math style={enhanced,
		colframe=red,colback=yellow!15!white,arc=2pt,boxrule=1pt}}
\usepackage{empheq}

\newcommand{\defi}{\colonequals}

\makeatother

\usepackage{babel}
\providecommand{\definitionname}{Definition}
\providecommand{\examplename}{Example}
\providecommand{\lemmaname}{Lemma}
\providecommand{\propositionname}{Proposition}
\providecommand{\remarkname}{Remark}
\providecommand{\theoremname}{Theorem}
\RequirePackage{xcolor}
\definecolor{tssteelblue}{RGB}{70,130,180}
\definecolor{tsorange}{RGB}{255,138,88}
\definecolor{tsblue}{RGB}{23,74,117}
\definecolor{tsforestgreen}{RGB}{21,122,81}
\definecolor{tsyellow}{RGB}{255,185,88}
\definecolor{tsgrey}{RGB}{200,200,200}
\definecolor{ttsorange}{RGB}{255,120,0}
\definecolor{forestgreen}{RGB}{0.13, 0.55, 0.13}
\definecolor{cof}{RGB}{219,144,71}
\definecolor{pur}{RGB}{186,146,162}
\definecolor{greeo}{RGB}{91,173,69}
\definecolor{greet}{RGB}{52,111,72}
\definecolor{myblue}{RGB}{80,191,227}
\definecolor{myblue2}{RGB}{80,141,227}
\definecolor{myblue3}{rgb}{0.0, 0.44, 1.0}
\definecolor{myblue4}{rgb}{0.01, 0.28, 1.0}
\RequirePackage{multirow}
\RequirePackage{array,makecell,cellspace}
\RequirePackage{booktabs,colortbl}
\RequirePackage{hhline}
\RequirePackage{capt-of}
\RequirePackage{calc}
\RequirePackage{adjustbox}
\RequirePackage{pdflscape}

\RequirePackage{rotating}
\RequirePackage{graphicx}
\graphicspath{{images/}}
\RequirePackage{tikz}
\RequirePackage[pages=some]{background}
\backgroundsetup{scale=1,color=black,opacity=1.0,angle=0,
	contents={\includegraphics[width=\paperwidth,height=\paperheight]{cover}}}
\RequirePackage{wrapfig}
\RequirePackage{tikz-cd}
\tikzcdset{
	arrow style=tikz,
	diagrams={>={Straight Barb[scale=0.8]}},math to/.tip={Glyph[glyph math command=rightarrow]},
	loop/.tip={Glyph[glyph math command=looparrowleft, swap]},
	loop'/.tip={Glyph[glyph math command=looparrowleft]},
	weird/.tip={Glyph[glyph math command=Rrightarrow, glyph length=1.5ex]},
	pi/.tip={Glyph[glyph math command=pi, glyph length=1.5ex, glyph axis=0pt]},
	white/.style={fill=white}
}

\RequirePackage{empheq}
\allowdisplaybreaks
\tikzset{
	math to/.tip={Glyph[glyph math command=rightarrow]},
	loop/.tip={Glyph[glyph math command=looparrowleft, swap]},
	loop'/.tip={Glyph[glyph math command=looparrowleft]},
	weird/.tip={Glyph[glyph math command=Rrightarrow, glyph length=1.5ex]},
	pi/.tip={Glyph[glyph math command=pi, glyph length=1.5ex, glyph axis=0pt]},
}
\RequirePackage{youngtab} 
\RequirePackage{young}
\newcolumntype{M}[1]{>{\centering\arraybackslash}m{#1}}
\newcolumntype{N}{@{}m{0pt}@{}}
\RequirePackage{array, boldline, rotating}
\RequirePackage{hhline,colortbl,bigstrut}
\RequirePackage{tabu}
\usepackage{tabularray}
\usetikzlibrary{shapes.geometric,shapes.multipart}
\usetikzlibrary{decorations.pathmorphing,decorations.text,decorations.pathreplacing,decorations.markings}
\usetikzlibrary{calc,intersections,through,backgrounds}
\usetikzlibrary{tikzmark}
\usetikzlibrary{matrix}
\pgfdeclarelayer{edgelayer,nodelayer}
\usetikzlibrary{patterns,shadows,calc,arrows}
\tikzstyle{none}=[inner sep=0pt]
\RequirePackage{pgfplots}\pgfplotsset{compat=1.13}
\usetikzlibrary{backgrounds,scopes,cd,positioning,shadows,shapes.multipart}
\RequirePackage[tikz]{bclogo} 
\usetikzlibrary{cd}
\RequirePackage{mathrsfs}
\RequirePackage[scr=boondoxo,scrscaled=1.05]{mathalfa}
\RequirePackage{mathtools}
\RequirePackage{amssymb}
\RequirePackage{dsfont}
\RequirePackage{extpfeil}
\usepackage{units}
\RequirePackage{blindtext}
\RequirePackage{booktabs}
\RequirePackage{url}
\usetikzlibrary{matrix}
\RequirePackage{multicol}
\usepackage{soul}
\usepackage[colorinlistoftodos]{todonotes}
\newcommand{\cor}[1]{{\color{black}#1}}

\newcommand{\scal}[2]{\left\langle#1,#2\right\rangle}
\newcommand{\scaln}[2]{\langle#1,#2\rangle}
\newcommand{\scalb}[2]{\big\langle#1,#2\big\rangle}

\newcommand{\parton}[1]{\left(#1\right)}
\newcommand{\partonn}[1]{(#1)}
\newcommand{\partonb}[1]{\big(#1\big)}
\newcommand{\partonB}[1]{\Big(#1\Big)}

\newcommand{\parq}[1]{\left[#1\right]}

\newcommand{\parqB}[1]{\Big[#1\Big]}

\newcommand{\restn}[1]{#1|}

\newcommand{\graf}[1]{\left\{#1\right\}}

\newcommand{\grafb}[1]{\big\{#1\big\}}
\newcommand{\grafB}[1]{\Big\{#1\Big\}}
\newcommand{\grafbb}[1]{\bigg\{#1\bigg\}}

\newcommand{\modu}[1]{\left|#1\right|}

\global\long\def\sB{\mathscr{B}}
\global\long\def\sC{\mathscr{C}}

\global\long\def\sG{\mathscr{G}}

\global\long\def\sP{\mathscr{P}}

\global\long\def\sU{\mathscr{U}}
\global\long\def\sV{\mathscr{V}}

\global\long\def\cA{\mathcal{A}}

\global\long\def\cC{\mathcal{C}}

\global\long\def\cG{\mathcal{G}}

\global\long\def\cH{\mathcal{H}}

\global\long\def\bC{\mathbb{C}}
\global\long\def\bD{\mathbb{D}}

\global\long\def\bG{\mathbb{G}}
\global\long\def\bH{\mathbb{H}}

\global\long\def\bL{\mathbb{L}}

\global\long\def\bN{\mathbb{N}}

\global\long\def\bR{\mathbb{R}}
\global\long\def\bS{\mathbb{S}}

\global\long\def\bV{\mathbb{V}}
\global\long\def\bW{\mathbb{W}}

\global\long\def\bZ{\mathbb{Z}}

\global\long\def\gL{\mathfrak{L}}
\global\long\def\gM{\mathfrak{M}}
\global\long\def\gN{\mathfrak{N}}

\global\long\def\gS{\mathfrak{S}}

\global\long\def\fC{\mathsf{C}}

\global\long\def\fO{\mathsf{O}}

\global\long\def\fU{\mathsf{U}}

\global\long\def\rc{\textrm{c}}
\global\long\def\rre{\textrm{e}}

\global\long\def\rm{\textrm{m}}
\global\long\def\rm{\textrm{micro}}

\global\long\def\minug{\leqslant}
		
\global\long\def\fr{\rightarrow}

\global\long\def\id{\mathds{1}}	
\global\long\def\defi{\vcentcolon =}

\global\long\def\Clo{\mathsf{Clo\,}}

\global\long\def\pr{\mathsf{pr}}

\global\long\def\and{\textrm{and}\,}

\global\long\def\Lin{\mathsf{Lin\,}}

\DeclareMathOperator{\Ker}{\mathsf{Ker}}

\DeclareMathOperator{\tr}{\textrm{tr}}

\global\long\def\div{\mathsf{div}\,}

\DeclareMathOperator{\curl}{\mathsf{curl}}
\DeclareMathOperator{\Curl}{\mathsf{Curl}}
\global\long\def\Sym{\mathsf{Sym}}
\global\long\def\sym{\textrm{sym}\,}
\global\long\def\skew{\textrm{skew}\,}

\global\long\def\dev{\textrm{dev}\,}
\global\long\def\Dev{\textrm{Dev}\,}

\global\long\def\SO{\textrm{SO}(3)}

\global\long\def\so{\mathfrak{so}\,(3)}

\DeclareMathOperator{\axl}{\mathsf{axl}}
\DeclareMathOperator{\anti}{\mathsf{anti}}

\DeclareMathOperator{\Hom}{\mathsf{Hom}}

\DeclareMathOperator{\Ela}{\mathfrak{Ela}}
\DeclareMathOperator{\High}{\mathfrak{High}}
\DeclareMathOperator{\Fix}{\mathsf{Fix}}

\DeclareMathOperator{\co}{\mathsf{co}}
\DeclareMathOperator{\Point}{\mathsf{Point}}

\DeclareMathOperator{\stt}{\;\left.\right|\;}
\DeclareMathOperator{\sttb}{\;\big.\big|\;}
\DeclareMathOperator{\sttB}{\;\Big.\Big|\;}

\global\long\def\pD{\text{D}}

\global\long\def\bCe{\mathbb{C}_{\textrm{e}}}
\global\long\def\bCm{\mathbb{C}_{\textrm{micro}}}

\global\long\def\R{\mathbb{R}}

\global\long\def\A{\mathscr{A}}

\global\long\def\ds{\textrm{dev sym}}

\global\long\def\cc{\mathbb{C}_{c}}

\global\long\def\So{\text{SO}(3)}

\global\long\def\C{\mathbb{C}}
\global\long\def\bL{\mathbb{L}}
\global\long\def\MM{\boldsymbol{\mathfrak{M}}}
\global\long\def\Sym{\textrm{Sym}\,(3)}
\global\long\def\G{\mathcal{G}}
\global\long\def\dx{\textrm{dx}}

\global\long\def\d{\textrm{d}}
\global\long\def\O{\textrm{O}}
\global\long\def\pr{\textrm{pr}}
\renewcommand{\nabla}{\textrm{D}} 

\global\long\def\Tr{\mathsf{Tr}}
\global\long\def\SOd{\text{SO}(2)}
\global\long\def\Tots{\mathfrak{Sym}}
\global\long\def\Harg{\mathfrak{Har}}
\global\long\def\Har{\mathsf{Har}}
\global\long\def\Isot{\mathsf{Isot}}

\newcommand{\vcenteredinclude}[1]{\begingroup
	\setbox0=\hbox{\includegraphics[width=0.06\textwidth]{#1}}%
	\parbox{\wd0}{\box0}\endgroup}

\title{On the representation of fourth and higher order anisotropic elasticity tensors
	in generalized continuum models}
\thanksmarkseries{arabic}
    \author{Marco Valerio d'Agostino\thanks{Marco Valerio d'Agostino, corresponding author, marco-valerio.dagostino@insa-lyon.fr, GEOMAS, INSA-Lyon, Universit\'{e} de Lyon, 20 avenue Albert Einstein, 69621, Villeurbanne cedex, France},  
	\, Robert J. Martin\thanks{Robert J. Martin, robert.martin@uni-due.de, Chair for Nonlinear
		Analysis and Modelling, Fakult\"{a}t f\"{u}r Mathematik, Universit\"{a}t Duisburg-Essen, Thea-Leymann-Stra\ss{}e 9, 45127 Essen, Germany.}, 
	\, Peter Lewintan\thanks{Peter Lewintan, peter.lewintan@uni-due.de, Chair for Nonlinear
		Analysis and Modelling, Fakult\"{a}t f\"{u}r Mathematik, Universit\"{a}t Duisburg-Essen, Thea-Leymann-Straße 9, 45127 Essen, Germany},
	\, Davide Bernardini\thanks{Davide Bernardini, dbernardini.didattica@uniroma1.it, Department of Structural and Geotechnical Engineering, Sapienza University of Rome, Rome, Italy}, 
	\\ 
	\, Alexandre Danescu\thanks{Alexandre Danescu, alexandre.danescu@ec-lyon.fr, Lyon Institute of Nanotechnology, \'{E}cole Centrale de Lyon, 69131 Ecully, France}  
	\, and Patrizio Neff\thanks{Patrizio Neff, patrizio.neff@uni-due.de, Head of Chair for Nonlinear Analysis and Modelling, Fakult\"{a}t f\"{u}r Mathematik, Universit\"{a}t Duisburg-Essen, Thea-Leymann-Stra\ss{}e 9, 45127 Essen, Germany}
	}

\begin{document}
	
	\maketitle
	\begin{abstract}
		The classification of all fourth-order anisotropic tensor classes for classical linear elasticity is well known. In this article, we review the related problem of explicitly computing the dimension and the expressions of the elements belonging to these classes, and we extend this computation to fourth-order elasticity tensors acting on non-symmetric matrices. These tensors naturally appear in generalized continuum models. Based on tensor symmetrization, we provide the most general forms of these tensors for orthotropic, transversely isotropic, cubic, and isotropic materials. We present a self-contained discussion and provide detailed calculations for simple examples.
	\end{abstract}
	\textbf{Keywords:} representation theory, group action, character, elasticity tensor, material symmetries, relaxed micromorphic model, generalized continua.

	\vspace{2mm}
	\textbf{}\\
	\textbf{AMS 2010 subject classification:}
	20C (representation theory of groups),
	58D19  (group actions and symmetry properties),
	20C15  	(ordinary representations and characters),
	74A30 (nonsimple materials),
	74A60 (micromechanical theories), 
	74B05 (classical linear elasticity),
	74M25 (micromechanics), 74Q15 (effective constitutive equations).
	
	\newpage{}
	
	\tableofcontents{}
	
	\section*{Introduction}
	
	Starting from the very early years of the linear theory of elasticity
	the question concerning the number of material parameters involved
	in the general theory and various particular symmetric situations was
	a matter of active scientific debate. A historical overview of the
	field is presented in the Historical Introduction chapter of Love's
	monograph \cite{love1944treatise} and the most fundamental debate
	concerned the opposition between the supporters and the opponents
	of the \textsl{rari-constant} and \textsl{multi-constant} elasticity
	theories. Introduced in the isotropic setting by Navier in 1821 \cite{Navier}
	and augmented in the general \ae{}olotropic (anisotropic) setting by Cauchy in 1828
	\cite{cauchy1827pression,cauchy1827condensation,cauchy1828equations}, 
	the rari-constant elasticity theory is based upon several assumptions,
	including central forces with arbitrary range of action (but small
	with respect to the macroscopic size of the body) and no residual
	stress. It predicts, in the general anisotropic case, only 15 elastic
	constants and in the particular case of isotropic materials $\nu=1/4$ (Poisson ratio).
	For a detailed historical account of the controversy involving the
	\textsl{rari-constant} and the \textsl{multi-constant} elasticity
	theories, we refer to \cite{todhunter2014history}.
	
	From a different phenomenological perspective, Green (and later Stokes)
	introduced what is called today the \textsl{strain-energy density}
	$W$ and assumed that it depends on the infinitesimal strain tensor $\varepsilon$.
	Neglecting higher order terms and assuming that the initial stress
	vanishes, the first non-trivial contribution comes from the second
	order polynomial terms which can be expressed as 
	\[
	    W(\varepsilon)
	    =
	    \frac{1}{2}
	    \scal{\mathbb{C}\,\varepsilon}{\varepsilon}.
	\]
	In this form and without additional assumptions, taking into account
	the symmetry of the strains one is left with 21 elastic coefficients.
	In the particular case of anisotropic materials, once the symmetry class of the body is established, the demand for the elastic constitutive law to remain invariant with respect to the elements of the accounted symmetry group turns into a large system of identities that must be satisfied by the components $\bC_{ijkl}$ of the $4^{\text{th}}$ order Hooke tensor $\mathbb{C}$.
	An illustration
	of this straightforward but very technical method is presented in
	the monograph by Love \cite{love1944treatise}. The monograph of Gurtin
	\cite{gurtin1973linear} attributes the results to Voigt \cite{voigt1908lehrbuch}
	and mentions that they can be obtained by using the results of Smith
	and Rivlin \cite{SmithRivlin} and Sirotin \cite{Sirotin}. Ting's monograph
	\cite{ting1996anisotropic} is dedicated entirely to anisotropic bodies and
	introduces an abstract formalism (see also \cite{bower2009applied})
	that associates to each transformation in the symmetry group of the
	material a second-order non-symmetric tensor $K$. In this framework,
	the invariance of the Hooke tensor with respect to the symmetry group
	is rewritten as the invariance of the fourth-order tensor $\mathbb{C}$
	with respect to the transformation $\mathbb{C}\rightarrow K\mathbb{C}K^{T}$. The Ting formalism
	is merely a condensed form of the classical invariance relation since
	now $K$ is, as expected, quadratic with respect to the element of
	the symmetry group. As noticed by Bower \cite{bower2009applied}, the
	Ting formalism is rather \textsl{convenient for computer applications}.
	
	To the best of our knowledge, the first attempt to unify the restrictions
	imposed by the symmetry group on the general form of the Hooke law
	and to extend them to a larger framework (piezo-elastic materials,
	higher gradients, higher-order polynomial constitutive relations)
	with group theoretical methods including representation theory results,
	the trace theorem and Haar integration were presented in \cite{danescu1997number}.
	The basic idea is to notice that \textsl{the invariance of the constitutive
		relation can be regarded as a question concerning the invariant vector
		sub-space of a general vector space with respect to a suitable defined
		action of the symmetry group}. In the classical case of linear
	elasticity theory the vector space is that of the Hooke tensors, the
	invariant vector sub-space is that of Hooke tensors satisfying the
	invariance with respect to the (fixed) symmetry group \cor{$\cG$ which is a closed subgroup of $\text{SO}(3)$}, while the action
	of the symmetry group associates to each symmetry element $Q\in{\cal G}$
	the eighth-order linear operator acting on Hooke tensors as 
	\[
	      \bC_{ijkl}
	      \rightarrow 
	      Q_{ia}Q_{jb}Q_{kc}Q_{ld}\bC_{abcd}\,.
	\]
	From classical group representation theory \textsl{the trace formula}
	provides the dimension of the invariant vector-subspace of Hooke tensors
	\textsl{as a function of the group only} and the explicit expression
	of an invariant Hooke tensor is obtained by symmetrization. Obviously,
	by definition, 
	\[
	     \widehat\bC_{ijkl}
	     =
	     \frac{1}{\textrm{card}({\cal G})}\sum_{Q\in{\cal G}}Q_{ia}Q_{jb}Q_{kc}Q_{ld}\bC_{abcd}
	\]
	is an invariant Hooke tensor with respect to the symmetry group \cor{$\cG$}.
	
	From this perspective, the result in \cite{danescu1997number} provides
	a unified setting, independent of the symmetry group, that both
	(i) determine the complexity of the theory (i.e.\ the number of elastic
		moduli) through the dimension of the
	invariant vector subspace of Hooke tensors and (ii) provide an effective method to find them. Obviously, the
	method works not only for quadratic forms with respect to strains
	but also for arbitrary homogeneous higher-order polynomials and, as
	a consequence, for arbitrary polynomial constitutive relations \cite{danescu1997number}.
	The results in \cite{danescu1997number} were subsequently extended
	to electro-elastic materials in \cite{DanescuTarantino1} and \cite{DanescuTarantino2}. In the context of piezoelectric materials, results concerning material invariance were also obtained in \cite{yang1995second,batra1995second}.
	Recently, a series of papers (cf.~\cite{auffray2013matrix,auffray2018complete} and the author's
	references within) rediscovered the symmetrization procedure in relation
	to the classical phenomenological higher gradient elasticity model of Mindlin. As noticed in
	\cite{DanescuTarantino1}, the symmetry properties of a classical
	(or generalized) constitutive relation depend strongly on the objects
	involved in its description so that as a consequence, higher-order
	tensors (involved in strain-gradient or second-gradient of strain)
	theories lead to more symmetry classes and, as expected, to a multitude
	of material parameters \cite{olive2014symmetry,auffray2013algebraic,olive2013symmetry,olive2017minimal,desmoratspace,auffray2017handbook,kolev2018characterization,forte1996symmetry,lazar2023toupin,lazar2022mathematical}. Along this path, the trace formula and its
	use in relation to the disjoint union decomposition of symmetry groups
	in \cite{danescu1997number} provide a way to explore the complexity
	of a theory without the explicit determination of the invariant forms
	of constitutive relation.
	
	In this paper, building upon the results initially presented in \cite{danescu1997number}, we explicitly calculate the expressions of the elasticity tensors used in enriched continuum models with respect to various symmetry classes. These models are, for example, currently employed to describe the mechanical properties of exotic metamaterials. Notably, in a series of papers \cite{voss2022modeling,neff2020identification,d2017effective,d2017panorama,barbagallo2017transparent,rizzi2021exploring,rizzi2022boundary,rizzi2022metamaterial,ramirez2023multi,demore2022unfolding,rizzi2022towards,chossat1991steady,azzi2023clips,olive2019effective}, it has been convincingly demonstrated how the relaxed micromorphic model effectively describes the band-gap phenomena occurring in the dynamic behavior of metamaterials. Our primary focus is on this generalized model.

	\section{Notation}
	
	Throughout this paper the Einstein convention of summation over repeated
	indices is used if not differently specified. We denote by $\R^{3\times3}$
	the set of real $3\times3$ second order tensors and by $\R^{3\times3\times3}$
	the set of real $3\times3\times3$ third order tensors. The standard
	Euclidean scalar product on $\R^{3\times3}$ is given by $\left\langle X,Y\right\rangle {}_{\R^{3\times3}}=\tr(X\, Y^{T})$
	and, thus, the Frobenius tensor norm is given by $\|X\|^{2}=\left\langle X,X\right\rangle {}_{\R^{3\times3}}$.
	Moreover, the identity tensor on $\R^{3\times3}$ will be denoted
	by $\mathds{1}$, so that $\tr(X)=\left\langle X,\mathds{1}\right\rangle $.
	We adopt the usual abbreviations of Lie-algebra theory, i.e. 
	\begin{itemize}
		\item $\Sym\defi\{X\in\R^{3\times3}\;|X^{T}=X\}$ denotes the vector-space
		of all symmetric $3\times3$ matrices,
		\item $\so\defi\{X\in\R^{3\times3}\;|X^{T}=-X\}$ is the Lie-algebra of skew
		symmetric tensors,
		\item $\mathfrak{sl}(3)\defi\{X\in\R^{3\times3}\;|\tr(X)=0\}$ is the Lie-algebra
		of traceless tensors, 
		\item $\R^{3\times3}\simeq\mathfrak{gl}(3)=\{\mathfrak{sl}(3)\cap\Sym\}\oplus\so\oplus\R\,\mathds{1}$
		is the \emph{orthogonal Cartan-decomposition of the Lie-algebra}.
	\end{itemize}
	In other words, for all $X\in\R^{3\times3}$, we consider the decomposition
	\begin{align}
		 X
		 =
		 \ds \,X
		 +
		 \skew X+\frac{1}{3}\mathrm{tr}(X)\,\mathds{1}\,, \label{eq:cartan lie}
	\end{align}
	where
	\begin{itemize}
		\item $\sym\,X=\frac{1}{2}(X^{T}+X)\in\Sym$ is the symmetric part of $X$, 
		\item $\skew\,X=\frac{1}{2}(X-X^{T})\in\so$ is the skew-symmetric part
		of $X$, 
		\item $\dev\,X=X-\frac{1}{3}\tr(X)\,\mathds{1}\in\mathfrak{sl}(3)$ is the
		deviatoric part of $X$. 
	\end{itemize}
	Throughout all the paper we indicate
	\begin{itemize}
		\item with an overline, e.g.\ $\overline{\C}$, the general fourth order tensors $\overline{\C}:\R^{3\times3}\rightarrow\R^{3\times3}$, 
		\item without superscripts, e.g.\ $\C$, classical fourth order tensors
		acting only on symmetric matrices \\
		($\C\colon\Sym\rightarrow\Sym$) or skew-symmetric ones ($\cc\colon\so\rightarrow\so$).
	\end{itemize}
	We also define
	\begin{itemize}
		\item $\text{GL}^+(3)\defi\{X\in\bR^{3\times3}\;|\;\det X>0\}$,
		\item $\id_2\in\bR^{3\times 3}$ the identity matrix, \cor{where the subscript 2 here pertains to the degree of the tensor (having 2 indices) and not to the dimension of $\bR^3$,} 
		\item  $\text{Sym}^+(3)\defi\{X\in\Sym\;|\;\scal{X\, v}{v}>0\quad\forall v\in\bR^3,\,v\neq0\}$,
		\item $\text{O}(3)\defi\{X\in\bR^{3\times3}\;|\;\det X=\pm1\}$,
		\item $\text{SO}(3)\defi\{X\in\bR^{3\times3}\;|\;\det X=1\}$.
	\end{itemize}
	We indicate by $\overline{\C} X$ the linear application of a tensor
	of $4^{\text{th}}$ order to a tensor of $2^{\text{nd}}$ order, i.e. 
	\begin{align}
		\left(\overline{\C}\:X\right)_{ij}=\overline{\C}_{ijhk}X_{hk}\,.
	\end{align}
	The operation of simple contraction between tensors of suitable order
	is also denoted by 
	\begin{align}
		(X  v)_{i}
		=
		X_{ij}v_{j}\,,
		\qquad
		(Z X)_{ij}
		=
		Z_{ih}X_{hj}\,.
	\end{align}
	
	\noindent Typical conventions for differential operations are employed,
	such as a comma followed by a subscript to denote the partial derivative
	with respect to the corresponding Cartesian coordinate:
	$\left(\cdot\right)_{,j}=\frac{\partial(\cdot)}{\partial x_{j}}$. The gradient operator will be denoted by $\pD$. Injective and surjective maps will be denoted by $\hookrightarrow$ and $\twoheadrightarrow$ respectively.

	\section{Preliminaries\label{Preliminaries}}
	
	First, we illustrate in a very simple and understandable way the ideas of the procedure (proposed in \cite{danescu1997number}) to find the symmetrized structure of a tensor if we are considering a particularly symmetry class. The technical tools we need are rigorously presented in 
	\begin{itemize}
		\item \cite{deitmar2014principles} for the Haar measure and invariant integration on locally compact topological groups like $\text{SO}(n)$,
		\item \cite{hall2013lie} for a general introduction to representation theory,
		\item \cite{golubitsky2012singularities} for the classification of the irreducible representations of $\So$ and the classification of its closed subgroups (which gives us the set of possible symmetries of a considered material), and the trace formula for finite-dimensional representations.
	\end{itemize}
In linearised elasticity, the elasticity tensor $\C$ is a linear map
$$ \C:\Sym\fr\Sym\,, $$
which gives the relation between the strain-tensor $\varepsilon\defi\sym\nabla u$ (the local deformation of the body) and the symmetric Cauchy stress-tensor $\sigma$ via
\begin{equation} 
	\sigma = \C\;\varepsilon.  \label{linear relation}
\end{equation}
Thus, the elasticity tensor is an element of the vector space $\text{Lin}(\Sym,\Sym)$, whose dimension is $36$. Moreover, according to the fact that we derive the relation \eqref{linear relation} from a variational principle, i.e.\ that the equilibrium configuration of the system is characterized as the stationary point of the functional
$$ 
     \frac{1}{2}\int_\Omega \left\langle \C\;\varepsilon,\varepsilon \right\rangle_{\R^{3\times3}}\,\dx\,,  
$$
the elasticity tensor has to be symmetric (and this symmetry is known as \textbf{major symmetry}) also with respect to the scalar product $\left\langle \C\;\varepsilon,\varepsilon \right\rangle_{\R^{3\times3}}$ which implies that $\C$ lives in the smaller space $\text{Sym}(\Sym,\Sym)$ of the symmetric applications from $\Sym$ to $\Sym$. The dimension of this space is 21 and so the full anisotropic elasticity tensor has at most 21 independent components instead of $9\times9=81$. 
According to the map $\underline{\MM}$ introduced in \eqref{M bar} (see the Appendix), we can represent $\C$ as a $6\times6$ symmetric matrix  $\widetilde{\mathbb{C}}$ with
\begin{equation}\label{C44}
\widetilde{\mathbb{C}}= \left(\begin{array}{cccccc}
\widetilde{\mathbb{C}}_{11} & \widetilde{\mathbb{C}}_{12} & \widetilde{\mathbb{C}}_{13} & \widetilde{\mathbb{C}}_{14} & \widetilde{\mathbb{C}}_{15} & \widetilde{\mathbb{C}}_{16}\\
& \widetilde{\mathbb{C}}_{22} & \widetilde{\mathbb{C}}_{23} & \widetilde{\mathbb{C}}_{24} & \widetilde{\mathbb{C}}_{25} & \widetilde{\mathbb{C}}_{26}\\
&  & \widetilde{\mathbb{C}}_{33} & \widetilde{\mathbb{C}}_{34} & \widetilde{\mathbb{C}}_{35} & \widetilde{\mathbb{C}}_{36}\\
&  &  & \widetilde{\mathbb{C}}_{44} & \widetilde{\mathbb{C}}_{45} & \widetilde{\mathbb{C}}_{46}\\
& \textrm{sym} &  &  & \widetilde{\mathbb{C}}_{55} & \widetilde{\mathbb{C}}_{56}\\
&  &  &  &  & \widetilde{\mathbb{C}}_{66}
\end{array}\right) = \left(\begin{array}{cccccc}
\C_{1111} & \C_{2211} & \C_{3311} & \C_{3211} & \C_{3111} & \C_{2111}\\
& \C_{2222} & \C_{3322} & \C_{3222} & \C_{3122} & \C_{2221}\\
&  & \C_{3333} & \C_{3332} & \C_{3331} & \C_{3321}\\
&  &  & \C_{3232} & \C_{3231} & \C_{3221}\\
& \textrm{sym} &  &  & \C_{3131} & \C_{3121}\\
&  &  &  &  & \C_{2121}
\end{array}\right).
\end{equation}
The question at hand is as follows: When we contemplate a material with a specific symmetry, how does this symmetry affect the structure of the elasticity tensor? To address this question, we must
define the material invariance and use it to derive conditions on $\C$, which are then interpreted within the framework of representation theory.

According to \cite{ciarlet1988mathematical}, considering a Lagrangian energy density $W\colon\textrm{GL}^+(3)\fr\R$ and a closed subgroup $\mathcal{G}$ of $\So$, we say that the material is $\mathcal{G}$-invariant if 
\begin{equation} 
	   W(F  Q)
	   = 
	   W(F) \qquad\forall\,Q\in\cG
	   \quad
	   \text{and}
	   \quad
	   \forall\,F\in\text{GL}^+\!(3).  \label{invariant condition}
\end{equation}
In continuum mechanics the matrix $F$ is the gradient of a deformation field $\varphi:\Omega\subseteq\R^3\fr \R^3$. 
Under the frame-invariance requirement for the energy density, setting $C=F^T  F$, the invariance condition \eqref{invariant condition} can be expressed in terms of the auxiliary function\footnote{The polar factorization of invertible matrices theorem \cite[Thm.3.2-2, pag.95]{ciarlet1988mathematical} establishes that $\text{GL}^+(3)\simeq\SO\times\text{Sym}^+(3)$. For a matrix $F\in\text{GL}^+(3)$, the unique pair $(R,U)\in\SO\times\text{Sym}^+(3)$ that yields $F=R \, U$ is determined by $U=\sqrt{F^T  F}$ and $R= F\, U^{-1}$ (see also \cite{fischle2017grioli,neff2014grioli}). 
 } 
$$
       \widehat{W}\colon\textrm{Sym}^+(3)\fr\R, \qquad \widehat{W}(F^T\!  F)=W(F) \quad\forall\,F\in\text{GL}^+\!(3),
$$ 
as
\begin{equation}\label{invariance C}
	\widehat{W}(Q^T\!  C \, Q)= \widehat{W}(C) \qquad\forall\,Q\in\mathcal{G}\quad\text{and}\quad\forall\,C\in\textrm{Sym}^+(3). 
\end{equation}
Defining the displacement field $u\colon\Omega\fr \R^3$ as $u=\varphi-\id$ (i.e. $u(x)=\varphi(x)-x$ for every $x\in\Omega$), and introducing the Green-St.Venant strain tensor
\begin{equation}
    E
    \defi 
    \frac{1}{2}(F^T\!  F-\id)=\frac{1}{2}\,(\nabla u^T+\nabla u)+\frac{1}{2}\,\nabla u^T\! \nabla u, 
\end{equation}
relation \eqref{invariance C} reads
\begin{equation}\label{invariance E}
     \widehat{W}(Q^T  (\id+2E) \, Q)= \widehat{W}(\id+2E)\qquad\forall\,Q\in\mathcal{G}\quad\text{and}\quad\forall\,E\in\Sym. 
\end{equation}
Expanding the energy density in a neighbourhood of the origin by
$$ \widehat{W}(\id+2E)=\widehat{W}(\id)+2\left\langle\nabla\widehat{W}(\id),E \right\rangle +2\left\langle\nabla^2\,\widehat{W}(\id)\, E,E \right\rangle +o\left( \left\|E\right\|^2 \right) ,  
$$
if we choose $\widehat{W}(\id)=0$ and the reference configuration is a natural state (i.e. $\nabla\widehat{W}(\id)=0$), then 
taking only the linear part $\varepsilon=\frac{1}{2}\,(\nabla u^T+\nabla u)=\sym\nabla u$ of $E$ (geometric linearity) and considering a homogeneous material,  setting $\C=4\nabla^2\,\widehat{W}(\id)$ the linearized energy can be expressed as a quadratic form in $\varepsilon$, i.e.\ $2\,\overline{W}(\varepsilon)=\left\langle \C\;\varepsilon,\varepsilon \right\rangle $. The invariance condition \eqref{invariance E} then reads
\begin{equation}
       \left\langle \C\;(Q^T\!\,\varepsilon\, Q),Q^T \varepsilon\, Q \right\rangle
       =
       \left\langle \C\;\varepsilon,\varepsilon \right\rangle \qquad\forall\,Q\in\mathcal{G}\quad\text{and}\quad\forall\,\varepsilon\in\Sym.
       \label{linear elasticity relation2}
\end{equation}
Writing the relation \eqref{linear elasticity relation2} component-wise, we obtain
\begin{equation}
	 \C_{abcd}\,Q_{ch}^{T}\,\varepsilon_{hk}\,Q_{kd}\,Q_{ai}^{T}\,\varepsilon_{ij}\,Q_{jb}
	 =
	 \C_{ijhk}\,\varepsilon_{hk}\,\varepsilon_{ij}
	 ,
\end{equation}
which is satisfied if and only if
\begin{equation}
     \C_{ijhk}=Q_{ia}Q_{jb}Q_{hc}Q_{kd}\,\C_{abcd} \qquad\forall\,Q\in\mathcal{G}. \label{relation invariance Hooke tensor}
\end{equation}
Therefore, the requirement of material invariance for the quadratic stored energy density translates into relation \eqref{relation invariance Hooke tensor} for the elasticity tensor which gives us relations between the components of $\C$. In this way, a part of the components of $\C$ can be expressed as linear combinations of a subset of its components.

We now ask for an algorithmic procedure which allows us to understand, once we fix the subgroup $\mathcal{G}$, the number of independent components of a tensor $\C$ which satisfy the relation \eqref{relation invariance Hooke tensor} and their position in its matrix representation $\widetilde{\mathbb{C}}$.

In order to do this, the idea is to interpret the relation \eqref{relation invariance Hooke tensor} in terms of an action of $\cG$ over the vector space $\text{Sym}(\Sym,\Sym)$ and characterize the tensors which respect the relation \eqref{relation invariance Hooke tensor} as those ones which are left fixed by the action. We give the formal definition of a linear action of a group over a vector space and we illustrate why we need this mathematical tool.

\begin{defn}[Linear action]\label{def:action}
	A \textbf{linear action} is a triple $(V,\cG,\varphi)$ where $V$ is a finite-dimensional real vector space, ${\cG}$ a topological group with unit element $e$ and $\varphi:{\cal G}\times V\rightarrow V$ is a map such that
\end{defn}
\begin{enumerate}
	\item $\varphi$ is a continuous map from $\mathcal{G}\times V $ to $V$, i.e. $\varphi\in\mathcal{C}^{\,0}\!\left({\cal G}\times V,V\right)$,
	\item $\varphi\left(e,v\right)=v$ for every $v\in V$,
	\item for fixed $g\in{\cal G}$ the application $\varphi_{g}:V\rightarrow V$,
	$\varphi_{g}( v):=\varphi\left(g,v\right)$ is linear,
	\item $\varphi(g_{1},\varphi(g_{2},v))=\varphi(g_{1}g_{2},v)$ for
	all $g_{1},g_{2}\in{\cal G}$ and $v\in V.$ 
\end{enumerate}
An action of a group on a vector space is therefore a way to move the elements inside the vector space along trajectories, called \textbf{orbits}, which have to respect the structure of the group. In the case of classical linear elasticity, the action we account for is given by the triple $\partonb{\!\Ela(3),\cG,\varphi}$ where 
\begin{itemize}
	\item $\Ela(3)$ is the space of the elasticity tensors, defined as \begin{align*}
	    \Ela(3)
	    \defi{}
	    &\text{Sym}(\Sym,\Sym)
	    \\
	    ={}
	    &\{\bC\in\Lin\partonb{\Sym,\Sym}\stt\scal{\bC\,A}{B}_{\bR^{3\times3}}
	    =\scal{A}{\bC\,B}_{\bR^{3\times3}}\quad\forall A,B\in\Sym\},
	\end{align*}
	\item $\cG$ is a proper subgroup of $\SO$ (and we will denote  the property of being a closed subgroup by $\cG\minug\SO$),
	\item $\varphi\in \cC^0(\cG\times\Ela(3),\Ela(3))$ is the map
	\begin{equation}\label{eq:main action}
	    \varphi\colon
	    {\cG}
	    \times 
	    \Ela(3)
	    \to
	    \Ela(3)
	    \quad\quad
	    (Q,\bC)
	    \longmapsto
	    \varphi(Q,\bC)\equalscolon\widehat\bC,
	    \qquad
	    \widehat\bC_{ijkl}
	    =
	    Q_{ia}Q_{jb}Q_{kc}Q_{ld}{\bC}_{abcd}\,.
	\end{equation}
\end{itemize}
The invariance condition established in \eqref{relation invariance Hooke tensor} reads in this new language as follows:
\[
     \bC\;\text{ satisfies} \; \eqref{relation invariance Hooke tensor} 
     \quad 
     \text{ if and only if }\;
     \quad \varphi(Q,\bC)
     =
     \bC
     \qquad
     \forall\,Q\in\cG.
\]
\begin{rem}
	The map $\varphi$ is well-defined; more specifically,
	\[
	    \varphi(Q,\bC)\in\Ela(3)
	    \qquad
	    \forall\,
	    \bC\in\Ela(3)\,.
	\]
	Indeed, let us consider $\bC\in\Ela(3)$. The minor symmetry $1\leftrightarrow2$ allows us to establish
	\begin{align*}
	    \widehat\bC_{jikl}
	     &
	     =
	     Q_{ja}Q_{ib}Q_{kc}Q_{ld}{\bC}_{abcd}
	     =
	     Q_{ja}Q_{ib}Q_{kc}Q_{ld}{\bC}_{bacd}
	     =
	     Q_{ib}Q_{ja}Q_{kc}Q_{ld}{\bC}_{bacd}
	     =
	    \widehat\bC_{ijkl}.
    \end{align*}
    It holds the same for the second minor symmetry $3\leftrightarrow4$ and the major symmetry $(12)\leftrightarrow(34)$.
\end{rem}

\begin{rem}
	In order to guarantee the well-posedness of the linearized elasticity problem, it is necessary to impose that the bilinear form $\scal{\bC\,X}{X}_{\bR^{3\times3}}$ is positive-definite, i.e., we need to require that
	\[
	    \bC
	    \in
	    \Ela^+(3)
	    \defi
	    \grafb{\bD\in\Ela^+(3)\stt\scal{\bD\,S}{S}_{\bR^{3\times3}}>0\quad\forall S\in\Sym\setminus\{0\}}.
	\]
	Note that $\Ela^+(3)$ is not a vector subspace of $\Ela(3)$; instead,  it is an open half-cone. Nevertheless, the action $\varphi$ will be defined on $\Ela(3)$ rather than $\Ela^+(3)$ because this allows us to handle linear actions. While it is conceivable to consider the action directly on $\Ela^+(3)$, doing so would complicate the general framework (by involving actions on smooth manifolds). Due to the technical constraint of ensuring $\bC \in \Ela^+(3)$, it becomes necessary to demonstrate that $\Ela^+(3)$ is an invariant subset of $\varphi$. In other words,
	%
	\[
	     \varphi(Q,\Ela^+(3))
	     \subseteq
	     \Ela^+(3)
	     \qquad
	     Q\in\SO.
	\]
	To show that, it suffices to note that for all $S\in\Sym\setminus\{0\}$ we have
	\begin{align}
	    \scal{\varphi(Q,\bC)\,S}{S}_{\bR^{3\times3}}
	    &
	    =
	    Q_{ia}Q_{jb}Q_{kc}Q_{ld}{\bC}_{abcd}\,S_{ij}\,S_{kl}
	    =
	    {\bC}_{abcd}\,\partonb{Q_{ia}\,S_{ij}\,Q_{jb}}\,\partonb{Q_{kc}S_{kl}\,Q_{ld}}
	    \nonumber
	    \\
	    &
	    =
	    {\bC}_{abcd}\,\partonb{Q_{ai}^T\,S_{ij}\,Q_{jb}}\,\partonb{Q_{ck}^T S_{kl}\,Q_{ld}}
	    =
	    \scalb{\bC\,(Q^T S \, Q)}{\underbrace{Q^T S \, Q}_{\mathclap{\in\,\Sym\setminus\{0\}}}}_{\bR^{3\times3}}
	    >0
	\end{align}
	for all $(Q,\bC)\in\SO\times\Ela^+(3)$. 
\end{rem}

Thus, we can say that an elasticity tensor $\bC$ respects the considered symmetry if it is left fixed by the action of the group on the vector space $V=\Ela(3)$. 
Therefore, the set of the tensors which verify \eqref{relation invariance Hooke tensor} is the subset of $\Ela(3)$ of tensors which do not move under the action of $\varphi$. 
This subset is a vector subspace\footnote{Indeed, as showed in formula \eqref{vector space fixed subspace}, $\textrm{Fix}_{\,{\cal G}}^{\, \varphi} V$ is the intersection of the vector subspaces $\{\Ker(\id_V-\varphi_Q)\}_{Q\in\cG}$ of $V$. } of $V$ and it is called the \textbf{fixed-point subspace}
\begin{equation}\label{vector space fixed subspace}
	  \textrm{Fix}_{\,{\cal G}}^{\, \varphi} V
	  \colonequals
	  \left\{ \C\in V\mid\varphi\,(Q,\C)=\C\ \textrm{for all}\ Q\in{\cal G}\right\}
	  =
	  \bigcap_{Q\in\cG}\underbrace{\Ker(\id_V-\varphi_Q)}_{\in\Lin(V,V)}. 
\end{equation}
\begin{figure}[H]
	\begin{centering}
		\begin{tikzpicture}
		\tikzset{
			nodeoformula/.style={rectangle,rounded corners=0.1cm,drop shadow={shadow xshift=0.5mm, shadow yshift=-0.5mm,opacity=1},draw=black, top color=white, bottom color=white, thick, inner sep=2mm, minimum size=2em, text centered},
			nodepoint/.style={circle,draw=gray,fill=gray,inner sep=0.8mm}
		}
		\node[anchor=south west,inner sep=0] at (0,0) {\includegraphics[scale=0.55]{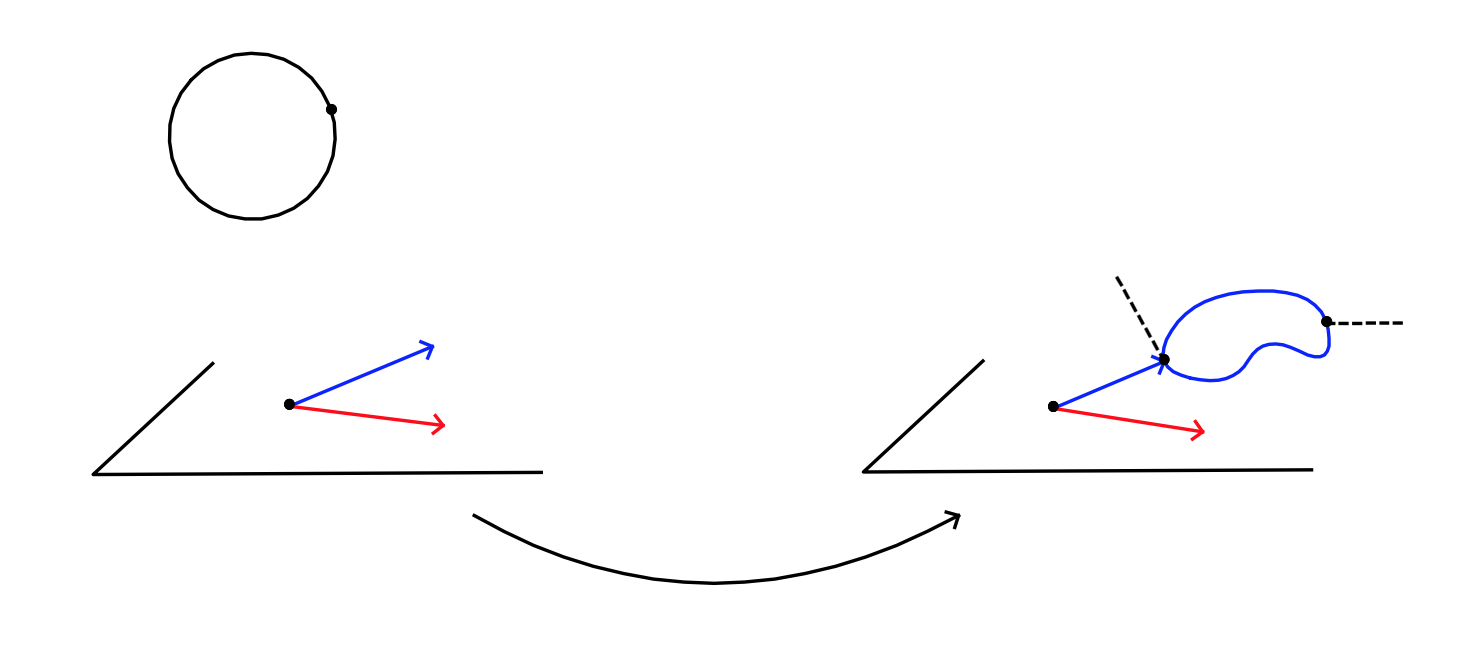}};
		\node[style=none](4)at(6.8, 1.2) {$\varphi$};
		\node[style=none](4)at(3.6, 5.5) {$g$};
		\node[style=none](4)at(3.8, 2) {$\textcolor{red}{u}$};
		\node[style=none](4)at(3.8, 3.2) {$\textcolor{blue}{v}$};
		\node[style=none](4)at(10.7, 2.9) {$\textcolor{blue}{v}$};
		\node[style=none](4)at(10.7, 4) {$\textcolor{blue}{\varphi(e,v)}$};
		\node[style=none](4)at(14.3, 3.25) {$\textcolor{blue}{\varphi(g,v)}$};
		\node[style=none](4)at(13, 3.9) {$\textcolor{blue}{\mathcal{O}_{v}}$};
		\node[style=none](4)at(12.8, 2.2) {$\textcolor{red}{u}\in\textrm{Fix}_{\,{\cal G}}^{\, \varphi} V$};
		\node[nodeoformula](2)at(2.45, 5.05) {$\mathcal{G}$};
		\end{tikzpicture}
		\par\end{centering}
	\caption{\label{fig:Left-invariance-of-a}Action, orbit and fixed point subspace.}
\end{figure}
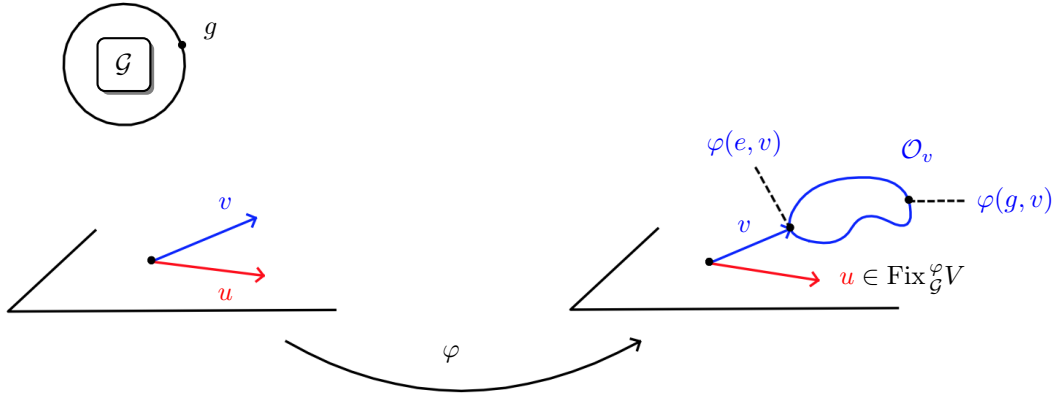 
The dimension of the fixed-point subspace provides the count of independent components of the invariant tensors. By applying the map $\underline{\MM}$ (see formula \eqref{M bar}) to the elements of $\textrm{Fix}_{\,\cG}^{\, \varphi} V$, we can derive the corresponding matrix representation.
Therefore, the crucial point lies in comprehending how to determine the set $\textrm{Fix}_{\,\cG}^{\, \varphi} V$. 
In elementary linear algebra, it is well-established that, in the context of a finite-dimensional vector space with a subspace, one can define a projection operator on it.
This operator enables us to isolate, from any arbitrary element in the vector space, only the component that belongs to the subspace. Consequently, the subspace can be obtained by projecting the entire vector space using the corresponding projection operator. In this context, working with a subspace is entirely equivalent to working with a projection operator.


In our specific scenario, we lack the direct capability to define $\textrm{Fix}_{\,{\cal G}}^{\, \varphi} V$. However, we can introduce a projection operator, and through this, we can ascertain that the resultant subspace precisely coincides with the set of fixed elements.

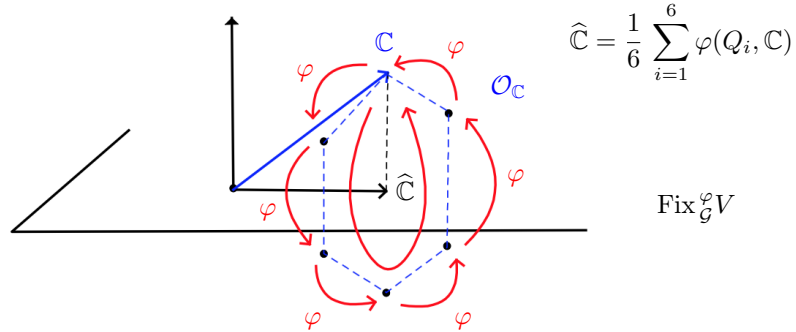
\begin{figure}[H]
	\begin{centering}
		\begin{tikzpicture}
		\tikzset{
			nodeoformula/.style={rectangle,rounded corners=0.1cm,drop shadow={shadow xshift=0.5mm, shadow yshift=-0.5mm,opacity=1},draw=black, top color=white, bottom color=white, thick, inner sep=2mm, minimum size=2em, text centered},
			nodepoint/.style={circle,draw=gray,fill=gray,inner sep=0.8mm}
		}
		\node[anchor=south west,inner sep=0] at (0,0) {\includegraphics[scale=0.55]{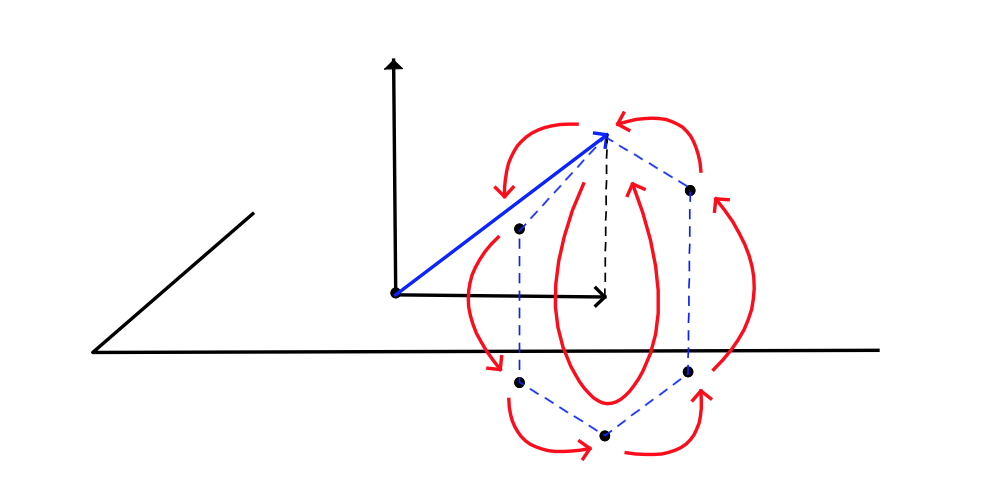}};
		\node[style=none](4)at(10, 1.6) {$\textrm{Fix}_{\,{\cal G}}^{\, \varphi} V$};
		\node[style=none](4)at(5.85, 3.8) {$\textcolor{blue}{\mathbb{C}}$};
		\node[style=none](4)at(6.13, 1.88) {$\textcolor{black}{\widehat{\mathbb{C}}}$};
		\node[style=none](4)at(9.8, 3.8) {$\displaystyle\widehat{\mathbb{C}}=\frac{1}{6}\,\sum_{i=1}^6\varphi(Q_i,\mathbb{C})$};
		\node[style=none](4)at(7.5, 3.2) {$\textcolor{blue}{\mathcal{O}_{\mathbb{C}}}$};
		\node[style=none](4)at(7.6, 2) {$\textcolor{red}{\varphi}$};
		\node[style=none](4)at(6.9, 0.1) {$\textcolor{red}{\varphi}$};
		\node[style=none](4)at(4.9, 0.1) {$\textcolor{red}{\varphi}$};
		\node[style=none](4)at(4.3, 1.55) {$\textcolor{red}{\varphi}$};
		\node[style=none](4)at(4.8, 3.4) {$\textcolor{red}{\varphi}$};
		\node[style=none](4)at(6.8, 3.7) {$\textcolor{red}{\varphi}$};
		\end{tikzpicture}
		\par\end{centering}
	\caption{\label{fig:average}Given an element $\mathbb{C}\in V$, its symmetrized $\widehat{\mathbb{C}}$ is obtained averaging over its orbit.}
\end{figure}

The concept is as follows: Let us take a generic element denoted as $\C$ belonging to the vector space $\Ela(3)$. We then track its trajectory  (its orbit) determined by the action of a discrete group $\mathcal{G}$ (we are making the hypothesis for the moment that $\cal{G}$ has a finite number of 6 elements). Now, if we perform the summation $\widehat{\mathbb{C}}=\frac{1}{6}\,\sum_{i=1}^6\varphi(Q_i,\mathbb{C})$, $Q_i\in\cG$, over all the elements in the orbit $\cal{O}_\C$ of $\C$, where
\begin{equation}
	\mathcal{O}_{\C}\colonequals\left\{ \mathbb{D}\in V\mid\exists\,Q\in\G\text{ such that}\,\mathbb{D}=\varphi(Q,\C)\right\}, 
\end{equation}  
 we obtain a tensor. This tensor automatically retains its invariance under the action of $\varphi$. As illustrated in Figure~\ref{fig:average}, this invariance can be interpreted as $\varphi$ simultaneously affecting all the terms in the sum, which are in turn interrelated such that each maps to another summand within the overall summation.
Indeed, denoting with $n$ the cardinality of the considered group, the tensor 
\begin{equation}\label{symmetrization}
	  \widehat{\bC}
	  =
	  \frac{1}{n}\,\sum_{i=1}^{n}\varphi(Q_{i},\C),
\end{equation}
i.e. the average of the elements of the orbit $\cal{O}_\C$ of $\C$, is invariant under the action: for all $\Lambda\in\mathcal{G}$ we have
\begin{equation}\label{invariance_symmetrized}
       \varphi\partonb{\Lambda,\widehat{\mathbb{C}}}
       =
       \varphi\parton{\Lambda,\frac{1}{n}\,\sum_{i=1}^{n}\varphi(Q_{i},\C)}
       =
       \frac{1}{n}\,\sum_{i=1}^{n}\varphi(\Lambda,\varphi(Q_{i},\C))
       =
       \frac{1}{n}\,\sum_{i=1}^{n}\varphi\left(\Lambda\, Q_{i},\C\right)\overset{(*)}{=}\frac{1}{n}\,\sum_{i=1}^{n}\varphi(Q_{i},\C)
       =
       \widehat{\mathbb{C}}.
\end{equation}
Another way to say this is: averaging the elements of the vector space with respect to the action of the group we obtain the elements of the fixed-point subspace.
The projection we are looking for has to do exactly this, i.e. it has to average the elements of the space of symmetric fourth order tensors.

From a technical point of view, there are a series of difficulties we have to deal with. Indeed, not all the closed subgroups of $\So$ are discrete and this means that we have to extend the sum on continuous group in a way in which we can still guarantee the validity of identities like (*) in equation \eqref{invariance_symmetrized} (which is obvious in the case of a discrete sum but it demands the notions of an invariant measure over the group $\cal G$ if we need to perform an integral). 

The technical tool we need here is the Haar measure over topological groups. Roughly speaking, the Haar measure has, in the context of topological groups, the same role which the Lebesgue measure plays in $\R^n$. One of the fundamental properties of the Lebesgue measure is its invariance with respect to rigid transformations, i.e. roto-translation maps of the form $f(x)=Q\, x +b$ with $Q\in\text{O}(n)$ and $b\in\R^n$. This property guarantees that the Lebesgue measure is a good notion to measure the size of subsets of $\R^n$ because if we move a subset with a rigid map its size does not change. On topological groups we do not have rotations and translations.

 However, we still have something which works in an analogous way: left- and right- translations. The \textbf{left-translation} is defined as follows: for any $h\in\cal{G}$, we set
 \begin{equation}
 	L_h\colon\mathcal{G}\fr\mathcal{G},\qquad L_{h}(g) \defi h\, g.
 \end{equation}
 In an analogous way we define \textbf{right-translations} by
 \begin{equation}
 R_h\colon\mathcal{G}\fr\mathcal{G},\qquad R_{h}(g) \defi g\, h.
 \end{equation}
  On a topological group, these movements precisely represent the transformations we wish to preserve. Luckily, as shown in \cite{deitmar2014principles}, every compact topological group (which are the groups of interest, like $\So$) has a measure $\mu$ which preserves both left- and right-translations such that the size of the group is 1 (i.e.\ $\mu(\mathcal{G})=1$); this is called the \textbf{normalized Haar measure} on $\mathcal{G}$. This means that if we are considering a measurable subset $\cA$ of $\mathcal{G}$, and we move this set according to $L_h$ or $R_h$, then $\mu(L_h(\cA))=\mu(R_h(\cA))=\mu(\cA)$. This property translates for the derived notion of integration (the \textbf{normalized Haar integral}) as follows: for any integrable function\footnote{The same invariance holds for both left and right-translations simultaneously. For instance, consider $h_1,h_2\in\cG$ and a function $f\colon\cG\to\bR$. We introduce the auxiliary function $\widetilde f(g)=f(g\,h_2)$. Then
  	\[
  	    \int_\cG f(h_1gh_2)\,\d\mu
  	    =
  	    \int_\cG f(h_1(gh_2))\,\d\mu
  	    =
  	    \int_\cG \widetilde f(h_1g)\,\d\mu
  	    =
  	    \int_\cG \widetilde f(g)\,\d\mu
  	    =
  	    \int_\cG  f(g h_2)\,\d\mu
  	    =
  	    \int_\cG  f(g)\,\d\mu\,.
  	\] } $f\colon\mathcal{G}\fr \R$, 
  \begin{equation}
  	   \intop_{\mathcal{G}}f\left(g\right)\d\mu
  	   =\intop_{\mathcal{G}}f\left(L_{h}\left(g\right)\right)\d\mu
  	   =\intop_{\mathcal{G}}f\left(hg\right)\d\mu
  	   =\intop_{\mathcal{G}}f\left(gh\right)\d\mu
  	   =\intop_{\mathcal{G}}f\left(R_{h}\left(g\right)\right)\d\mu
  	   \qquad\forall\,h\in\mathcal{G}\,.
  \end{equation}

\begin{figure}[H]
	\begin{centering}
			\includegraphics[scale=0.75]{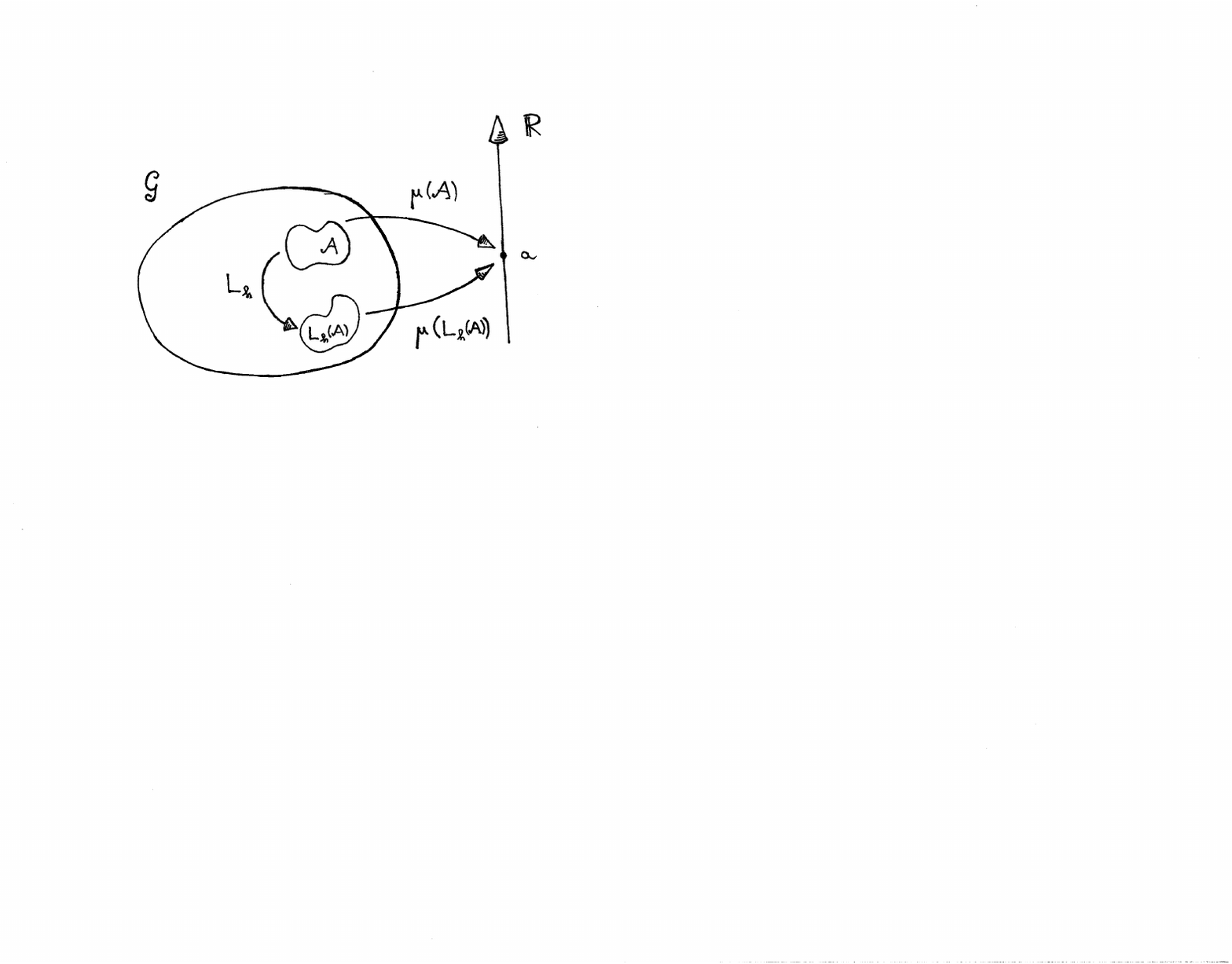} 
	\par\end{centering}
\caption{\label{fig:Left-invariance-of-a}Left-invariance of a measure.}
\end{figure} 
Therefore, the average in \eqref{symmetrization} can be expressed as follows for continuous groups:

\begin{equation}\label{symmetrization_continous}
	\widehat{\mathbb{C}}=\int_{\cal{G}}\varphi(Q,\C)\,\d\mu\,,
\end{equation}
where $\mu$ is the normalized Haar measure on $\mathcal{G}$. It is the invariance property of the Haar measure with respect to both left- and right-translations which guarantees also in the continuous case that the averaged tensor $\widehat{\mathbb{C}}$ is left fixed by the action. Indeed, exactly as in \eqref{invariance_symmetrized}, for $\Lambda\in\So$ we have
\begin{align}\label{invariance_symmetrized_continous}
	   \varphi\partonb{\Lambda,\widehat{\mathbb{C}}}
	   &
	   =
	   \varphi\parton{\Lambda,\int_{\textrm{SO}\left(3\right)}\varphi(Q,\C)\,\d\mu}
	   =\int_{\textrm{SO}\left(3\right)}\varphi\left(\Lambda,\varphi(Q,\C)\right)\d\mu
	   =
	   \int_{\textrm{SO}(3)}\varphi(\Lambda\,Q,\C)\,\d\mu
	   =
	   \int_{\textrm{SO}\left(3\right)}\varphi(Q,\C)\,\d\mu
	   =
	   \widehat{\mathbb{C}}. 
\end{align}	
This motivates to define the projection
\begin{equation*}
	\sP\colon\Ela(3)
	\fr
	\textrm{Fix}_{\,{\cal G}}^{\, \varphi} \Ela(3),
\end{equation*}
where
\begin{equation}\label{definition P}
	\tcbhighmath[drop fuzzy shadow]{
		\begin{aligned}
		    \sP(\C)
		    &
		    \defi
		    \int_{{\cal G}}\varphi(Q,\C)\,\d\mu
		    =
		    \int_{{\cal G}}\varphi_{Q}(\C)\,\d\mu
		    &
		    \qquad
		    \textrm{for the continuous case},
		    \\[2mm]
		    \sP(\C)
		    &
		    \defi
		    \frac{1}{\textrm{card}\,\mathcal{G}}\,\sum_{Q\in\cG}\varphi(Q,\bC)
		    =
		    \frac{1}{\textrm{card}\,\mathcal{G}}\,\sum_{Q\in\cG}\varphi_Q(\bC)
		    &
		    \qquad
		    \textrm{for the discrete case}.
		\end{aligned}
		}
\end{equation}
\begin{rem}
	For a discrete group $\mathcal{G}$, the normalized Haar measure is simply a weighted counting measure; more specifically,
	\[
		\mu(\mathcal{A})=\frac{\textrm{card}\,\mathcal{A}}{\textrm{card}\,\mathcal{G}}
	\]
	for all measurable $\mathcal{A}\subset\mathcal{G}$. In this case,
	\[
		\int_{{\cal G}}\varphi_Q(\bC)\,\d\mu
		= \frac{1}{\textrm{card}\,\mathcal{G}}\,\sum_{Q\in\cG}\varphi_Q(\bC)
		\,,
	\]
	thus the discrete case in \eqref{definition P} can be considered a special case of the general Haar-measure based formula.
\end{rem}
\begin{rem}
	Note that $\sP$ is surjective, i.e. $\sP\colon\Ela(3)
	\twoheadrightarrow
	\textrm{Fix}_{\,{\cal G}}^{\, \varphi} \Ela(3)$. Indeed, if $\bC\in\textrm{Fix}_{\,{\cal G}}^{\, \varphi} \Ela(3) $, then by definition of the fixed point subspace, $\varphi(Q,\C)=\bC$ for every $Q\in\cG$ and thus
	\[
	    \sP(\bC)
	    =
	    \int_{\cG}\underbrace{\varphi(Q,\C)}_{\mathclap{~~=\,\bC\;\;\forall\,Q\in\cG}}\,\d\mu
	    =
	    \int_{\cG}\bC\,\d\mu
	    =
	    \bC
	    \underbrace{\int_{\cG}\d\mu}_{=1}
	    =
	    \bC,
	\]
	i.e., we obtain the inclusion $\textrm{Fix}_{\,{\cal G}}^{\, \varphi} \Ela(3)\subseteq\text{Im}(\sP)$. Analogously for the discrete case,
	\[
	\sP(\bC)
	=
	\frac{1}{\text{card}\,\cG}\sum_{\cG}\underbrace{\varphi(Q,\C)}_{\mathclap{~~=\,\bC\;\;\forall\,Q\in\cG}}
	=
	\frac{1}{\text{card}\,\cG}\sum_{\cG}\bC
	=
	\frac{1}{\text{card}\,\cG} (\text{card}\,\cG)\,\bC
	=
	\bC.
	\] 
\end{rem}

\subsection{The trace formula}

We have successfully acquired the skills to compute the structure of symmetrized tensors. However, there remains the task of determining the dimension of $\textrm{Fix}_{\,\cG}^{\, \varphi} V$. To confront this secondary challenge, we introduce a pivotal tool into our arsenal: the \textbf{trace formula}. Indeed, the general result establishes the following (see \cite[Thm.2.3, p.76]{golubitsky2012singularities}).
\begin{thm}\label{thm1}
        Consider a linear action $(V,\cH,\varphi)$ where $\cH$ is a compact Lie group. Then for a Lie subgroup $\cG\minug\cH$, we have
        \[
             \dim\Fix_{\,\cG}^{\,\varphi} V
             =
             \int_{g\in\cG} \tr\varphi_g\,\d\mu\,,
        \]
        where the measure $\d\mu$ is the normalized Haar measure on $\cG$ and the maps $\{\varphi_g\}$ are the linear maps as defined in Definition \ref{def:action}.
\end{thm}
\begin{proof}
	   Let us consider the projection operator $\sP$ associated to $\Fix_{\,\cG}^{\,\varphi} V$. Then
	   \[
	         \tr\sP
	         =
	         \dim\Fix_{\,\cG}^{\,\varphi} V\,.
	   \]
	   This is an elementary consequence of the fact that it is possible to find an isomorphism\footnote{Indeed, it is possible to choose a basis $\{v_i\}_{i=1}^{m}$ of $V$ ($m=\dim V$) such that, denoting by $d$ the dimension of the subspace $\text{Im}(\sP)$, $v_1,\ldots,v_d$ is a basis of $\text{Im}(\sP)$. Hence the isomorphism $\Upsilon$ is obtained via the identifications $v_i\mapsto e_i$ for each $i$, where $\{e_i\}_i^{m}$ is the canonical basis of  $\bR^m$.} $\Upsilon:V\fr \R^m$, for a suitable $m\in\mathbb{N}$, such that the projection $\sP$ can be represented as a matrix operator $A_{\sP}$ with
	   \begin{figure}[htbp!]
	   		\centering
	   	\begin{equation}
	   		\begin{minipage}[c]{.30\textwidth}
	   				\begin{equation*}
	   					\hspace{-1cm}\xymatrix{ 
	   							 V \ar[r]^{\sP} \ar@<2pt>[d]_{\Upsilon} 
	   							 &     
	   							 \text{Im}(\sP)
	   							 =
	   							 \textrm{Fix}_{\,{\cG}}^{\, \varphi}
	   							 V
	   							 \subseteq 
	   							 V \hspace{-24mm} \ar@<2pt>[d]^{\Upsilon} 
	   							 \\ 
	   							 \bR^m  \ar[r]_{A_{\sP}}  
	   							 & 
	   							 \R^p\subseteq\bR^m \hspace{-13mm}
	   					}
	   				\end{equation*}
	   		\end{minipage}%
	   		\hspace{1cm}
	   			\begin{minipage}[c]{.30\textwidth}
	   					\begin{equation*}
	   						A_{\mathscr{P}}=\left( \begin{array}{c|c}   \raisebox{0pt}{{\large\mbox{{$\id_{\R^p}$}}}}  &  \raisebox{0pt}{{$0$}} \\[-4ex]   &  \\[-0.5ex]     &  \\[-0.5ex] \hline   \raisebox{-3pt}{{$0$}}  & \raisebox{-3pt}{{$0$}} \\  \end{array} \right).
	   						\end{equation*}
	   				\end{minipage}
	   		\end{equation}
	   		\end{figure}
	   		
	   		The construction of $\sP$ we derived in eq.\eqref{definition P} can be straightforwardly generalized to any linear action $(V,\cH,\varphi)$, where $\cH$ is a compact Lie group and $\cG$ is a Lie subgroup. Hence we obtain
	   		\begin{align*}
	   			  \hspace{5cm}
	   		      \dim\Fix_{\,\cG}^{\,\varphi} V
	   		      =
	   		      \tr\sP
	   		      \overset{\eqref{definition P}}{=}
	   		      \tr
	   		      \int_\cG\varphi_g\,\d\mu
	   		      =
	   		      \int_\cG \tr\varphi_g\,\d\mu.
	   		      \hspace{5cm}\qedhere
	   		\end{align*}
\end{proof}

Therefore, from the knowledge of the traces $\{\text{tr}\,\varphi_Q\}_{Q\in\cG}$, simply computing an integral we can finally obtain $\dim\textrm{Fix}_{\,{\cG}}^{\, \varphi} V$. The quantities $\{\text{tr}\,\varphi_Q\}_{Q\in\cG}$ are known in the literature as \textbf{characters} of $\{\varphi_Q\}_{Q\in\cG}$ and they will be denoted by $\{\chi(Q)\}_{Q\in\cG}$, $\chi(Q)\in\bR$. 

\begin{example}
	  Let us illustrate the content of Theorem \ref{thm1} with an example. Consider $V=\text{Sym}(2)$ and $\cG$, $\varphi$ and $\sP$ as in Example \ref{Ex1} below. Then $\dim V=3$ and $\dim\Fix_{\,\cG}^{\,\varphi} V=1$. Hence, after identifying $V$ with $\bR^3$ and $\Fix_{\,\cG}^{\,\varphi}V$ with the subspace of $V$ generated by $(1,0,0)$ (denoted by $\langle(1,0,0)\rangle$), $A_\sP\colon\bR^3\to\bR^3$ is such that $\text{Im}(A_\sP)=\langle(1,0,0)\rangle$ and 
	  $
	  A_\sP
	  =
	  \left(
	  \begin{smallmatrix}     
	  	1 & 0 & 0\\
	  	0 & 0 & 0\\
	  	0 & 0 & 0
	  \end{smallmatrix}\right)$.
\end{example}

\bigskip

%
Equipping the space $\Lin(V,V)$ with a scalar product $\scal{\cdot}{\cdot}_{\Lin(V,V)}$, the trace of a linear map $\varphi\colon V\to V$ can be introduced as
\[
    \tr\varphi
    \defi
    \scal{\varphi}{\id_{\Lin(V,V)}}_{\Lin(V,V)}.
\]
From the definition of the action, it follows that the identity of $\Lin(\Ela(3),\Ela(3))$ is $\varphi_{\id_2}$ 
which acts as
\[
    (\varphi(\id_2,\bC))_{ijkl}
    =
    (\otimes^4\id_2)_{iajbkcld}\,\bC_{abcd}
    =
    \delta_{ia}\delta_{jb}\delta_{kc}\delta_{ld}
    \bC_{abcd}
    =
    \bC_{ijkl}
    \qquad
    \forall i,j,k,l\in\{1,2,3\}.
\]
Then it would be natural to set
\begin{equation}\label{trace}
   \tr\varphi_Q
   =
   \chi(Q)
   =
   \scal{\varphi_Q}{\otimes^4\id_2}_{\otimes^8\bR^3}.
\end{equation}
Nevertheless this is incorrect and the reason is at the same time very subtle and very simple. The crucial remark here is the fact that both $\otimes^4\id_2$ and $\varphi_Q$ can act as linear maps over the bigger vector space $\Lin\partonb{\bR^{3\times3},\bR^{3\times3}}$ and formally these maps have the same expressions. Hence, in eq.\eqref{trace} we need to impose that we are considering $\varphi_Q$ as a map restricted to the subspace $\Ela(3)\subset\Lin\partonb{\bR^{3\times3},\bR^{3\times3}}$. 
The simplest way to achieve this is considering the projection operator associated to $\Ela(3)$. This projection \enquote{has to symmetrize} elements belonging to $\Lin\partonb{\bR^{3\times3},\bR^{3\times3}}$. For this reason it is denoted as \textbf{symmetrization identity}. In order to get an element of $\Ela(3)$ from an element of $\Lin\partonb{\bR^{3\times3},\bR^{3\times3}}$ we need to symmetrize it w.r.t.\ the symmetries possessed by the elements of $\Ela(3)$. Let us introduce the symmetrization identity
\begin{equation}\label{Symmetrization identity}
     \Pi^{\Sym}\in\Lin\partonb{\bR^{3\times3},\Sym},
     \qquad
     \text{where}
     \qquad
     \Pi^{\Sym}_{iajb}
     =
     \frac{1}{2}\partonb{\delta_{ia}\delta_{jb}+\delta_{ib}\delta_{ja}}
     \,,
\end{equation}
which acts as
\[
     \Pi^{\Sym}:
     \bR^{3\times3}
     \longrightarrow
     \bR^{3\times3},
     \qquad
     S
     \longmapsto
     \Pi^{\Sym}\,S,
     \quad
     \text{where}
     \quad
     (\Pi^{\Sym}\,S)_{ij}
     =
     \frac{1}{2}
     \partonb{\delta_{ia}\delta_{jb}+\delta_{ib}\delta_{ja}}
     S_{ab}.
\]
Clearly, $\Pi^{\Sym}\,S\in\Sym$ because
\begin{align*}
	(\Pi^{\Sym}\,S)_{ji}
	&
	=
	\frac{1}{2}
	\partonb{\delta_{ja}\delta_{ib}+\delta_{jb}\delta_{ia}}
	S_{ab}
	=
	\frac{1}{2}
	\partonb{\delta_{ja}\delta_{ib}S_{ab}+\delta_{jb}\delta_{ia}S_{ab}}
	=
	\frac{1}{2}
	\partonb{\delta_{ja}\delta_{ib}S_{ab}+\delta_{ia}\delta_{jb}S_{ab}}
	\\
	&
	=
	\frac{1}{2}
	\partonb{\delta_{ia}\delta_{jb}S_{ab}+\delta_{ja}\delta_{ib}S_{ab}}
	=
	(\Pi^{\Sym}\,S)_{ij}.
\end{align*}
Hence, to build up the projection from $\Lin\partonb{\bR^{3\times3},\bR^{3\times3}}$ to $\Ela(3)$ we need to symmetrize w.r.t. both the arguments via $\Pi^{\Sym}$ and w.r.t. the exchange of the arguments to obtain the major symmetry. This means that the projection from $\Lin\partonb{\bR^{3\times3},\bR^{3\times3}}$ to $\Ela(3)$ will be
\begin{equation}\label{eq:sym id tensor}
   \Pi^{\Ela(3)}:
   \Lin\partonb{\bR^{3\times3},\bR^{3\times3}}
   \xtwoheadrightarrow{}
   \Ela(3),
   \qquad
   \Pi^{\Ela(3)}
   \defi
   \sym\partonb{\Pi^{\Sym}\otimes\Pi^{\Sym}},
\end{equation}
where the symmetrization operation $\sym$ in the previous definition is defined as usual, i.e.,
\[
   \sym\partonb{\Pi^{\Sym}\otimes\Pi^{\Sym}}_{iajbkcld}
   =
   \frac{1}{2}
   \partonB{
   \partonb{\Pi^{\Sym}\otimes\Pi^{\Sym}}_{iajbkcld}
   +
   \partonb{\Pi^{\Sym}\otimes\Pi^{\Sym}}_{kcldiajb}}
\]
for every $
i,a,j,b,k,c,l,d
\in\{1,2,3\}$. Thus, finally, the component-wise expression of $\Pi^{\Ela(3)}$ is
\begin{align}
	\Pi^{\Ela(3)}_{iajbkcld} 
	& =
	\frac{1}{8}\Big(\delta_{ia}\delta_{jb}\delta_{kc}\delta_{ld}+\delta_{ia}\delta_{jb}\delta_{lc}\delta_{kd}+\delta_{ja}\delta_{ib}\delta_{lc}\delta_{kd}+\delta_{ja}\delta_{ib}\delta_{kc}\delta_{ld}\nonumber
	\\
	& \quad+\delta_{ka}\delta_{lb}\delta_{ic}\delta_{jd}+\delta_{ka}\delta_{lb}\delta_{jc}\delta_{id}+\delta_{la}\delta_{kb}\delta_{jc}\delta_{id}+\delta_{la}\delta_{kb}\delta_{ic}\delta_{jd}\Big).
\end{align}
Now we can compute the trace of the linear maps $\{\varphi_Q\}_{Q\in\cG}$ as maps from $\Ela(3)$ to $\Ela(3)$ considering the composition maps $\{\varphi_Q\circ\Pi^{\Ela{3}}\}_{Q\in\cG}$, obtaining
\begin{align*}
	\chi(Q) 
	& 
	=
	\textrm{tr}\,\big(\varphi_Q\circ\Pi^{\Ela(3)}\big)
	=
	\scal{\varphi_Q\circ\Pi^{\Ela(3)}}{\otimes^4\id_2}_{\Ela(3)}
	=
	Q_{ia}Q_{jb}Q_{kc}Q_{ld}
	\Pi^{\Ela(3)}_{a \alpha b \beta c \gamma d \delta}
	\delta_{i\alpha}\delta_{j\beta}\delta_{k\gamma}\delta_{l\delta}
	\\
	&
	=
	Q_{ia}Q_{jb}Q_{kc}Q_{ld}\Pi_{aibjckdl}
	=
	\Pi_{iajbkcld}Q_{ia}Q_{jb}Q_{kc}Q_{ld}.
\end{align*}
Expanding, we find
\begin{align*}
	\chi(Q)
	& =\Pi_{iajbkcld}Q_{ia}Q_{jb}Q_{kc}Q_{ld}
	\\
	& =
	\frac{1}{8}\big(\delta_{ia}\delta_{jb}\delta_{kc}\delta_{ld}+\delta_{ia}\delta_{jb}\delta_{lc}\delta_{kd}+\delta_{ja}\delta_{ib}\delta_{lc}\delta_{kd}+\delta_{ja}\delta_{ib}\delta_{kc}\delta_{ld}
	\\
	& \quad+\delta_{ka}\delta_{lb}\delta_{ic}\delta_{jd}+\delta_{ka}\delta_{lb}\delta_{jc}\delta_{id}+\delta_{la}\delta_{kb}\delta_{jc}\delta_{id}+\delta_{la}\delta_{kb}\delta_{ic}\delta_{jd}\big)Q_{ia}Q_{jb}Q_{kc}Q_{ld},
	\\
	& 
	=
	\frac{1}{8}\big(\delta_{ia}\delta_{jb}\delta_{kc}\delta_{ld}Q_{ia}Q_{jb}Q_{kc}Q_{ld}+\delta_{ia}\delta_{jb}\delta_{lc}\delta_{kd}Q_{ia}Q_{jb}Q_{kc}Q_{ld}
	\\
	& 
	\quad+\delta_{ja}\delta_{ib}\delta_{lc}\delta_{kd}Q_{ia}Q_{jb}Q_{kc}Q_{ld}+\delta_{ja}\delta_{ib}\delta_{kc}\delta_{ld}Q_{ia}Q_{jb}Q_{kc}Q_{ld}\\
	& 
	\quad+\delta_{ka}\delta_{lb}\delta_{ic}\delta_{jd}Q_{ia}Q_{jb}Q_{kc}Q_{ld}+\delta_{ka}\delta_{lb}\delta_{jc}\delta_{id}Q_{ia}Q_{jb}Q_{kc}Q_{ld}\\
	& 
	\quad+\delta_{la}\delta_{kb}\delta_{jc}\delta_{id}Q_{ia}Q_{jb}Q_{kc}Q_{ld}+\delta_{la}\delta_{kb}\delta_{ic}\delta_{jd}Q_{ia}Q_{jb}Q_{kc}Q_{ld}\big)\\
	& =\frac{1}{8}\big(\underbrace{\delta_{ia}Q_{ia}}_{\textrm{tr}Q}\underbrace{\delta_{jb}Q_{jb}}_{\textrm{tr}Q}\underbrace{\delta_{kc}Q_{kc}}_{\textrm{tr}Q}\underbrace{\delta_{ld}Q_{ld}}_{\textrm{tr}Q}+\underbrace{\delta_{ia}Q_{ia}}_{\textrm{tr}Q}\underbrace{\delta_{jb}Q_{jb}}_{\textrm{tr}Q}\underbrace{\delta_{cl}Q_{ld}\delta_{dk}Q_{kc}}_{Q_{cd}Q_{dc}=\,\textrm{tr}Q^{2}}\\
	& \quad+\underbrace{\delta_{bi}Q_{ia}\delta_{aj}Q_{jb}}_{\textrm{tr}Q^{2}}\underbrace{\delta_{kd}Q_{kc}\delta_{cl}Q_{ld}}_{\textrm{tr}Q^{2}}+\underbrace{\delta_{aj}Q_{jb}\delta_{bi}Q_{ia}}_{\textrm{tr}Q^{2}}\underbrace{\delta_{kc}Q_{kc}}_{\textrm{tr}Q}\underbrace{\delta_{ld}Q_{ld}}_{\textrm{tr}Q}\\
	& \quad+\underbrace{\delta_{ak}Q_{kc}\delta_{ci}Q_{ia}}_{\textrm{tr}Q^{2}}\underbrace{\delta_{dj}Q_{jb}\delta_{bl}Q_{ld}}_{\textrm{tr}Q^{2}}+\underbrace{\delta_{ak}Q_{kc}\delta_{cj}Q_{jb}\delta_{bl}Q_{ld}\delta_{di}Q_{ia}}_{\textrm{tr}Q^{4}}\\
	& \quad+\underbrace{\delta_{al}Q_{ld}\delta_{di}Q_{ia}}_{\textrm{tr}Q^{2}}\underbrace{\delta_{cj}Q_{jb}\delta_{bk}Q_{kc}}_{\textrm{tr}Q^{2}}+\underbrace{\delta_{al}Q_{ld}\delta_{dj}Q_{jb}\delta_{bk}Q_{kc}\delta_{ci}Q_{ia}}_{\textrm{tr}Q^{4}}\big)\\
	& =\frac{1}{8}\Big((\textrm{tr}\,Q)^{4}+(\textrm{tr}\,Q)^{2}\textrm{tr}\,Q^{2}+(\textrm{tr}\,Q^{2})^{2}+(\textrm{tr}\,Q)^{2}\textrm{tr}\,Q^{2}
	+
	(\textrm{tr}\,Q^{2})^{2}+\textrm{tr}\,Q^{4}+(\textrm{tr}\,Q^{2})^{2}+\textrm{tr}\,Q^{4}\Big)
\end{align*}
and thus
\begin{equation}
	   \chi(Q)
	   =
	   \frac{1}{8}\partonB{(\textrm{tr}\,Q)^{4}+2(\textrm{tr}\,Q)^{2}\textrm{tr}\,Q^{2}+2\,\textrm{tr}\,Q^{4}+3(\textrm{tr}\,Q^{2})^{2}}.
	   \label{charac_elasticity}
\end{equation}
Moreover, using the Cayley-Hamilton theorem (see the Appendix) for proper
orthogonal rotations, the character $\chi$ can be also written as
a polynomial in the first invariant of $Q,$ i.e. $\textrm{tr}\,Q$,
as\footnote{
		In the following section, we will analyze generalized elasticity models that accommodate ambient space dimensions different from $3$. This exploration delves into the spaces of elasticity tensors $\Ela(n)$. It's noteworthy that the trace formula remains consistent in the $n$-dimensional case up to \eqref{charac_elasticity}. However, the Cayley-Hamilton formula for $Q\in\text{SO}(n)$ is dimension-dependent, impacting the expression of the characters as polynomials in $\tr Q$. Consequently, we will derive appropriate expressions based on $n$. For instance, when $n=2$, \eqref{charac_elasticity_CH} would undergo a transformation to
		\[
		\chi(Q)=(\textrm{tr}\,Q)^4-3(\textrm{tr}\,Q)^2+2.
		\]
} 
\begin{equation}
\label{charac_elasticity_CH}
	\chi(Q)=(\textrm{tr}\,Q)^{4}-3(\textrm{tr}\,Q)^{3}+2(\textrm{tr}\,Q)^{2}+\textrm{tr}\,Q\,.
\end{equation}
Finally, we can summarize the proposed method according to the developed scheme:
\begin{itemize}
	\item The material invariance condition requires the relations
	$$ 
	    \C_{ijhk}=Q_{ia}Q_{jb}Q_{hc}Q_{kd}\,\C_{abcd} 
	    \qquad
	    \qquad
	    \forall
	    \,
	    Q\in
	    \cG
	$$
	between the components of the elasticity tensor. We express this in terms of the action $\varphi$ as
	$$ 
	    \varphi(Q,\C)=\mathbb{C}
	    \qquad
	    \qquad
	    \forall
	    \,
	    Q\in
	    \cG, 
	$$
	which translates the problem of looking for constitutive tensors satisfying the material invariance condition to the search for a particular subspace of $V=\Ela(3) $, the fixed point subspace $\textrm{Fix}_{\,{\cal G}}^{\, \varphi} V$.
	\item The fixed point subspace can be obtained computing the projection  $\mathscr{P}$ which acts by averaging a considered tensor over its orbit.
    \item The dimension of the fixed-point subspace, which gives the number of independent components of the invariant tensors, is calculated via the trace formula, i.e.\ by evaluating
    $$ \int_{{\cal G}}\textrm{tr}\,\varphi(Q,\C)\,\d\mu\quad\text{or}\quad\frac{1}{\textrm{card}\,\mathcal{G}}\,\sum_{Q\in\mathcal{G}}\textrm{tr}\,\varphi(Q,\C), $$
    where the trace of the linear applications is computed by virtue of the symmetrization identity $\Pi^{\Ela(3)}$, i.e., the projection $\Pi^{\Ela(3)}:\textrm{Lin}\left(\R^{3\times3},\R^{3\times3}\right)\twoheadrightarrow\Ela(3)$. 
     
\end{itemize}

\begin{example}\label{Ex1}
	
	We consider the illustrative variational problem
	$$ 
	     \mathscr{A}\left[ f \right] =\int_\Omega W(\nabla f(x))\,\dx\,, 
	$$
	where $\Omega\subseteq\R^2$ is a Lipschitz domain,  $f\colon\Omega\subseteq\R^2\fr\R$ are the admissible functions and $W\colon\R^2\fr\R$ is the Lagrangian energy density. From a physical standpoint, we are dealing with a variational problem concerning a scalar quantity (such as the temperature) on a two-dimensional  body. In this situation, we say that the Lagrangian energy density is invariant with respect to the action of a closed subgroup $\mathcal{G}$ of $\textrm{SO}(2)$ if
	$$ W(Q\,\zeta)=W(\zeta) \qquad\forall\,Q\in\mathcal{G}\;\text{and}\;\zeta\in\R^2. $$
	If we consider the simpler case in which $W$ is a quadratic form in $\zeta$, we have the linear problem
	$$ 
	       \mathscr{A}_{\text{lin}}\left[ f \right] 
	       =
	       \frac{1}{2}\int_\Omega \left\langle \mathbb{C}\,\nabla f(x), \nabla f(x)\right\rangle \,\dx, 
	$$
	where
	$\mathbb{C}\in\textrm{Sym}\!\left(2\right) $, and the invariance condition becomes
	$$
	\left\langle \mathbb{C}\,(Q\,\zeta), Q\,\zeta\right\rangle =\left\langle \mathbb{C}\,\zeta, \zeta\right\rangle \qquad\forall\,Q\in\mathcal{G}\quad\text{and}\quad\forall\,\zeta\in\R^2
	$$
	i.e.
	$$
	\mathbb{C}_{ab}\,Q_{bj}\,\zeta_j\,Q_{ai}\,\zeta_i=\mathbb{C}_{ij}\,\zeta_j\,\zeta_i \qquad\forall\,Q\in\mathcal{G}\quad\text{and}\quad\forall\,\zeta\in\R^2
	$$
	which gives
	$$ 
	     \bC_{ab}\,Q_{ai}\,Q_{bj}=\bC_{ij}
	     \qquad
	     \forall\,Q\in\cG
	     \quad
	     \text{and}
	     \quad
	     \forall\, i,j. 
	$$
	Thus, in this case, the symmetry invariance condition reads as follows: let us consider the triple $\left(\textrm{Sym}(2),\cG,\varphi\right)$,
	where the action $\varphi$ is defined by
	\begin{equation}
	      \varphi:\cG\times\textrm{Sym}(2)\fr\textrm{Sym}(2),
	      \qquad
	      \varphi(Q,\bC)
	      =
	      \widehat{\bC}
	      \quad
	      \textrm{with}
	      \quad
	      \widehat{\bC}_{ij}
	      \defi
	      \bC_{ab}\,Q_{ai}\,Q_{bj}\,.
	      \label{eq:azione matrix}
	\end{equation}
	Then $\mathbb{C}$ respects the symmetry if and only if $\mathbb{C}=\varphi(Q,\mathbb{C})$ for every $Q\in\mathcal{G}$. We are going now to study the class of tensors $\mathbb{C}$ which are invariant with respect to the action of the full group $\textrm{SO}\!\left(2\right)$ following the stated scheme. Setting
	\begin{align*}
	       \varphi_Q:\textrm{Sym}(2)\fr\textrm{Sym}(2),
	       \qquad 	
	       \varphi_Q(\C)_{ij}=\bC_{ab}\,Q_{ai}\,Q_{bj}, 
	\end{align*}
	and denoting with $\mu$ the Haar measure on $\textrm{SO}\!\left(2\right)$, the projection $\mathscr{P}$ is given by
	\begin{align*}
	      \sP(\C)=\left( \int_{Q\in\textrm{SO}(2)}\varphi_Q\,\d\mu\right) \,\C
	      =
	      \int_{\textrm{SO}(2)}\varphi_Q\,\C\,\d\mu\,,
	\end{align*}
	or component-wise by
	\begin{align*}
	       \left( \sP(\C)\right)_{ij} 
	       =
	       \int_{Q\in\textrm{SO}(2)}\C_{ab}Q_{ai}Q_{bj} \,\d\mu\,.
	\end{align*}
	Setting 
	$$ 
	    \psi_\C^{ij}:\textrm{SO}(2)\fr\R,\qquad\psi_\C^{ij}(Q)\defi\C_{ab}Q_{ai}Q_{bj}
	$$
	we can use the formula \eqref{Haar circle} to calculate such integrals.
	 Indeed, we have that
	$$
	     \int_{\textrm{SO}\left(2\right)}\psi_\C^{ij}(Q) \,\d\mu
	     =
	     \frac{1}{2\pi}\int_0^{2\pi}\psi_{\C\,\textrm{per}}^{ij}(\vartheta)\,\d\vartheta,
	$$ 
	where we use the parametrization
	$$
	\vartheta\mapsto Q(\vartheta)=\begin{pmatrix}\cos\,\vartheta & -\sin\,\vartheta\\
	\sin\,\vartheta & \cos\,\vartheta
	\end{pmatrix},\qquad\vartheta\in\left[0,2\pi \right) 
	$$
	of $\textrm{SO}\!\left(2\right)$.
	Thanks to the considered parametrization, we obtain the following expressions for the functions $\psi_{\C\,\textrm{per}}^{ij}(\vartheta)$:
	$$ 
	\psi_{\C\,\textrm{per}}^{ij}(\vartheta)=\C_{ab}Q_{ai}(\vartheta)Q_{bj}(\vartheta)=\left( Q^T(\vartheta)\,(\C\, Q(\vartheta))\right) _{ij}=\left( \begin{pmatrix}\cos\,\vartheta & \sin\,\vartheta\\
	-\sin\,\vartheta & \cos\,\vartheta
	\end{pmatrix}\,\left[ \begin{pmatrix}\C_{11} & \C_{12}\\
	\C_{12} & \C_{22}
	\end{pmatrix}\,\begin{pmatrix}\cos\,\vartheta & -\sin\,\vartheta\\
	\sin\,\vartheta & \cos\,\vartheta
	\end{pmatrix}\right]\right)_{ij}  ,
	$$
	and thus
	\begin{equation*}
	\left\{
	\begin{aligned}
	\psi_{\C\,\textrm{per}}^{11}(\vartheta) &=\C_{11}\,\cos^2\vartheta+\C_{22}\,\sin^2\vartheta+\C_{12}\,\sin\,2\vartheta,
	\\
	\psi_{\C\,\textrm{per}}^{22}(\vartheta) &=\C_{11}\,\sin^2\vartheta+\C_{22}\,\cos^2\vartheta-\C_{12}\,\sin\,2\vartheta,
	\\
	\psi_{\C\,\textrm{per}}^{12}(\vartheta) 
	&
	=
	\frac{1}{2}\left( \C_{22}-\C_{11}\right)\sin\,2\vartheta +\C_{12}\,\cos\,2\vartheta.
	\end{aligned}
	\right. 
	\end{equation*}
	Integrating,
	 we find
	\begin{align*}
	      \left( \sP(\C)\right)_{11} &=\frac{\C_{11}+\C_{22}}{2},&\left( \sP(\C)\right)_{12} &=0,&\left( \sP(\C)\right)_{22} &=\frac{\C_{11}+\C_{22}}{2}.
	\end{align*}
	Thus, recalling that
	$$
	     \textrm{Fix}_{\,{\textrm{SO}(2)}}^{\,\varphi} \,\textrm{Sym}(2)
	     =
	     \graf{
	     	\C\in\textrm{Sym}(2) \stt \sP(\C)=\C 
	     }, 
	$$
	i.e.\ that
	$$ 
	     \sP(\C)
	     =
	     \begin{pmatrix}
	     	\frac{\C_{11}+\C_{22}}{2} & 0
	     	\\
	        0 & \frac{\C_{11}+\C_{22}}{2}
	     \end{pmatrix}
	     =
	     \begin{pmatrix}
	     	\C_{11} & \C_{12}
	     	\\
	        \C_{12} & \C_{22}
	     \end{pmatrix}
	     =
	     \C,
	$$
	and therefore
	$$ \frac{\C_{11}+\C_{22}}{2}=\C_{11}\quad \Leftrightarrow\quad\C_{11}=\C_{22},\qquad\text{and}\qquad\C_{12}=0 $$
	for $\C\in\textrm{Fix}_{\,{\textrm{SO}(2)}}^{\,\varphi} \,\textrm{Sym}(2)$, we find
	$$
	   \textrm{Fix}_{\,{\textrm{SO}(2)}}^{\,\varphi} \,\textrm{Sym}(2)
	   =
	   \grafB{
	   	   \C\in\textrm{Sym}(2)\sttb
	   	       \C
	   	       =
	   	       \begin{pmatrix}
	   	       	   a & 0
	   	       	   \\
	               0& a
	           \end{pmatrix}, \, a\in\R  
       }. 
	$$
	In this case, the dimension of the fixed subspace (namely 1) can be obtained directly, but let us calculate it with the trace formula in order to show how the theoretical machinery works. To do this (as shown in eq.\eqref{Symmetrization identity}), we write down the
	symmetrization identity of $\textrm{Sym}\left(2\right)$,
	\[
	    \Pi^{\text{Sym}(2)}
	    =
	    \frac{1}{2}(\delta_{ia}\delta_{jb}+\delta_{ja}\delta_{ib})\,.
	\]
	Thus, the character $\chi(Q)$ of $Q\in\textrm{SO}\!\left(2\right)$ is
	\begin{align*}
	     \chi\left(Q\right) 
	     & 
	     =
	     \scal{\varphi_Q\circ\Pi^{\text{Sym}(2)}}{\id_2\otimes\id_2}_{\R^{2\times2}}
	     =
	     Q_{ia}Q_{jb}\frac{1}{2}\left(\delta_{ia}\delta_{jb}+\delta_{ja}\delta_{ib}\right)
	     =\frac{1}{2}\left(Q_{ia}Q_{jb}\delta_{ia}\delta_{jb}+Q_{ia}Q_{jb}\delta_{ja}\delta_{ib}\right)
	     \\
	     &
	     =\frac{1}{2}\left((\textrm{tr}\,Q)^{2}+\textrm{tr}\,Q^{2}\right)
	     =\frac{1}{2}\left[\left(2\,\cos\vartheta\right)^{2}+2\,\cos^{2}\vartheta-2\,\sin^{2}\vartheta\right]
	     =3\,\cos^{2}\vartheta-\sin^{2}\vartheta
	\end{align*}
	and therefore
	\begin{align*}
	    \dim\left(\textrm{Fix}_{\,\textrm{SO}(2)}^{\,\varphi}\textrm{Sym}(2)\right) 
	    & 
	    =
	    \underset{Q\in\textrm{SO}(2)}{\int}\chi(Q)\,\d Q
	    =\frac{1}{2\pi}\int_{0}^{2\pi}\left(3\,\cos^{2}\vartheta-\sin^{2}\vartheta\right)\,\d\vartheta
	    =
	    1.
	\end{align*}
	We can also visualize the fixed point subspace in $\R^3$ identifying it with $\textrm{Sym}\left(2\right)$ via the following isometry:
	$$ \Lambda:\textrm{Sym}\left(2\right)\fr \R^3, \qquad  \begin{pmatrix}\C_{11} & \C_{12}\\
	\C_{12} & \C_{22}
	\end{pmatrix}=\C\mapsto\xi_{\,\C}\defi\begin{pmatrix}\C_{11} \\ \C_{22} \\ \sqrt{2}\,\C_{12}
	\end{pmatrix}.$$
	This is analogous to the Mandel-notation (see \cite{mandel1962plastic}) for this simple example. Via the introduced isometry, 
	
	\noindent
	$\Lambda\partonb{ \textrm{Fix}_{\,\textrm{SO}(2)}^{\,\varphi}\textrm{Sym}\left(2\right)}$ is the line in $\R^3$ generated by the vector $(1,1,0)$.
	
	With this example, we want also to illustrate a general result valid in linear representation theory when $\mathcal{G}$ is a closed subgroup of $\textrm{SO}\!\left(3\right)$: the orbit $\mathcal{O}_{\C}$ of an element is contained in the intersection between the sphere of radius $\left\| \C\right\| $ (this is because the matrices $Q$ considered in the action are orthogonal and preserve the norm) and the affine plane centred in its projection $\sP(\C)$ and parallel to the kernel of the projection $\mathscr{P}$ (this intersection is a circle).
	
	Let us so determine $\ker\mathscr{P}$ using the characterization of it as the orthogonal subspace to the image of $\mathscr{P}$. We find that
	$$ \C\in\ker\mathscr{P}\quad \Leftrightarrow\quad \left\langle\begin{pmatrix}\C_{11} & \C_{12}\\
	\C_{12} & \C_{22}
	\end{pmatrix},\begin{pmatrix}1 & 0\\
	0 & 1
	\end{pmatrix} \right\rangle_{\textrm{Sym}\left(2\right)}=0 $$
	is satisfied for
	$\C_{11}=-\C_{22}$ and $\C_{12}\in\R$.
	Therefore
	$$
	    \ker\sP
	    =
	    \grafB{  
	         \C\in\textrm{Sym}(2)\sttB\C=
	              \begin{pmatrix}
	              	 \C_{11} & \C_{12}
	              	 \\
	                 \C_{12} & -\C_{11}
	         \end{pmatrix} 
	    }. 
	$$
	The image of $\ker\mathscr{P}$ in $\R^3$ via $\Lambda$ is thus the plane generated by the vectors $(1,-1,0)$ and $(0,0,1)$.
	
	In order to verify that $\Lambda(\mathcal{O}_\C)$ is contained in the affine plane $\Lambda(\sP(\C)\,+\,\ker\mathscr{P})$, remarking that $\Lambda\left( \textrm{Fix}_{\,\textrm{SO}\left(2\right)}^{\,\varphi}\textrm{Sym}\left(2\right)\right)$ is generated by the vector $(1,1,0)$, it is sufficient to observe that 
	$$ \left\langle \Lambda\,(\varphi(Q,\C))-\Lambda\,(\sP(\C)),\begin{pmatrix}1 \\ 1 \\ 0
	\end{pmatrix} \right\rangle_{\R^3}=0\qquad\forall\,Q\in\textrm{SO}(2).  $$
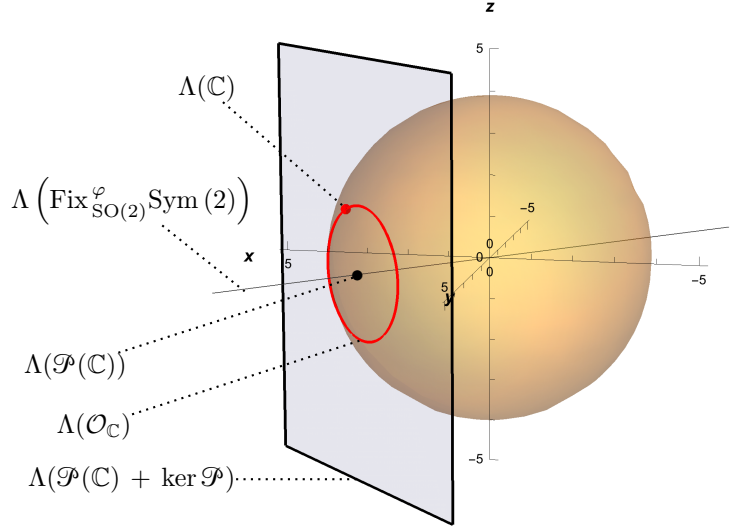
\begin{figure}[H]
	\begin{centering}
		\begin{tikzpicture}
		\tikzset{
			nodeoformula/.style={rectangle,rounded corners=0.1cm,drop shadow={shadow xshift=0.5mm, shadow yshift=-0.5mm,opacity=1},draw=black, top color=white, bottom color=white, thick, inner sep=2mm, minimum size=2em, text centered},
			nodepoint/.style={circle,draw=gray,fill=gray,inner sep=0.8mm}
		}
		\node[anchor=south west,inner sep=0] at (0,0) {\includegraphics[scale=0.55]{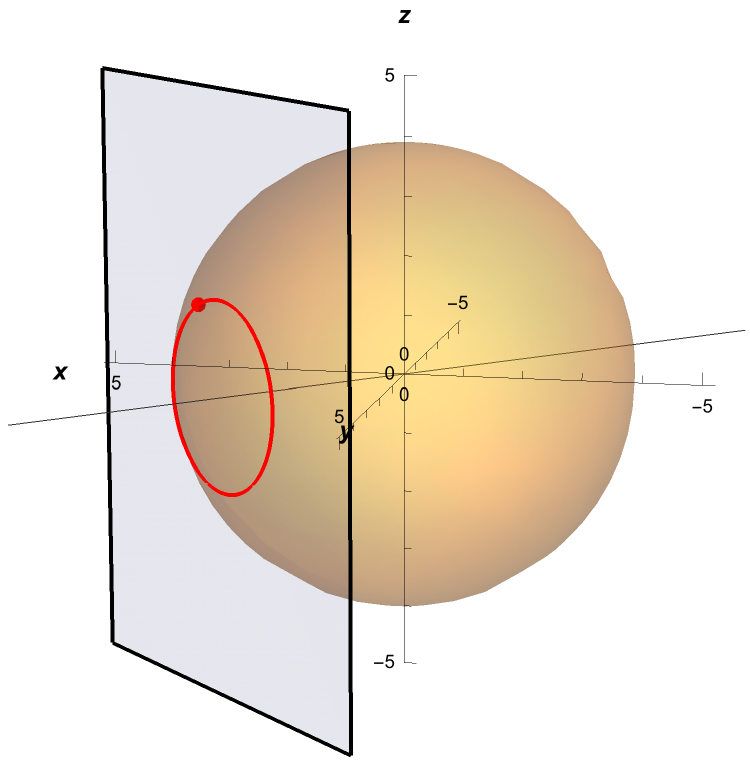}};
		\node[style=none](1)at(0, 6) {$\Lambda(\C)$};
		\node[style=none](2)at(-1.5, 1.5) {$\Lambda(\mathcal{O}_\C)$};
		\node[style=none](3)at(-1, 4.5) {$\Lambda\left( \textrm{Fix}_{\,\textrm{SO}\left(2\right)}^{\,\varphi}\textrm{Sym}\left(2\right)\right)$};
		\node[style=none](4)at(-1, 0.8) {$\Lambda(\sP(\C)\,+\,\ker\mathscr{P})$};
		\node[style=none](5)at(-1.75,2.3) {$\Lambda(\sP(\C))$};
		\draw[thick,dotted] (1) -- (1.85, 4.4);
		\draw[thick,dotted] (2) -- (2.05, 2.65);
		\draw[thick,dotted] (3) -- (0.5, 3.3);
		\draw[thick,dotted] (4) -- (2, 0.8);
		\foreach \Point in {(2, 3.5)}{
			\node at \Point {\textbullet};
		}
	   \draw[thick,dotted] (5) -- (2, 3.5);
		\end{tikzpicture}
		\par\end{centering}
	\caption{\label{Example}Action, orbit and fixed point subspace.}
\end{figure} 	
\end{example}

\subsection{Explicit expression for the Hooke tensor in classical elasticity}
 Applying the previous schema, some of the most common symmetries in material science are obtained. 
\begin{itemize}
	\item \textbf{Orthotropic materials} (9 elastic constants) The symmetry
	group contains all reflections with respect to three orthogonal planes
	and the Hooke tensor is 
	\begin{equation}
		\left(\begin{array}{cccccc}
			\widetilde{\mathbb{C}}_{11} & \widetilde{\mathbb{C}}_{12} & \widetilde{\mathbb{C}}_{13} & 0 & 0 & 0\\
			& \widetilde{\mathbb{C}}_{22} & \widetilde{\mathbb{C}}_{23} & 0 & 0 & 0\\
			&  & \widetilde{\mathbb{C}}_{33} & 0 & 0 & 0\\
			&  &  & \widetilde{\mathbb{C}}_{44} & 0 & 0\\
			& \textrm{sym} &  &  & \widetilde{\mathbb{C}}_{55} & 0\\
			&  &  &  &  & \widetilde{\mathbb{C}}_{66}
		\end{array}\right)
		\,.
	\end{equation}
	\item \textbf{Transversely isotropic materials} (5 elastic constants) The
	symmetry group contains all rotations of a fixed given axis (here
	$e_{3}.$) The general form of the Hooke tensor is
	\begin{equation}
		\left(\begin{array}{cccccc}
			\widetilde{\mathbb{C}}_{11} & \widetilde{\mathbb{C}}_{12} & \widetilde{\mathbb{C}}_{13} & 0 & 0 & 0\\
			& \widetilde{\mathbb{C}}_{11} & \widetilde{\mathbb{C}}_{13} & 0 & 0 & 0\\
			&  & \widetilde{\mathbb{C}}_{33} & 0 & 0 & 0\\
			&  &  & \widetilde{\mathbb{C}}_{44} & 0 & 0\\
			& \textrm{sym} &  &  & \widetilde{\mathbb{C}}_{44} & 0\\
			&  &  &  &  & \widetilde{\mathbb{C}}_{66}
		\end{array}\right)
	\end{equation}
	with $\widetilde{\mathbb{C}}_{11}=\widetilde{\mathbb{C}}_{12}+2\,\widetilde{\mathbb{C}}_{66}.$
	\item \textbf{Cubic materials} (3 elastic constants) The symmetry group
	is that of the symmetries of the cube. The general form of the Hooke
	tensor is
	\begin{equation}
		\left(\begin{array}{cccccc}
			\widetilde{\mathbb{C}}_{11} & \widetilde{\mathbb{C}}_{12} & \widetilde{\mathbb{C}}_{12} & 0 & 0 & 0\\
			& \widetilde{\mathbb{C}}_{11} & \widetilde{\mathbb{C}}_{12} & 0 & 0 & 0\\
			&  & \widetilde{\mathbb{C}}_{11} & 0 & 0 & 0\\
			&  &  & \widetilde{\mathbb{C}}_{44} & 0 & 0\\
			& \textrm{sym} &  &  & \widetilde{\mathbb{C}}_{44} & 0\\
			&  &  &  &  & \widetilde{\mathbb{C}}_{44}
		\end{array}\right)
		\,.
	\end{equation}
	\item \textbf{Isotropic material} (2 elastic constants); the symmetry group contains
	all proper orthogonal rotations. Using the notation of Voigt, the Hooke
	tensor is that of a cubic material but the additional relation $\widetilde{\mathbb{C}}_{11}=\widetilde{\mathbb{C}}_{12}+2\,\widetilde{\mathbb{C}}_{44}$
	holds. 
\end{itemize}

\begin{rem}
	It is possible to classify and obtain all the admissible symmetry classes for 
	$\Ela(3)$. This problem was first successfully addressed in \cite{forte1996symmetry} and subsequently explored in a series of papers \cite{auffray2017handbook,olive2017minimal,olive2013symmetry,olive2014symmetry,abramian2020recovering,azzi2023clips,olive2019effective}, where the authors proposed and developed new methods. These methods allow the generalization of the results to other classes of tensors, such as those arising from second- or strain-gradient theories, piezoelasticity, photoelasticity, etc. In Appendix \ref{Geometry of the space of elasticity}, a shortcut to this problem is presented. Nevertheless, knowledge of representation theory, including irreducible representations, is indispensable.
\end{rem}

\begin{rem}
	The full group $\cG$ that we should consider in the context of studying the symmetry classes in elasticity is, a priori, the orthogonal group $\textrm{O}(3)$, as it is the subgroup of the isometry group of $\bR^3$ that has a fixed point ($\textrm{Isom}(\bR^3) \simeq \bR^3 \rtimes \textrm{O}(3)$). As explained in \cite[par. 2.1]{forte1996symmetry}, since we are dealing with even-order tensors, any tensor in $\Ela(3)$ is invariant w.r.t.\ $-\id \in \textrm{O}(3) \setminus \textrm{SO}(3)$. This, in particular, implies that we can replace $\textrm{O}(3) \simeq \textrm{SO}(3) \times \{\id, - \id\}$ with its connected component,  $\textrm{SO}(3)$. In the 2D case, such reduction is not possible because\footnote{In more precise terms, the subgroup  $\{\pm\id\}$ of $\textrm{O}(n)$, when $n$ is even, is also a subgroup of $\textrm{SO}(n)$.} $-\id \in \textrm{SO}(2)$, and therefore we consider the orthogonal group $\textrm{O}(2)$ as the full group $\cG$.
\end{rem}

\begin{rem}
	There are also other possible approaches to determine the number of independent components and the structure of symmetrized tensors, such as the Clebsch-Gordan formulae. However, this approach requires knowledge of irreducible representations.
\end{rem}

\section{Extended continuum models}

We now extend the previous setting to more general
situations which can occur in mathematical modeling of mechanical
phenomena.

\subsection{Non-symmetric theories}\label{Non symmetric theories2}

In a more general framework including additional degrees of freedom
(see for example \cite{d2017effective,madeo2014band,neff2014unifying}),
a constitutive relation between an arbitrary second-order tensor,
denoted by $P$ and a non-symmetric stress tensor will involve,
in a general case, 45 elastic constants. Using the classical notation
we have 
\begin{equation}
	\sigma_{ij}=\overline{\mathbb{C}}{}_{ijkl}P_{kl},
\end{equation}
and only the major symmetries $\overline{\mathbb{C}}_{ijkl}=\overline{\mathbb{C}}_{klij}$ are assumed.
Hereafter, it will be more convenient to use an extended notation
of Voigt
\begin{equation}\label{geberal Voigt}
	\left(\begin{array}{c}
		\sigma_{11}\\
		\sigma_{22}\\
		\sigma_{33}\\
		\sigma_{23}\\
		\sigma_{32}\\
		\sigma_{13}\\
		\sigma_{31}\\
		\sigma_{12}\\
		\sigma_{21}
	\end{array}\right)=\left(\begin{array}{ccccccccc}
		\widetilde{\mathbb{C}}_{11} & \widetilde{\mathbb{C}}_{12} & \widetilde{\mathbb{C}}_{13} & \widetilde{\mathbb{C}}_{14} & \widetilde{\mathbb{C}}_{15} & \widetilde{\mathbb{C}}_{16} & \widetilde{\mathbb{C}}_{17} & \widetilde{\mathbb{C}}_{18} & \widetilde{\mathbb{C}}_{19}\\
		& \widetilde{\mathbb{C}}_{22} & \widetilde{\mathbb{C}}_{23} & \widetilde{\mathbb{C}}_{24} & \widetilde{\mathbb{C}}_{25} & \widetilde{\mathbb{C}}_{26} & \widetilde{\mathbb{C}}_{27} & \widetilde{\mathbb{C}}_{28} & \widetilde{\mathbb{C}}_{29}\\
		&  & \widetilde{\mathbb{C}}_{33} & \widetilde{\mathbb{C}}_{34} & \widetilde{\mathbb{C}}_{35} & \widetilde{\mathbb{C}}_{36} & \widetilde{\mathbb{C}}_{37} & \widetilde{\mathbb{C}}_{38} & \widetilde{\mathbb{C}}_{39}\\
		&  &  & \widetilde{\mathbb{C}}_{44} & \widetilde{\mathbb{C}}_{45} & \widetilde{\mathbb{C}}_{46} & \widetilde{\mathbb{C}}_{47} & \widetilde{\mathbb{C}}_{48} & \widetilde{\mathbb{C}}_{49}\\
		&  &  &  & \widetilde{\mathbb{C}}_{55} & \widetilde{\mathbb{C}}_{56} & \widetilde{\mathbb{C}}_{57} & \widetilde{\mathbb{C}}_{58} & \widetilde{\mathbb{C}}_{59}\\
		&  &  &  &  & \widetilde{\mathbb{C}}_{66} & \widetilde{\mathbb{C}}_{67} & \widetilde{\mathbb{C}}_{68} & \widetilde{\mathbb{C}}_{69}\\
		&  & \textrm{sym} &  &  &  & \widetilde{\mathbb{C}}_{77} & \widetilde{\mathbb{C}}_{78} & \widetilde{\mathbb{C}}_{79}\\
		&  &  &  &  &  &  & \widetilde{\mathbb{C}}_{88} & \widetilde{\mathbb{C}}_{89}\\
		&  &  &  &  &  &  &  & \widetilde{\mathbb{C}}_{99}
	\end{array}\right)\left(\begin{array}{c}
		P_{11}\\
		P_{22}\\
		P_{33}\\
		P_{23}\\
		P_{32}\\
		P_{13}\\
		P_{31}\\
		P_{12}\\
		P_{21}
	\end{array}\right).
\end{equation}

We note that, in this extended framework, the physical meaning of all $\widetilde{C}_{ij}$ above is obviously different from that of $\widetilde{C}_{ij}$ in \eqref{C44}) for subscripts with $4 \leq i \leq 6$ or $4\leq j \leq 6$.


The character of the representation used in the extended theory can
again be computed using the symmetry properties of the Hooke tensor.
In this case, since only the major symmetry of the Hooke tensor is
involved, the symmetrization identity $\overline{\Pi}$ which projects $\textrm{Lin}(\R^{3\times3},\R^{3\times3})$ to $\textrm{Sym}(\R^{3\times3},\R^{3\times3})$
is 
\begin{equation}
	\overline{\Pi}=\frac{1}{2}\left(\delta_{ia}\delta_{jb}\delta_{kc}\delta_{ld}+\delta_{ka}\delta_{lb}\delta_{ic}\delta_{jd}\right),
\end{equation}
so that the corresponding character is 
$
	\chi(Q)=\frac{1}{2}\partonB{(\textrm{tr}\,Q)^{4}+\partonb{\textrm{tr}(Q^{2})}^{2}}
$
(see the calculations in Appendix \ref{Non symmetric theories}). 
Notice that for $Q=\id_{\R^3}$ we have $\textrm{tr}\,\id_{\R^3}=3$ so
that $\chi(\id_{\R^3})=\frac{1}{2}(3^{4}+3^{2})=45$ which is, as
expected, the number of elastic constants without any additional symmetry.

In the following, we give the details of the computations that provide
the general forms of the Hooke tensor for orthotropic, transversely
isotropic, cubic and isotropic materials. Although the computations
are performed using the fourth-order notations for the Hooke tensor,
the results are presented in the following using the extended Voigt
notation. For each of the symmetry classes above we start with a generic
(45 independent elastic constants) Hooke tensor $\overline\bC$ and use the following procedure.
\begin{enumerate}
	\item[Step 1:] Using the trace formula and the normalized Haar integral (or average
	over the group in the discrete case) we compute the number of independent
	parameters after symmetrization as 
	\begin{equation}
		\dim(\textrm{Fix}_{\,{\mathcal{G}}}^{\,\varphi}\,\textrm{Sym}(\R^{3\times3},\R^{3\times3}))=\int_{\mathcal{G}}\chi(Q)\,\d\mu.
	\end{equation}
	\item[Step 2:] We compute the symmetrization $\mathscr{P}(\overline{\C})$ over the group $\mathcal{G}$.
	Notice that, by definition, the components of $\mathscr{P}(\overline{\C})$
	are linear combinations of elements of ${\overline{\C}}.$ 
	\item[Step 3:] By using standard linear algebra computations, we choose among the
	components of $\mathscr{P}(\overline{\C})$ a basis and
	we check that it contains exactly 
	\[
	\textrm{dim}(\textrm{Fix}_{\,{\cal G}}^{\,\varphi}\,\textrm{Sym}(\R^{3\times3},\R^{3\times3}))
	\]
	elements. 
\end{enumerate}
\begin{itemize}
	\item \textbf{Orthotropic materials:} From the trace formula, the number
	of elastic constants is 15. The general form of the Hooke tensor is
	
	\begin{equation}
		\left(\begin{array}{ccccccccc}
			\widetilde{\mathbb{C}}_{11} & \widetilde{\mathbb{C}}_{12} & \widetilde{\mathbb{C}}_{13} & 0 & 0 & 0 & 0 & 0 & 0\\
			& \widetilde{\mathbb{C}}_{22} & \widetilde{\mathbb{C}}_{23} & 0 & 0 & 0 & 0 & 0 & 0\\
			&  & \widetilde{\mathbb{C}}_{33} & 0 & 0 & 0 & 0 & 0 & 0\\
			&  &  & \widetilde{\mathbb{C}}_{44} & \widetilde{\mathbb{C}}_{45} & 0 & 0 & 0 & 0\\
			&  &  &  & \widetilde{\mathbb{C}}_{55} & 0 & 0 & 0 & 0\\
			&  &  &  &  & \widetilde{\mathbb{C}}_{66} & \widetilde{\mathbb{C}}_{67} & 0 & 0\\
			&  & \textrm{sym} &  &  &  & \widetilde{\mathbb{C}}_{77} & 0 & 0\\
			&  &  &  &  &  &  & \widetilde{\mathbb{C}}_{88} & \widetilde{\mathbb{C}}_{89}\\
			&  &  &  &  &  &  &  & \widetilde{\mathbb{C}}_{99}
		\end{array}\right)
		\,.
	\end{equation}
	
	\item \textbf{Transversely isotropic materials:} In contrast to the case of classical elasticity, when the involved 4-order tensor has only the major symmetry, the actions with respect to the two closed subgroups $\textrm{SO}(2;e_3)$ and $\textrm{O}(2;e_3)$ of $\So$ are not equivalent. In other words, $\textrm{SO}(2;e_3)$ and $\textrm{O}(2;e_3)$ are different symmetry groups for the vector space $\textrm{Sym}(\R^{3\times3},\R^{3\times3})$, see for example \cite{olive2013symmetry}. Recall the definitions of these two groups,
	\begin{align}
	\label{transversal hemitropy}
	    \textrm{SO}(2;e_3)
	    \defi
	    \graf{
	         \begin{pmatrix}
	         	 \cos\theta & -\sin\theta & 0
	         	 \\
	             \sin\theta & \cos\theta & 0
	             \\
	             0 & 0 & 1
	         \end{pmatrix}
	         \sttB
	         \theta\in[0,2\pi)
	    }
	    \quad
	    \text{transversal hemitropic}
	    &, 
	    \\[2mm]
	    \label{transversal isotropy}
	    \textrm{O}(2;e_3)
	    \defi
	    \textrm{SO}(2;e_3)\cup
	    \graf{
	        \begin{pmatrix}
	        	\cos\theta & \sin\theta & 0
	        	\\
	        	\sin\theta & -\cos\theta & 0
	        	\\
	        	0 & 0 & -1
	        \end{pmatrix}
	        \sttB
	        \theta\in[0,2\pi) 
        }
        \quad
        \text{transversal isotropic}&. 
	\end{align}
	From a geometrical point of view \eqref{transversal hemitropy} reflects the invariance with respect to the rotations keeping fixed the axis $e_3$ of $\R^3$ (proper rotations), while the group given in \eqref{transversal isotropy} accounts also for the inversions with respect to the plane $\left\langle e_1,e_2 \right\rangle$ (improper rotations)\footnote{$e_1,e_2,e_3$ are the elements of the canonical orthonormal basis of $\R^3$.}. 
	Via the trace formula and the explicit expression for the Haar measure on $\textrm{SO}(2)$ and $\textrm{O}(2)$ derived in \eqref{Haar circle} and \eqref{Haar O(2)} respectively, we compute the following numbers of independent components for the two considered symmetries:
	\begin{equation}\label{Dimensions transversal}
	    \dim\,\textrm{Fix}_{\,\textrm{SO}(2,e_3)}^{\,\varphi}\textrm{Sym}(\R^{3\times3},\R^{3\times3})
	    =
	    11,
	    \qquad
	    \qquad
	    \quad
	    \dim\,\textrm{Fix}_{\,\textrm{O}(2,e_3)}\textrm{Sym}(\R^{3\times3},\R^{3\times3})
	    =8. 
	\end{equation}
	
	The symmetrization process, for the invariance with respect to $\textrm{SO}(2,e_3)$, gives 
	\begin{equation}
		\left(\begin{array}{ccccccccc}
			\widetilde{\mathbb{C}}_{11} & \widetilde{\mathbb{C}}_{12} & \widetilde{\mathbb{C}}_{13} & 0 & 0 & 0 & 0 & \widetilde{\mathbb{C}}_{18} & -\widetilde{\mathbb{C}}_{18}\\
			& \widetilde{\mathbb{C}}_{11} & \widetilde{\mathbb{C}}_{13} & 0 & 0 & 0 & 0 & \widetilde{\mathbb{C}}_{18} & -\widetilde{\mathbb{C}}_{18}\\
			&  & \widetilde{\mathbb{C}}_{33} & 0 & 0 & 0 & 0 & \widetilde{\mathbb{C}}_{38} & -\widetilde{\mathbb{C}}_{38}\\
			&  &  & \widetilde{\mathbb{C}}_{44} & \widetilde{\mathbb{C}}_{45} & 0 & -\widetilde{\mathbb{C}}_{56} & 0 & 0\\
			&  &  &  & \widetilde{\mathbb{C}}_{55} & \widetilde{\mathbb{C}}_{56} & 0 & 0 & 0\\
			&  &  &  &  & \widetilde{\mathbb{C}}_{44} & \widetilde{\mathbb{C}}_{45} & 0 & 0\\
			&  & \textrm{sym} &  &  &  & \widetilde{\mathbb{C}}_{55} & 0 & 0\\
			&  &  &  &  &  &  & \widetilde{\mathbb{C}}_{88} & \widetilde{\mathbb{C}}_{89}\\
			&  &  &  &  &  &  &  & \widetilde{\mathbb{C}}_{88}
		\end{array}\right)
		\,,
	\end{equation}
	where $\widetilde{\mathbb{C}}_{11}=\widetilde{\mathbb{C}}_{12}+\widetilde{\mathbb{C}}_{88}+\widetilde{\mathbb{C}}_{89}$, while for the invariance with respect to $\textrm{O}(2,e_3)$ we obtain
	\begin{equation}
	\left(\begin{array}{ccccccccc}
	\widetilde{\mathbb{C}}_{11} & \widetilde{\mathbb{C}}_{12} & \widetilde{\mathbb{C}}_{13} & 0 & 0 & 0 & 0 & 0 & 0\\
	& \widetilde{\mathbb{C}}_{11} & \widetilde{\mathbb{C}}_{13} & 0 & 0 & 0 & 0 & 0 & 0\\
	&  & \widetilde{\mathbb{C}}_{33} & 0 & 0 & 0 & 0 & 0 & 0\\
	&  &  & \widetilde{\mathbb{C}}_{44} & \widetilde{\mathbb{C}}_{45} & 0 & 0 & 0 & 0\\
	&  &  &  & \widetilde{\mathbb{C}}_{55} & 0 & 0 & 0 & 0\\
	&  &  &  &  & \widetilde{\mathbb{C}}_{44} & \widetilde{\mathbb{C}}_{45} & 0 & 0\\
	&  & \textrm{sym} &  &  &  & \widetilde{\mathbb{C}}_{55} & 0 & 0\\
	&  &  &  &  &  &  & \widetilde{\mathbb{C}}_{88} & \widetilde{\mathbb{C}}_{89}\\
	&  &  &  &  &  &  &  & \widetilde{\mathbb{C}}_{88}
	\end{array}\right)
	\,,
	\end{equation}
	again with $\widetilde{\mathbb{C}}_{11}=\widetilde{\mathbb{C}}_{12}+\widetilde{\mathbb{C}}_{88}+\widetilde{\mathbb{C}}_{89}$.
	\item \textbf{Cubic materials:} From the trace formula, the number of elastic constants
	in the extended theory is 4. Using the notation of Voigt, the general
	form of the Hooke tensor is 
	\begin{equation}
		\left(\begin{array}{ccccccccc}
			\widetilde{\mathbb{C}}_{11} & \widetilde{\mathbb{C}}_{12} & \widetilde{\mathbb{C}}_{12} & 0 & 0 & 0 & 0 & 0 & 0\\
			& \widetilde{\mathbb{C}}_{11} & \widetilde{\mathbb{C}}_{12} & 0 & 0 & 0 & 0 & 0 & 0\\
			&  & \widetilde{\mathbb{C}}_{11} & 0 & 0 & 0 & 0 & 0 & 0\\
			&  &  & \widetilde{\mathbb{C}}_{44} & \widetilde{\mathbb{C}}_{45} & 0 & 0 & 0 & 0\\
			&  &  &  & \widetilde{\mathbb{C}}_{44} & 0 & 0 & 0 & 0\\
			&  &  &  &  & \widetilde{\mathbb{C}}_{44} & \widetilde{\mathbb{C}}_{45} & 0 & 0\\
			&  & \textrm{sym} &  &  &  & \widetilde{\mathbb{C}}_{44} & 0 & 0\\
			&  &  &  &  &  &  & \widetilde{\mathbb{C}}_{44} & \widetilde{\mathbb{C}}_{45}\\
			&  &  &  &  &  &  &  & \widetilde{\mathbb{C}}_{44}
		\end{array}\right)
	\end{equation}
	\item \textbf{Isotropic materials:} The number of elastic moduli is 3; the
	Hooke tensor using Voigt notation is 
	\begin{equation}\label{isotropic full}
		\left(\begin{array}{ccccccccc}
			\widetilde{\mathbb{C}}_{11} & \widetilde{\mathbb{C}}_{12} & \widetilde{\mathbb{C}}_{12} & 0 & 0 & 0 & 0 & 0 & 0\\
			& \widetilde{\mathbb{C}}_{11} & \widetilde{\mathbb{C}}_{12} & 0 & 0 & 0 & 0 & 0 & 0\\
			&  & \widetilde{\mathbb{C}}_{11} & 0 & 0 & 0 & 0 & 0 & 0\\
			&  &  & \widetilde{\mathbb{C}}_{44} & \widetilde{\mathbb{C}}_{45} & 0 & 0 & 0 & 0\\
			&  &  &  & \widetilde{\mathbb{C}}_{44} & 0 & 0 & 0 & 0\\
			&  &  &  &  & \widetilde{\mathbb{C}}_{44} & \widetilde{\mathbb{C}}_{45} & 0 & 0\\
			&  & \textrm{sym} &  &  &  & \widetilde{\mathbb{C}}_{44} & 0 & 0\\
			&  &  &  &  &  &  & \widetilde{\mathbb{C}}_{44} & \widetilde{\mathbb{C}}_{45}\\
			&  &  &  &  &  &  &  & \widetilde{\mathbb{C}}_{44}
		\end{array}\right)
	\end{equation}
	with $\widetilde{\mathbb{C}}_{11}=\widetilde{\mathbb{C}}_{12}+\widetilde{\mathbb{C}}_{44}+\widetilde{\mathbb{C}}_{45}.$
	This relation generalizes the well-known formula $\widetilde{\mathbb{C}}_{11}=\widetilde{\mathbb{C}}_{12}+2\,\widetilde{\mathbb{C}}_{44}$
	from classical elasticity. We can also write the constitutive
	relation as 
	\begin{equation}
		\textrm{sym}\,\sigma
		=
		2\,\mu\,\textrm{sym}\,P
		+
		\lambda(\textrm{tr}\,P)\id,
		\qquad
		\textrm{skew}\,\sigma
		=
		2\,\mu_{c}\,\textrm{skew}\,P
		\,,
	\end{equation}
	where 
	\begin{equation}\label{new parameters}
	\lambda=\widetilde{\mathbb{C}}_{12},\quad \mu=\widetilde{\mathbb{C}}_{11}-\widetilde{\mathbb{C}}_{12}-\frac{(\widetilde{\mathbb{C}}_{44}+\widetilde{\mathbb{C}}_{55})}{2}=\frac{(\widetilde{\mathbb{C}}_{44}+\widetilde{\mathbb{C}}_{45})}{2}\quad
	\text{and}\quad\mu_{c}=\frac{(\widetilde{\mathbb{C}}_{44}-\widetilde{\mathbb{C}}_{45})}{2}.
	\end{equation} 
	Indeed, thanks to the relations in \eqref{new parameters}, we have
	\begin{equation}\label{matrix new parameters}
		\left(\begin{array}{c}
	\sigma_{11}\\
	\sigma_{22}\\
	\sigma_{33}\\
	\sigma_{23}\\
	\sigma_{32}\\
	\sigma_{13}\\
	\sigma_{31}\\
	\sigma_{12}\\
	\sigma_{21}
	\end{array}\right)=	\left(\begin{array}{ccccccccc}
	2\mu+\lambda & \lambda & \lambda & 0 & 0 & 0 & 0 & 0 & 0\\
	\lambda& 2\mu+\lambda & \lambda & 0 & 0 & 0 & 0 & 0 & 0\\
	\lambda& \lambda & 2\mu+\lambda & 0 & 0 & 0 & 0 & 0 & 0\\
	0& 0 & 0 & \mu+\mu_c & \mu-\mu_c & 0 & 0 & 0 & 0\\
	0& 0 & 0 & \mu-\mu_c & \mu+\mu_c & 0 & 0 & 0 & 0\\
	0& 0 & 0 & 0 & 0 & \mu+\mu_c  & \mu-\mu_c & 0 & 0\\
	0& 0 & 0 & 0 & 0 & \mu-\mu_c & \mu+\mu_c  & 0 & 0\\
	0& 0 & 0 & 0 & 0 & 0 & 0 & \mu+\mu_c & \mu-\mu_c\\
	0& 0 & 0 & 0 & 0 & 0 & 0 & \mu-\mu_c & \mu+\mu_c
	\end{array}\right)\left(\begin{array}{c}
	P_{11}\\
	P_{22}\\
	P_{33}\\
	P_{23}\\
	P_{32}\\
	P_{13}\\
	P_{31}\\
	P_{12}\\
	P_{21}
	\end{array}\right),
	\end{equation}
	which gives
	\begin{align}
		\left(\begin{array}{c}
	\sigma_{11}\\
	\sigma_{22}\\
	\sigma_{33}\\
	\sigma_{23}\\
	\sigma_{32}\\
	\sigma_{13}\\
	\sigma_{31}\\
	\sigma_{12}\\
	\sigma_{21}
	\end{array}\right)&=
	\left(\begin{array}{c}
	(2\mu+\lambda) P_{11}+\lambda P_{22}+\lambda P_{33}\\
	\lambda P_{11}+(2\mu+\lambda) P_{22}+\lambda P_{33}\\
	\lambda P_{11}+\lambda P_{22}+(2\mu+\lambda) P_{33}\\
	\mu(P_{23}+P_{32})+\mu_c(P_{23}-P_{32})\\
	\mu(P_{23}+P_{32})+\mu_c(P_{23}-P_{32})\\
	\mu(P_{13}+P_{31})+\mu_c(P_{13}-P_{31})\\
	\mu(P_{13}+P_{31})+\mu_c(P_{13}-P_{31})\\
	\mu(P_{12}+P_{21})+\mu_c(P_{12}-P_{21})\\
	\mu(P_{12}+P_{21})+\mu_c(P_{12}-P_{21})
	\end{array}\right)=
	\left(\begin{array}{c}
	2\mu P_{11}+\lambda \tr\,P\\
	2\mu P_{22}+\lambda \tr\,P\\
	2\mu P_{33}+\lambda \tr\,P\\
	\mu(P_{23}+P_{32})+\mu_c(P_{23}-P_{32})\\
	\mu(P_{23}+P_{32})+\mu_c(P_{23}-P_{32})\\
	\mu(P_{13}+P_{31})+\mu_c(P_{13}-P_{31})\\
	\mu(P_{13}+P_{31})+\mu_c(P_{13}-P_{31})\\
	\mu(P_{12}+P_{21})+\mu_c(P_{12}-P_{21})\\
	\mu(P_{12}+P_{21})+\mu_c(P_{12}-P_{21})
	\end{array}\right)\\
	&=2\mu\left(\begin{array}{c}
	P_{11}\\
	P_{22}\\
	P_{33}\\
	\frac{P_{23}+P_{32}}{2}\\
	\frac{P_{23}+P_{32}}{2}\\
	\frac{P_{13}+P_{31}}{2}\\
	\frac{P_{13}+P_{31}}{2}\\
	\frac{P_{12}+P_{21}}{2}\\
	\frac{P_{12}+P_{21}}{2}
	\end{array}\right)+\lambda \tr P\left(\begin{array}{c}
	1\\
	1\\
	1\\
	0\\
	0\\
	0\\
	0\\
	0\\
	0
	\end{array}\right)+2\mu_c\left(\begin{array}{c}
	0\\
	0\\
	0\\
	\frac{P_{23}-P_{32}}{2}\\
	\frac{P_{23}-P_{32}}{2}\\
	\frac{P_{13}-P_{31}}{2}\\
	\frac{P_{13}-P_{31}}{2}\\
	\frac{P_{12}-P_{21}}{2}\\
	\frac{P_{12}-P_{21}}{2}
	\end{array}\right),
	\end{align}
	i.e.
	\begin{equation}
	     \sigma
	     =
	     2\,\mu\,\sym P
	     +
	     2\,\mu_c\,\skew P
	     + 
	     \lambda\tr(P)\,\id\,.
	\end{equation}
	Thus, in this case, the elastic energy can be expressed as
	\begin{align}
	    \frac{1}{2}\langle \overline{\bC}\,P, P\rangle
	    &
	    =
	    \frac{1}{2}\langle \sigma, P\rangle
	    =
	    \frac{1}{2}
	    \langle\,2\,\mu\,\sym P
	    + \lambda\tr(P)\,\id
	    + 2\mu_c\,\skew P, P
	    \rangle
	    \\
	    &
	    = 
	    \mu\left\langle \,\sym P, P\right\rangle 
	    + \mu_c\left\langle \skew P, P\right\rangle  
	    + \frac{\lambda}{2}\,\tr(P)\left\langle \id,P\right\rangle
	    =
	    \mu\left\| \sym P \right\|^2
	    + \mu_c\,\left\| \skew P \right\|^2
	    + \frac{\lambda}{2}\tr(P)^2.
	    \nonumber 
	\end{align}
	
\end{itemize}

\subsection{Second gradient theory}

In the following, we want to show how the general framework proposed
in this paper can be easily applied to the second gradient theories which, in recent years, have found wide use in modeling physical phenomena (see\cite{batra1975heat,huang1996energy,ferretti2014modeling,madeo2016continuum,yang1995second}, for example). First, however, we would like to dedicate some words to a more detailed discussion on how the invariance conditions are derived from the Lagrangian energy density in this case. Indeed, concerning the second gradient theory, some authors derived an incorrect invariance law \cite{elzanowski1992symmetry,murdoch1979symmetry} and, to our knowledge, the correct answer is given in \cite{munch2018rotational}. Thus, we follow \cite{munch2018rotational}  to explain the situation:  let us consider a (non-linear) energy density
$$
W(\nabla\varphi,\nabla^2\varphi)
$$
depending also on the second gradient of the displacement field $\varphi:\Omega\subseteq\R^n\fr\R^n$.
Considering a generic orientation preserving diffeomorphism $\zeta\colon\R^n\fr\R^n$, $\xi\mapsto\zeta(\xi)$, we say that the considered body is invariant with respect to $\zeta$ if the total energy is preserved under a diffeomorphic change of variables, i.e.
\begin{equation} \label{invariance condition second gradient}
      \int_{\xi\in\zeta^{-1}(\Omega)}W\left( \nabla_\xi\varphi(\zeta(\xi)),\nabla^2_\xi\varphi(\zeta(\xi)) \right) \left| \det\nabla_\xi\zeta(\xi)\right| \textrm{d}\xi
      = 
      \int_{x\in\Omega}W\left( \nabla_x\varphi(x),\nabla^2_x\varphi(x) \right) \dx. 
\end{equation}
Since material invariance means invariance of the elastic response with respect to rotations of the specimen with respect to the machine test, 
we need to ask for the following constraint for the change of variables:
$$ \nabla_\xi\zeta(\xi)\in\So \qquad\forall\,\xi\in\zeta^{-1}(\Omega)\,.  $$
Under this hypothesis, we can firstly remark that
$ 
     \det\nabla_\xi\zeta(\xi)=1
$
for all $\xi\in\zeta^{-1}(\Omega)$.
Secondly, setting
$ 
    \varphi^\flat(\xi)\defi\varphi(\zeta(\xi)) 
$
for all $\xi\in\zeta^{-1}(\Omega)$, 
and observing that from the identity
$ 
    x=\zeta(\zeta^{-1}(x))  
$
we obtain the identity
$$ \id=\nabla_\xi\big[ \zeta(\underbrace{\zeta^{-1}(x)}_{=\,\xi} )\big]\,  \nabla_x\zeta^{-1}(x)\quad\Longleftrightarrow\quad\nabla_x\zeta^{-1}(x)=\left[ \nabla_\xi\zeta(\xi) \right]^{-1},  $$
expanding the expressions $\nabla_\xi\varphi(\zeta(\xi))$ and $\nabla^2_\xi\varphi(\zeta(\xi))$  we arrive at
$
     \nabla_\xi\varphi(\zeta(\xi))
     = 
     \nabla_x\varphi(\zeta(\xi))\,\nabla_\xi\zeta(\xi)= \nabla_x\varphi(x)\,\nabla_\xi\zeta(\xi)
$
and
\begin{align}
     \nabla^2_\xi\varphi(\zeta(\xi))
     \nonumber
     &=
     \nabla_\xi\left[ \nabla_x\varphi(\zeta(\xi))\,\nabla_\xi\zeta(\xi)\right]= \left[ \nabla_\xi\nabla_x\varphi(\zeta(\xi))\right]\,\nabla_\xi\zeta(\xi)+\nabla_x\varphi(\zeta(\xi))\,\nabla^2_\xi\zeta(\xi)
     \nonumber
     \\
     &=\left[ \nabla^2_x\varphi(\zeta(\xi))\,\nabla_\xi\zeta(\xi)\right]\odot\nabla_\xi\zeta(\xi)+\nabla_x\varphi(x)\,\nabla^2_\xi\zeta(\xi)
     \nonumber
     \\
     &=\left[ \nabla^2_x\varphi(x)\,\nabla_\xi\zeta(\xi)\right]\odot\nabla_\xi\zeta(\xi)+\nabla_x\varphi(x)\,\nabla^2_\xi\zeta(\xi),
     \nonumber
\end{align}
where the contraction $\odot$ is defined by
$ 
        (A\odot B)_{ijk}=A_{imk}B_{mj}. 
$
for $A\in\R^{3\times3\times3}$ and $B\in\R^{3\times3}$.
Thus in our case, we have
$$ 
       \left( \left[ \nabla^2_x\varphi(x)\,\nabla_\xi\zeta(\xi)\right]\odot\nabla_\xi\zeta(\xi)\right)_{ijk}
       =
       \frac{\partial^2\varphi_i}{\partial x_n\partial x_m}\left(\zeta(\xi) \right)  \frac{\partial\zeta_n}{\partial \xi_k}(\xi)\frac{\partial\zeta_m}{\partial \xi_j}(\xi) \qquad\forall\,\xi\in\zeta^{-1}(\Omega).  
$$
Due to a rigidity result (see \cite{john1961rotation} and, for a modern version, \cite{neff2008curl}) asserting that if $\zeta$ is sufficiently regular $\big($for example in $W^{1,2}(\zeta^{-1}(\Omega))\big)$ and such that $\nabla_\xi\zeta(\xi)\in\So$ for every $\xi$, then there exists one constant $\overline{Q}\in\So$ such that $\nabla_\xi\zeta(\xi)=\overline{Q}$ for all $\xi$. This means that the term $ \nabla^2_\xi\zeta(\xi) $ is $0$ and the invariance condition \eqref{invariance condition second gradient} reduces to 
\begin{equation}\label{non linear invariance condition second gradient}
     \int_{x\in\Omega}W\left( \nabla_x\varphi(x),\nabla^2_x\varphi(x) \right) \dx
     =
     \cor{
     \int_{\xi\in\zeta^{-1}(\Omega)}
         W
         \big( 
         \nabla_x\varphi(\zeta(\xi))\,\overline{Q}
         ,
         \left[ \nabla^2_x\varphi(\zeta(\xi))\,\overline{Q}\right]\odot\overline{Q} 
         \big)  
     \textrm{d}\xi\,.}
\end{equation}
In the linearized boundary value homogeneous problem, the internal energy is given as the functional
\begin{equation}
	\tcbhighmath[drop fuzzy shadow]{
		\mathscr{A}\left[u\right]
		\defi
		\frac{1}{2}
		\int_{\Omega}\left(
		    \left\langle \C\,\varepsilon,\varepsilon\right\rangle 
		    +2\left\langle \mathbb{H}\,\varepsilon,\nabla^{2}u\right\rangle +\left\langle \mathbb{G}\,\nabla^{2}u,\nabla^{2}u\right\rangle 
		    \right)\dx,
		\label{eq:pb1}
		}
\end{equation}
where $\varepsilon\defi\sym\,\nabla u$ and 
\begin{align*}
	\C & \in\textrm{Sym}\left(\Sym,\Sym\right), & \mathbb{H} & \in\textrm{Lin}\left(\Sym,\R^{3}\otimes\Sym\right), & \mathbb{G} & \in\textrm{Sym}\left(\R^{3}\otimes\Sym,\R^{3}\otimes\Sym\right).
\end{align*}
We shall specify the invariance condition \eqref{non linear invariance condition second gradient} to the linearized case in \eqref{eq:pb1}.
Denoting with $\omega\colon\Omega\fr\so$ the skew-symmetric part of $\nabla u$,
we find that
\[
\nabla^{2}u\colon\Omega\fr\R^{3}\otimes\Sym\subseteq\R^{3\times3\times3}
\]
can be represented as the sum
\begin{equation}
	\nabla^{2}u=\nabla\left(\nabla u\right)=\nabla\left(\varepsilon+\omega\right)=\nabla\varepsilon+\nabla\omega,\label{eq:decomposition sum}
\end{equation}
with
\begin{align*}
	\nabla\varepsilon & \colon\Omega\fr\Sym\otimes\R^{3}\subseteq\R^{3\times3\times3}, & \nabla\omega & \colon\Omega\fr\so\otimes\R^{3}\subseteq\R^{3\times3\times3}, & \R^{3\times3\times3} & \simeq\left(\Sym\otimes\R^{3}\right)\bigoplus\left(\so\otimes\R^{3}\right),
\end{align*}
and 
\begin{align*}
	\dim\left(\Sym\otimes\R^{3}\right) & =18, & \dim\left(\so\otimes\R^{3}\right) & =9.
\end{align*}

\begin{rem}\label{rem:strain gradient}
	There exists a linear operator $L$ such that $\text{D}^2u=L(\text{D}\varepsilon)$. Indeed,
	$
	    u_{k,ij}
	    =
	    \varepsilon_{ik,j}
	    +
	    \varepsilon_{kj,i}
	    -
	    \varepsilon_{ji,k}
	$
	for every $ i,j,k\in\{1,2,3\}$.
\end{rem}

\begin{rem}
	The tensor product $\otimes$ is not symmetric, so the two spaces $\Sym\otimes\R^{3}$
	and $\R^{3}\otimes\Sym$ are not the same space (but they are isomorphic
	having the same dimension). Denoting with $\textrm{Sym}\left(3,3\right)$
	the space of fully symmetric third-order tensors, i.e. 
	\[
	\textrm{Sym}\left(3,3\right)\defi\left\{ T\in\R^{3\times3\times3}\left.\right|T_{ijh}=T_{jih}=T_{ihj}\:\:\forall i,j,h\in\left\{ 1,2,3\right\} \right\} ,
	\]
	which is a 10-dimensional vector space, we have that $\textrm{Sym}\left(3,3\right)=\left(\Sym\otimes\R^{3}\right)\cap\left(\R^{3}\otimes\Sym\right).$
	Decomposing the identity of $\R^{3\times3\times3}$ using the
	two projections $\pr_{\,\Sym\otimes\R^{3}}$ and $\pr_{\,\text{\ensuremath{\so}}\otimes\R^{3}}$
	associated respectively to the two orthogonal subspaces $\Sym\otimes\R^{3}$
	and $\so\otimes\R^{3}$, i.e.
	\[
	    \id_{\R^{3\times3\times3}}
	    =
	    \pr_{\,\Sym\otimes\R^{3}}
	    +
	    \pr_{\,\text{\ensuremath{\so}}\otimes\R^{3}},
	\]
	we have that 
	\[
	    \nabla^{2}u
	    =
	    \nabla\varepsilon+\nabla\omega
	    =
	    \pr_{\,\Sym\otimes\R^{3}}(\nabla^{2}u)
	    +
	    \pr_{\,\text{\ensuremath{\so}}\otimes\R^{3}}(\nabla^{2}u).
	\]
\end{rem}
The second gradient model can be also formulated considering the gradient
of $\varepsilon$ (strain gradient theories, see Remark \ref{rem:strain gradient} (cf.~\cite{neff2009subgrid})).
In this case the action functional defining the problem is given by
\begin{equation}
	  \tcbhighmath[drop fuzzy shadow]{
	  	     \underline{\mathscr{A}}\left[u\right]:=\frac{1}{2}\int_{\Omega}
	  	     \partonb{
	  	     	     \left\langle \C\,\varepsilon,\varepsilon\right\rangle 
	  	     	     +2\left\langle \underline{\mathbb{H}}\,\varepsilon,\nabla\varepsilon\right\rangle 
	  	     	     +\left\langle \underline{\mathbb{G}}\,\nabla\varepsilon,\nabla\varepsilon\right\rangle 
	  	     } \dx,
	  	     \label{eq:pb2}
	  	     }
\end{equation}
with
\begin{align*}
	\underline{\mathbb{H}} & :\Sym\fr\Sym\otimes\R^{3}, & \underline{\mathbb{G}} & :\Sym\otimes\R^{3}\fr\Sym\otimes\R^{3}.
\end{align*}
We will study the symmetrizations of the tensors $\mathbb{H},\mathbb{G},\underline{\mathbb{H}},\underline{\mathbb{G}}$.
In order to apply the mathematical tools introduced in Section \ref{Preliminaries},
we have to identify, in this setting, the elements of the
triple $\left(V,\G,\varphi\right)$. Let us introduce
\begin{align*}
	V_{0} & \defi\textrm{Sym}\left(\Sym,\Sym\right) & V_{1} & \defi\textrm{Lin}\left(\Sym,\R^{3}\otimes\Sym\right) & V_{2} & \defi\textrm{Sym}\left(\R^{3}\otimes\Sym,\R^{3}\otimes\Sym\right)\\
	&  & \underline{V}_{\,1} & \defi\textrm{Lin}\left(\Sym,\Sym\otimes\R^{3}\right) & \underline{V}_{\,2} & \defi\textrm{Sym}\left(\Sym\otimes\R^{3},\Sym\otimes\R^{3}\right),
\end{align*}
and define
$
	V  \defi V_{0}\times V_{1}\times V_{2}
$
and
$
    \underline{V} \defi V_{0}\times\underline{V}_{\,1}\times\underline{V}_{\,2}.
$
Introducing the actions
\begin{align}\label{second gradient invariance}
	\varphi_{0} & :{\cal G}\times V_{0}\rightarrow V_{0}, & \varphi_{0}(Q,\mathbb{C}) & =\widehat{\mathbb{C}}, & \widehat{\mathbb{C}}_{abcd} & =Q_{ai}Q_{bj}Q_{ck}Q_{dl}\mathbb{C}_{ijkl},\nonumber \\
	\varphi_{1} & :{\cal G}\times V_{1}\rightarrow V_{1}, & \varphi_{1}(Q,\mathbb{H}) & =\widehat{\mathbb{H}}, & \widehat{\mathbb{H}}_{abcde} & =Q_{ai}Q_{bj}Q_{ck}Q_{dl}Q_{em}\mathbb{H}_{ijklm},\nonumber \\
	\varphi_{2} & :{\cal G}\times V_{2}\rightarrow V_{2}, & \varphi_{2}(Q,\mathbb{G}) & =\widehat{\mathbb{G}}, & \widehat{\mathbb{G}}_{abcdef} & =Q_{ai}Q_{bj}Q_{ck}Q_{dl}Q_{em}Q_{fn}\mathbb{G}_{ijklmn}\\
	\underline{\varphi}_{\,1} & :{\cal G}\times\underline{V}_{\,1}\rightarrow\underline{V}_{\,1}, & \underline{\varphi}_{1}(Q,\underline{\mathbb{H}}) & =\widehat{\underline{\mathbb{H}}}, & \widehat{\underline{\mathbb{H}}}_{abcde} & =Q_{ai}Q_{bj}Q_{ck}Q_{dl}Q_{em}\underline{\mathbb{H}}_{ijklm},\nonumber \\
	\underline{\varphi}_{\,2} & :{\cal G}\times\underline{V}_{\,2}\rightarrow\underline{V}_{\,2}, & \underline{\varphi}_{2}(Q,\underline{\mathbb{G}}) & =\widehat{\underline{\mathbb{G}}}, & \widehat{\underline{\mathbb{G}}}_{abcdef} & =Q_{ai}Q_{bj}Q_{ck}Q_{dl}Q_{em}Q_{fn}\underline{\mathbb{G}}_{ijklmn},\nonumber 
\end{align}
and setting 
\begin{align*}
	\varphi 
	& \defi\varphi_{0}\times\varphi_{1}\times\varphi_{2}, 
	& \varphi:{\cal G}\times V\rightarrow V,\quad\varphi(Q,\left(\mathbb{C},\mathbb{H},\mathbb{G}\right))\defi\left(\varphi_{0}(Q,\mathbb{C}),\varphi_{1}(Q,\mathbb{H}),\varphi_{2}(Q,\mathbb{G})\right)=\left(\widehat{\mathbb{C}},\widehat{\mathbb{H}},\widehat{\mathbb{G}}\right),
	\\
	\underline{\varphi} 
	& 
	\defi\varphi_{0}\times\underline{\varphi}_{\,1}\times\underline{\varphi}_{\,2}, 
	& 
	\underline{\varphi}:{\cal G}\times\underline{V}\rightarrow\underline{V},\quad\underline{\varphi}(Q,\left(\mathbb{C},\underline{\mathbb{H}},\underline{\mathbb{G}}\right))\defi\left(\varphi_{0}(Q,\mathbb{C}),\varphi_{1}(Q,\underline{\mathbb{H}}),\varphi_{2}(Q,\underline{\mathbb{G}})\right)=\left(\widehat{\mathbb{C}},\widehat{\underline{\mathbb{H}}},\widehat{\underline{\mathbb{G}}}\right),
\end{align*}
we consider the two following triples for problems (\ref{eq:pb1})
and (\ref{eq:pb2}) respectively:
$
    \left(V,\G,\varphi\right)
$
and
$
    (\underline{V},\G,\underline{\varphi}).
$
Initially, we will ascertain the number of independent components for the tensors in question, namely,  $\mathbb{H},\mathbb{G},\underline{\mathbb{H}},\underline{\mathbb{G}}$,
when subjected to their respective actions $\varphi_{1},\varphi_{2},\underline{\varphi}_{1},\underline{\varphi}_{2}$.
Subsequently, we will determine the matrix representations of these symmetrized tensors, focusing on the cubic symmetry,
extending the classical Voigt notation to the case of  third order tensors.

\paragraph{{\Large Case} $\left(V_{\,1},\protect\G,\varphi\right)$\,:}

The number of independent components of $\nabla^{2}u$ is $18$, the
symmetries of which are 
\[
u_{i,jk}=u_{i,kj},\qquad\textrm{with }i,j,k=1,\ldots,3\,.
\]
In this way the tensor $\mathbb{H}$ has the symmetries
\begin{equation}
	\mathbb{H}_{abcde}=\mathbb{H}_{acbde}=\mathbb{H}_{abced},\label{symmetries H}
\end{equation}
and the set of index permutations which leaves $\mathbb{H}$ invariant
is equivalent to $S_{2}\times S_{2}$ ($S_{2}$ is the group of the
permutations of two indices), so it has 4 different elements. In
order to determine the character $\chi(Q)$ of an element $Q\in\G$
we have to specify the identity tensor $\id_{V_{\,1}}$ of $V_{1}$.
Due to the symmetries (\ref{symmetries H}), the symmetrization identity
$\Pi^{\,V_{1}}$ is
\begin{align*}
	\Pi^{\,V_{1}}_{iajbkcldme} & =\frac{1}{4}(\delta_{ia}\delta_{jb}\delta_{kc}\delta_{ld}\delta_{me}+\delta_{ia}\delta_{kb}\delta_{jc}\delta_{ld}\delta_{me}+\delta_{ia}\delta_{jb}\delta_{kc}\delta_{md}\delta_{le}+\delta_{ia}\delta_{kb}\delta_{jc}\delta_{md}\delta_{le})
	\,.
\end{align*}
The character $\chi(Q)$, being defined as $\left\langle \varphi_{Q}\circ\Pi^{\,V_{1}},\otimes^5\id_2\right\rangle $
with $\varphi_{Q}=\otimes^5 Q$ is (see the Appendix for detailed calculations)
\begin{equation}
	 \tcbhighmath[drop fuzzy shadow]{
	 	     \scaln{\varphi_{Q}\circ\Pi^{\,V_{1}}}{\otimes^5\id_2} 
	 	     =
	 	     \frac{1}{4}\left((\textrm{tr}\,Q)^{5}+2\,(\textrm{tr}\,Q)^{3}\textrm{tr}\,Q^{2}+\textrm{tr}\,Q\left(\textrm{tr}\,Q^{2}\right)^{2}\right),
	 }
\end{equation}
which, due to the Hamilton-Cayley transformation, becomes 
\begin{equation}
	\chi(Q)=(\textrm{tr}\,Q)^{5}-2\,(\textrm{tr}\,Q)^{4}+(\textrm{tr}\,Q)^{3}.\label{H-1}
\end{equation}
A very useful check to verify the accuracy of the performed calculations
consists in computing the character of the matrix $Q=\id$. In this
case, if the derived formula (\ref{H-1}) is correct, we have to find
the dimension of the vector space $V_{1}$. In our case, having that
$\dim\,\Sym\otimes\R^{3}=18$, the dimension of $V_{1}$ is $18\times6=108$,
and indeed
\begin{align*}
	\chi(\id) & =(\textrm{tr}\,\id)^{5}-2\,(\textrm{tr}\,\id)^{4}+(\textrm{tr}\,\id)^{3}=108\,.
\end{align*}

\paragraph{{\Large Case} $\left(\underline{V}_{\,1},\protect\G,\underline{\varphi}\right)$\,:}

The number of independent components of $\nabla\varepsilon$ is $18$,
the symmetries of which are 
\[
\varepsilon_{ij,k}=\varepsilon_{ji,k},\qquad\textrm{with }i,j,k=1,\ldots,3\,.
\]
The tensor $\underline{\mathbb{H}}$ has the symmetries
\[
\underline{\mathbb{H}}_{abcde}=\underline{\mathbb{H}}_{bacde}=\underline{\mathbb{H}}_{abced},
\]
and the set of index permutations which leaves $\underline{\mathbb{H}}$
invariant is still equivalent to $S_{2}\times S_{2}$. The symmetrized
identity tensor $\Pi\,^{\underline{V}_{\,1}}$ of $\underline{V}_{\,1}$
is
\begin{align*}
	\Pi\,^{\underline{V}_{\,1}}_{iajbkcldme} & =\frac{1}{4}(\delta_{ia}\delta_{jb}\delta_{kc}\delta_{ld}\delta_{me}+\delta_{ja}\delta_{ib}\delta_{kc}\delta_{ld}\delta_{me}+\delta_{ia}\delta_{jb}\delta_{kc}\delta_{md}\delta_{le}+\delta_{ja}\delta_{ib}\delta_{kc}\delta_{md}\delta_{le}.
\end{align*}
The character $\chi(Q)$ is 
\begin{equation}
	   \tcbhighmath[drop fuzzy shadow]{
	   	\chi_Q
	   	=
	   	\scaln{\varphi_{Q}\circ\Pi\,^{\underline{V}_{\,1}}}{\otimes^5\id_2}   
	   	=
	   	\frac{1}{4}\left((\textrm{tr}\,Q)^{5}+2\,(\textrm{tr}\,Q)^{3}\textrm{tr}\,Q^{2}+\textrm{tr}\,Q\left(\textrm{tr}\,Q^{2}\right)^{2}\right),}
\end{equation}
which, using to the Hamilton-Cayley transformation, can be expressed as
\begin{equation}
	\chi(Q)=(\textrm{tr}\,Q)^{5}-2\,(\textrm{tr}\,Q)^{4}+(\textrm{tr}\,Q)^{3}.\label{Hbar}
\end{equation}

\paragraph{{\Large Case} $\left(V_{2},\protect\G,\varphi_{2}\right)$\,:}

The tensor $\mathbb{G}$ has the symmetries

\begin{equation}
	\mathbb{G}_{abcdef}=\mathbb{G}_{acbdef}=\mathbb{G}_{abcdfe}=\mathbb{G}_{defabc},\label{eq:simmetrie G-1}
\end{equation}
and the set of index permutations which leaves $\mathbb{G}$ invariant
is equivalent to $S_{2}\times S_{2}\times S_{2}$ and has 8
different elements. The symmetrization identity tensor $\Pi^{\,V_{2}}$
is
\begin{align*}
	\Pi^{\,V_{2}}_{iajbkcldmenf} & =\frac{1}{8}(\delta_{ia}\delta_{jb}\delta_{kc}\delta_{ld}\delta_{me}\delta_{nf}+\delta_{ia}\delta_{kb}\delta_{jc}\delta_{ld}\delta_{me}\delta_{nf}+\delta_{ia}\delta_{jb}\delta_{kc}\delta_{ld}\delta_{ne}\delta_{mf}+\delta_{ia}\delta_{kb}\delta_{jc}\delta_{ld}\delta_{ne}\delta_{mf}\\
	& \quad+\delta_{la}\delta_{mb}\delta_{nc}\delta_{id}\delta_{je}\delta_{kf}+\delta_{la}\delta_{nb}\delta_{mc}\delta_{id}\delta_{je}\delta_{kf}+\delta_{la}\delta_{mb}\delta_{nc}\delta_{id}\delta_{ke}\delta_{jf}+\delta_{la}\delta_{nb}\delta_{mc}\delta_{id}\delta_{ke}\delta_{jf}).
\end{align*}
The character $\chi(Q)$, defined as 
$\scaln{\varphi_{Q}\circ \Pi^{\,V_{2}}}{\otimes^6\id_2}   $
with $\varphi_{Q}=\otimes^6 Q$,
is
\begin{equation}
	\tcbhighmath[drop fuzzy shadow]{
		\scaln{\varphi_{Q}\circ \Pi^{\,V_{2}}}{\otimes^6\id_2} 
		 =
		 \frac{1}{8}\left((\textrm{tr}\,Q)^{6}+2(\textrm{tr}\,Q)^{4}\textrm{tr}\,Q^{2}+(\textrm{tr}\,Q)^{2}\left(\textrm{tr}\,Q^{2}\right)^{2}+2\,\textrm{tr}\,Q^{4}\textrm{tr}\,Q^{2}+2\left(\textrm{tr}\,Q^{2}\right)^{3}\right)}
\end{equation}
which, with to the Hamilton-Cayley transformation, becomes 
\begin{equation}
	\chi(Q)=(\textrm{tr}\,Q)^{6}-4\left(\textrm{tr}\,Q\right)^{5}+6\left(\textrm{tr}\,Q\right)^{4}-2\left(\textrm{tr}\,Q\right)^{3}-2\left(\textrm{tr}\,Q\right)^{2}.\label{eq:H_C2G-1}
\end{equation}
A quick check yields
\begin{align*}
	\chi(\id) & =(\textrm{tr}\,\id)^{6}-4\left(\textrm{tr}\,\id\right)^{5}+6\left(\textrm{tr}\,\id\right)^{4}-2\left(\textrm{tr}\,\id\right)^{3}-2\left(\textrm{tr}\,\id\right)^{2}=171,
\end{align*}
which is exactly the dimension of $\underline{V}_{\,2}$.

\paragraph{{\Large Case} $\left(\underline{V}_{\,2},\protect\G,\underline{\varphi}_{2}\right)$\,:}

The tensor $\underline{\mathbb{G}}$ has the symmetries

\begin{equation}
	\underline{\mathbb{G}}_{\,abcdef}=\underline{\mathbb{G}}_{\,bacdef}=\underline{\mathbb{G}}_{\,abcedf}=\underline{\mathbb{G}}_{\,defabc},\label{eq:simmetrie G}
\end{equation}
and the set of index permutations which leaves $\underline{\mathbb{G}}$
invariant is again equivalent to $S_{2}\times S_{2}\times S_{2}$ with 8 different elements. The symmetrization identity tensor $\Pi^{\,\underline{V}_{\,2}}$
is
\begin{align*}
	\Pi^{\,\underline{V}_{\,2}}_{iajbkcldmenf} & =\frac{1}{8}(\delta_{ia}\delta_{jb}\delta_{kc}\delta_{ld}\delta_{me}\delta_{nf}+\delta_{ja}\delta_{ib}\delta_{kc}\delta_{ld}\delta_{me}\delta_{nf}+\delta_{ia}\delta_{jb}\delta_{kc}\delta_{md}\delta_{le}\delta_{nf}+\delta_{ja}\delta_{ib}\delta_{kc}\delta_{md}\delta_{le}\delta_{nf}\\
	& \quad+\delta_{la}\delta_{mb}\delta_{nc}\delta_{id}\delta_{je}\delta_{kf}+\delta_{ma}\delta_{lb}\delta_{nc}\delta_{id}\delta_{je}\delta_{kf}+\delta_{la}\delta_{mb}\delta_{nc}\delta_{jd}\delta_{ie}\delta_{kf}+\delta_{ma}\delta_{lb}\delta_{nc}\delta_{jd}\delta_{ie}\delta_{kf}).
\end{align*}
The character $\chi(Q)$, defined as 
$\scaln{\varphi_{Q}\circ \Pi^{\,\underline{V}_{\,2}}}{\otimes^6\id_2} $
with 
$\varphi_{Q}=\otimes^6 Q$,
is
\begin{equation}
	\tcbhighmath[drop fuzzy shadow]{
		  \scaln{\varphi_{Q}\circ \Pi^{\,\underline{V}_{\,2}}}{\otimes^6\id_2}
		   =
		   \frac{1}{8}\left((\textrm{tr}\,Q)^{6}+2(\textrm{tr}\,Q)^{4}\textrm{tr}\,Q^{2}+(\textrm{tr}\,Q)^{2}\left(\textrm{tr}\,Q^{2}\right)^{2}+2\,\textrm{tr}\,Q^{4}\textrm{tr}\,Q^{2}+2\left(\textrm{tr}\,Q^{2}\right)^{3}\right)}
\end{equation}
which, again using the Hamilton-Cayley transformation, can be written as
\begin{equation}
	\chi(Q)=(\textrm{tr}\,Q)^{6}-4\left(\textrm{tr}\,Q\right)^{5}+6\left(\textrm{tr}\,Q\right)^{4}-2\left(\textrm{tr}\,Q\right)^{3}-2\left(\textrm{tr}\,Q\right)^{2}.\label{eq:H_C2Gbar}
\end{equation}
Checking the character of the identity $\id$ results in
\begin{align*}
	\chi(\id) & =(\textrm{tr}\,\id_2)^{6}-4\left(\textrm{tr}\,\id_2\right)^{5}+6\left(\textrm{tr}\,\id_2\right)^{4}-2\left(\textrm{tr}\,\id_2\right)^{3}-2\left(\textrm{tr}\,\id_2\right)^{2}=171,
\end{align*}
which is again equal to the dimension of $\underline{V}_{\,2}$.

\subsubsection{Extended Voigt isomorphisms and structure of the symmetrized tensors}

As for classical elasticity with to the Voigt representation
isomorphism, we now want to define an analogous isomorphism to represent
the tensors $\underline{\mathbb{H}}$ and $\underline{\mathbb{G}}$
as matrices. In \cite{auffray2013matrix}, following reasonable criteria (i.e.\ organizing the components in a matrix in order to group together the
zero entries when accounting for material symmetries), the Voigt isomorphism
is generalized as follows: denoting by $\left\{ \tau_{\alpha}\right\} _{\alpha=1}^{18}$
the canonical basis of $\R^{18}$ and considering the orthonormal
basis
\[
\left\{ \varsigma_{ijk}\defi\left(\frac{1-\delta_{ij}}{\sqrt{2}}+\frac{\delta_{ij}}{2}\right)\left(e_{i}\otimes e_{j}+e_{j}\otimes e_{i}\right)\otimes e_{k}\right\} _{i,j,k=1}^{3}
\]
of $\textrm{Sym}\left(3\right)\otimes\R^{3}$ we set
\[
\boldsymbol{\mathfrak{N}}:\textrm{Sym}\left(3\right)\otimes\R^{3}\fr\R^{18}
\]
as follows:
\begin{align*}
	\boldsymbol{\mathfrak{N}}\,\varsigma_{111} & =\tau_{1}, & \boldsymbol{\mathfrak{N}}\,\varsigma_{221} & =\tau_{2}, & \boldsymbol{\mathfrak{N}}\,\varsigma_{122} & =\tau_{3}, & \boldsymbol{\mathfrak{N}}\,\varsigma_{331} & =\tau_{4}, & \boldsymbol{\mathfrak{N}}\,\varsigma_{133} & =\tau_{5},\\
	\boldsymbol{\mathfrak{N}}\,\varsigma_{222} & =\tau_{6}, & \boldsymbol{\mathfrak{N}}\,\varsigma_{112} & =\tau_{7}, & \boldsymbol{\mathfrak{N}}\,\varsigma_{121} & =\tau_{8}, & \boldsymbol{\mathfrak{N}}\,\varsigma_{332} & =\tau_{9}, & \boldsymbol{\mathfrak{N}}\,\varsigma_{233} & =\tau_{10},\\
	\boldsymbol{\mathfrak{N}}\,\varsigma_{333} & =\tau_{11}, & \boldsymbol{\mathfrak{N}}\,\varsigma_{113} & =\tau_{12}, & \boldsymbol{\mathfrak{N}}\,\varsigma_{131} & =\tau_{13}, & \boldsymbol{\mathfrak{N}}\,\varsigma_{223} & =\tau_{14}, & \boldsymbol{\mathfrak{N}}\,\varsigma_{232} & =\tau_{15},\\
	\boldsymbol{\mathfrak{N}}\,\varsigma_{123} & =\tau_{16}, & \boldsymbol{\mathfrak{N}}\,\varsigma_{132} & =\tau_{17}, & \boldsymbol{\mathfrak{N}}\,\varsigma_{231} & =\tau_{18}.
\end{align*}
In this way we can define
\begin{align*}
	\boldsymbol{\mathfrak{M}}\boldsymbol{\mathfrak{N}} & :\textrm{Lin}\left(\textrm{Sym}\left(3\right),\textrm{Sym}\left(3\right)\otimes\R^{3}\right)\fr\textrm{Lin}\left(\R^{6},\R^{18}\right),\\
	\underline{\boldsymbol{\mathfrak{N}}} & :\textrm{Sym}\left(\textrm{Sym}\left(3\right)\otimes\R^{3},\textrm{Sym}\left(3\right)\otimes\R^{3}\right)\fr\textrm{Sym}\left(\R^{18},\R^{18}\right)
\end{align*}
by
\begin{align}
	\left\langle \boldsymbol{\gM\gN}(\underline{\bH})\, a,b\right\rangle _{\R^{18}} 
	& 
	=
	\left\langle \underline{\bH}\:\MM^{-1}\, a,\boldsymbol{\gN}^{-1}\, b\right\rangle _{\R^{3\times3}} 
	& \forall a\in\R^{6},b\in\R^{18},
	\label{transformation1}
	\\
	\left\langle \underline{\boldsymbol{\gN}}(\underline{\bG})\, p,q\right\rangle _{\R^{18}} 
	& 
	=
	\left\langle \underline{\bG}\:\boldsymbol{\gN}^{-1}\, p,\boldsymbol{\gN}^{-1}\, q\right\rangle _{\R^{3\times3\times3}} 
	& 
	\forall p,q\in\R^{18}.
	\label{transformation2}
\end{align}

\subsubsection{Cubic case}
Now, we want to study in detail the particular case
in which the symmetry group is the cubic group $\mathbb{O}$ which
has 24 elements. We want to calculate the dimension of the fixed subspaces
for $V_{1},V_{2},\underline{V}_{\,1},\underline{V}_{\,2},$ and determine
the structure of the symmetrized tensors. We have
\[
    \dim\,\textrm{Fix}_{\,\mathbb{O}}^{\,\varphi} V_{j}
    =
    \frac{1}{24}\sum_{Q\in\mathbb{O}}\chi_{\varphi_{j}}\left(Q\right),\qquad\dim\,\textrm{Fix}_{\,\mathbb{O}}^{\,\varphi}\underline{V}_{\,j}
    =
    \frac{1}{24}\sum_{Q\in\mathbb{O}}\chi_{\underline{\varphi}_{\,j}}\left(Q\right),\qquad j\in\left\{ 1,2\right\} ,
\]
thus, thanks to the character formulas (\ref{Hbar}),(\ref{H-1}),(\ref{eq:H_C2G-1}),(\ref{eq:H_C2Gbar})
we find 
\begin{itemize}
	\item for 
	$
	      V_{1},\underline{V}_{\,1}
			:\quad
	      \dim\,\textrm{Fix}_{\,\mathbb{O}}^{\,\varphi}V_{1}
	      =
	      \dim\,\textrm{Fix}_{\,\mathbb{O}}^{\,\varphi}\underline{V}_{\,1}=3
	$,
	which gives 3 independent constants for $\mathbb{H}$ and $\underline{\mathbb{H}}$,
	\item for 
	$
	     V_{2},\underline{V}_{\,2}
			:\quad
	     \dim\,\textrm{Fix}_{\,\mathbb{O}}^{\,\varphi}V_{2}
	     =
	     \dim\,\textrm{Fix}_{\,\mathbb{O}}^{\,\varphi}\underline{V}_{\,2}
	     =
	     11
	$
	which gives 11 independent constants for $\mathbb{G}$ and $\underline{\mathbb{G}}$.
\end{itemize}
Thanks to the transformations defined in (\ref{transformation1})
and (\ref{transformation2}), the matrix structures of the symmetrized
tensors are (these results were already derived in \cite{auffray2013matrix})
\[
\underline{\mathbb{G}}=\begin{pmatrix}\underline{G}_{\,1} & 0 & 0 & 0\\
0 & \underline{G}_{\,1} & 0 & 0\\
0 & 0 & \underline{G}_{\,1} & 0\\
0 & 0 & 0 & \underline{G}_{\,2}
\end{pmatrix}\in\textrm{Sym}(18),
\]
where
\begin{align*}
	\underline{G}_{\,1} & =\begin{pmatrix}\eta_{11} & \eta_{12} & \eta_{13} & \eta_{12} & \eta_{13}\\
		& \eta_{22} & \eta_{23} & \eta_{24} & \eta_{25}\\
		&  & \eta_{33} & \eta_{25} & \eta_{35}\\
		& \textrm{sym} &  & \eta_{22} & \eta_{23}\\
		&  &  &  & \eta_{33}
	\end{pmatrix}\in\textrm{Sym}(5), & \underline{G}_{\,2} & =\begin{pmatrix}\gamma_{11} & \gamma_{12} & \gamma_{12}\\
		& \gamma_{11} & \gamma_{12}\\
		\textrm{sym} &  & \gamma_{11}
	\end{pmatrix}\in\textrm{Sym}(3),
\end{align*}
with
\begin{align*}
	\eta_{11} & =\frac{1}{3}\,\left(\underline{\mathbb{G}}_{\,111111}+\underline{\mathbb{G}}_{\,222222}+\underline{\mathbb{G}}_{\,333333}\right), & \eta_{22} & =\frac{1}{6}\,\left(\underline{\mathbb{G}}_{\,112112}+\underline{\mathbb{G}}_{\,113113}+\underline{\mathbb{G}}_{\,221221}+\underline{\mathbb{G}}_{\,223223}+\underline{\mathbb{G}}_{\,331331}+\underline{\mathbb{G}}_{\,332332}\right),\\
	\eta_{24} & =\frac{1}{3}\,\left(\underline{\mathbb{G}}_{\,112332}+\underline{\mathbb{G}}_{\,113223}+\underline{\mathbb{G}}_{\,221331}\right), & \eta_{33} & =\frac{1}{6}\,\left(\underline{\mathbb{G}}_{\,121121}+\underline{\mathbb{G}}_{\,122122}+\underline{\mathbb{G}}_{\,131131}+\underline{\mathbb{G}}_{\,133133}+\underline{\mathbb{G}}_{\,232232}+\underline{\mathbb{G}}_{\,233233}\right),\\
	\eta_{35} & =\frac{1}{3}\,\left(\underline{\mathbb{G}}_{\,121233}+\underline{\mathbb{G}}_{\,122133}+\underline{\mathbb{G}}_{\,131232}\right), & \eta_{12} & =\frac{1}{6}\,\left(\underline{\mathbb{G}}_{\,111221}+\underline{\mathbb{G}}_{\,111331}+\underline{\mathbb{G}}_{\,112222}+\underline{\mathbb{G}}_{\,113333}+\underline{\mathbb{G}}_{\,222332}+\underline{\mathbb{G}}_{\,223333}\right),\\
	\gamma_{11} & =\frac{1}{3}\,\left(\underline{\mathbb{G}}_{\,123123}+\underline{\mathbb{G}}_{\,132132}+\underline{\mathbb{G}}_{\,231231}\right), & \eta_{13} & =\frac{1}{6}\,\left(\underline{\mathbb{G}}_{\,111122}+\underline{\mathbb{G}}_{\,111133}+\underline{\mathbb{G}}_{\,121222}+\underline{\mathbb{G}}_{\,131333}+\underline{\mathbb{G}}_{\,222233}+\underline{\mathbb{G}}_{\,232333}\right),\\
	\gamma_{12} & =\frac{1}{3}\,\left(\underline{\mathbb{G}}_{\,123132}+\underline{\mathbb{G}}_{\,123231}+\underline{\mathbb{G}}_{\,132231}\right), & \eta_{23} & =\frac{1}{6}\,\left(\underline{\mathbb{G}}_{\,112121}+\underline{\mathbb{G}}_{\,113131}+\underline{\mathbb{G}}_{\,122221}+\underline{\mathbb{G}}_{\,133331}+\underline{\mathbb{G}}_{\,223232}+\underline{\mathbb{G}}_{\,233332}\right),\\
	&  & \eta_{25} & =\frac{1}{6}\,\left(\underline{\mathbb{G}}_{\,112233}+\underline{\mathbb{G}}_{\,113232}+\underline{\mathbb{G}}_{\,121332}+\underline{\mathbb{G}}_{\,122331}+\underline{\mathbb{G}}_{\,131223}+\underline{\mathbb{G}}_{\,133221}\right),
\end{align*}
and
\[
    \underline{\mathbb{H}}=\begin{pmatrix}0 & 0 & 0 & 0\\
    	0 & \underline{H}_{\,2} & 0 & 0\\
    	0 & 0 & 0 & 0\\
    	0 & 0 & \underline{H}_{\,2} & 0\\
    	0 & 0 & 0 & 0\\
    	0 & 0 & 0 & \underline{H}_{\,2}\\
    	\underline{H}_{\,1} & 0 & 0 & 0
    \end{pmatrix}\in\R^{6\times18},
    \quad
    \text{where}
    \quad
	\underline{H}_{\,1} 
	=
	\begin{pmatrix}
		\zeta_{1} & -\zeta_{1} & 0\\
		-\zeta_{1} & 0 & \zeta_{1}\\
		0 & \zeta_{1} & -\zeta_{1}
	\end{pmatrix}\in\R^{3\times3},
	\quad  
	\underline{H}_{\,2}  
	=\begin{pmatrix}
		\zeta_{2}\\
		\zeta_{3}\\
		-\zeta_{2}\\
		-\zeta_{3}
	\end{pmatrix}
	\in\R^{3\times3},
\]
with
\begin{align*}
	\zeta_{1} & =\frac{1}{6}\,\left(-\underline{\mathbb{H}}_{\,11213}+\underline{\mathbb{H}}_{\,11312}+\underline{\mathbb{H}}_{\,22123}-\underline{\mathbb{H}}_{\,22312}-\underline{\mathbb{H}}_{\,33123}+\underline{\mathbb{H}}_{\,33213}\right),\\
	\zeta_{2} & =\frac{1}{6}\,\left(\underline{\mathbb{H}}_{\,12311}-\underline{\mathbb{H}}_{\,12322}-\underline{\mathbb{H}}_{\,13211}+\underline{\mathbb{H}}_{\,13233}+\underline{\mathbb{H}}_{\,23122}-\underline{\mathbb{H}}_{\,23133}\right),\\
	\zeta_{3} & =\frac{1}{6}\,\left(\underline{\mathbb{H}}_{\,12113}-\underline{\mathbb{H}}_{\,12223}-\underline{\mathbb{H}}_{\,13112}+\underline{\mathbb{H}}_{\,13323}+\underline{\mathbb{H}}_{\,23212}-\underline{\mathbb{H}}_{\,23313}\right).
\end{align*}

\subsubsection{Transversal hemitropic and transversal isotropic case}\label{transversal isotropic} 

Next, we study the two particular cases in which the invariance of the elasticity tensor $\underline{\mathbb{G}}$ is taken with respect to  $
\textrm{SO}(2;e_3)$ and $\textrm{O}(2;e_3)$. These results are not new; indeed, the structure for the symmetrized tensors was already obtained in \cite{auffray2013matrix}. Via the trace formula and the explicit expression for the Haar measure on $\textrm{SO}(2)$ and $\textrm{O}(2)$ derived in \eqref{Haar circle} and \eqref{Haar O(2)} respectively, we compute the following numbers of independent components for the two considered symmetries:
\begin{equation}\label{Dimensions transversal}
\dim\,\textrm{Fix}_{\,\textrm{SO}(2,e_3)}^{\,\varphi}\underline{V}_{\,2}=31,\qquad\dim\,\textrm{Fix}_{\,\textrm{O}(2,e_3)}^{\,\varphi}\underline{V}_{\,2}=21. 
\end{equation}
The correct number of independent components for the transversal isotropic case was also obtained with another approach in an unpublished note \cite{batra-private}.
Denoting with $\underline{\mathbb{G}}_{\;\textrm{SO}(2;e_3)}$ and $\underline{\mathbb{G}}_{\;\textrm{O}(2;e_3)}$ the symmetrized tensors with respect to the actions of $\textrm{SO}(2;e_3)$ and $\textrm{O}(2;e_3)$ respectively, their matrix representations via the isomorphism $\boldsymbol{\mathfrak{N}}$ are 
\[
\underline{\mathbb{G}}_{\;\textrm{SO}(2;e_3)}=\begin{pmatrix}
\underline{G}_{\,1} & \underline{G}_{\,2} & 0 & 0\\
\underline{G}_{\,2} & \underline{G}_{\,1} & 0 & 0\\
0 & 0 & \underline{G}_{\,3} & \underline{G}_{\,4}\\
0 & 0 & \underline{G}_{\,4} & \underline{G}_{\,5}
\end{pmatrix}\in\text{Sym}(18),\qquad\underline{\mathbb{G}}_{\;\textrm{O}(2;e_3)}=\begin{pmatrix}
\underline{G}_{\,1} & 0 & 0 & 0\\
0 & \underline{G}_{\,1} & 0 & 0\\
0 & 0 & \underline{G}_{\,3} & 0\\
0 & 0 & 0 & \underline{G}_{\,5}
\end{pmatrix}\in\text{Sym}(18),
\]
where
\begin{align*}
\underline{G}_{\,1} & =\begin{pmatrix}\alpha_{11} & \alpha_{12} & \alpha_{13} & \alpha_{14} & \alpha_{15}{\color{white}\Bigr|}\\
& \alpha_{22} & -\alpha_{13}+\displaystyle{\frac{\sqrt{2}(\alpha_{11}-\alpha_{22})}{2}}{\color{white}\biggr|} & \alpha_{14}-\sqrt{2}\,\alpha_{34} & \alpha_{15}-\sqrt{2}\,\alpha_{35} \\
 &  & -\alpha_{12}+\displaystyle{\frac{\alpha_{11}+\eta_{22}}{2}}{\color{white}\biggr|} & \alpha_{34} & \alpha_{35}\\
 &  \textrm{sym} &  & \alpha_{44} & \alpha_{45}\\
 &  &  &  & \alpha_{55} {\color{white}\biggr|}
\end{pmatrix}\in\text{Sym}(5) &  11\;\text{independent components},\\
\underline{G}_{\,2} & =\begin{pmatrix}0 & \beta_{12} & \displaystyle{-\frac{\sqrt{2}\,\beta_{12}}{2}}{\color{white}\biggr|} & \beta_{24}+\sqrt{2}\,\beta_{34} & \beta_{25}+\sqrt{2}\,\beta_{35}\\
& 0 & \displaystyle{-\frac{\sqrt{2}\,\beta_{12}}{2}}{\color{white}\biggr|} & \beta_{24} & \beta_{25}\\
&  & 0 & \beta_{34} & \beta_{35}{\color{white}\biggr|}\\
& \textrm{skew} &  & 0 & \beta_{45}{\color{white}\biggr|}\\
&  &  &  & 0{\color{white}\biggr|}
\end{pmatrix}\in\mathfrak{so}(5) & 6\;\text{independent components},\\
\underline{G}_{\,3} & 
=
\begin{pmatrix}\gamma_{11} & \gamma_{12} & \gamma_{13} & \gamma_{12} & \gamma_{13}\\
& \gamma_{22} & \gamma_{23} & \gamma_{22} & \gamma_{23}\\
&  & \gamma_{33} & \gamma_{23} & \gamma_{33}\\
& \textrm{sym} &  & \gamma_{22} & \gamma_{23}\\
&  &  &  & \gamma_{33}
\end{pmatrix}
+
\begin{pmatrix}0 & 0 & 0 & 0 & 0\\
 & 0 & 0 & -\eta_{11} & -\sqrt{2}\,\eta_{12} \\
 &  & 0 & -\sqrt{2}\,\eta_{12} & -(\eta_{22}+\eta_{33})\\
& \textrm{sym} &  & 0 & 0\\
&  &  &  & 0
\end{pmatrix}\in\text{Sym}(5) & 6\;\text{independent components},\\
\underline{G}_{\,4} & =\begin{pmatrix} 0 & \zeta_{12} & -\zeta_{12} \\
0 & \zeta_{22} & -\zeta_{22}-\sqrt{2}\,\zeta_{31} \\
\zeta_{31} & \zeta_{32}  & -\zeta_{32} \\
0 & \zeta_{22}+\sqrt{2}\,\zeta_{31} & -\zeta_{22} \\
-\zeta_{31} & \zeta_{32}  &  -\zeta_{32}
\end{pmatrix}\in\R^{5\times3} &  4\;\text{independent components},\\
\underline{G}_{\,5} & =\begin{pmatrix}\eta_{11} & \eta_{12} & \eta_{12}\\
& \eta_{22} & \eta_{23} \\
\textrm{sym}&  & \eta_{22}  
\end{pmatrix}\in\text{Sym}(3) &  4\;\text{independent components},\\
\end{align*}
and 

\begin{align*}
      \alpha_{11}&=\frac{1}{64}\,\Big( 5\,\bG_{111111} + 4\,\bG_{111122} + 2\,\bG_{111221} + \bG_{112112} + 4\,\bG_{112121} + 2\,\bG_{112222} \\
      &\qquad\;+ 4\,\bG_{121121} + 4\,(\bG_{121222} + \bG_{122122} + \bG_{122221}) + 
      \bG_{221221} + 5\,\bG_{222222}\Big), \\
      \alpha_{12}&= \frac{1}{64}\,\Big(\bG_{111111} + 6\,\bG_{111221} + \bG_{112112} + 6\,\bG_{112222} - 4\,\bG_{121121} - 
      4\,\bG_{122122} + \bG_{221221} + \bG_{222222}\Big), \\
      \alpha_{13}&=\frac{1}{32\,\sqrt{2}}\,\Big(\bG_{111111} + 6\,\bG_{111122} - \bG_{112112} - 2\,\bG_{112121} + 6\,\bG_{121222} - 
      2\,\bG_{122221} - \bG_{221221} + \bG_{222222}\Big),\\
      \alpha_{14}&=\frac{1}{32}\, \Big(3\,\bG_{111331} + \bG_{112332} + 2\,\bG_{121332} + 2\,\bG_{122331} + \bG_{221331} + 
      3\,\bG_{222332}\Big),\\
      \alpha_{15}&=\frac{1}{16 \sqrt{2}}\,\Big(3\,\bG_{111133} + \bG_{112233} + 2\bG_{121233} + 2\,\bG_{122133} + \bG_{133221} + 
      3\,\bG_{222233}\Big),\\
      \alpha_{22}&=\frac{1}{64}\,\Big(\bG_{111111} - 4\,\bG_{111122} + 2\,\bG_{111221} + 5\,\bG_{112112} - 4\,\bG_{112121} + 
      2\,\bG_{112222}\\
       &\qquad\;+ 4\,\bG_{121121} - 4\,\bG_{121222} + 4\,\bG_{122122} - 4\,\bG_{122221} + 
      5\,\bG_{221221} + \,\bG_{222222}\Big),\\
       \alpha_{34}&=\frac{1}{16 \sqrt{2}}\Big(\bG_{111331} - \bG_{112332} + 2\,\bG_{121332} + 
       2\,\bG_{122331} - \bG_{221331} + \bG_{222332}\Big),\\
        \alpha_{35}&=\frac{1}{16} \Big(\bG_{111133} - \bG_{112233} + 2\,\bG_{121233} + 2\,\bG_{122133} - \bG_{133221} + \bG_{222233}\Big),\\
         \alpha_{44}&=\frac{1}{8}\Big(\bG_{331331} + \bG_{332332}\Big),
         \quad
          \alpha_{45}=\frac{1}{4\sqrt{2}}\Big(\bG_{133331} + \bG_{233332}\Big),
          \quad
           \alpha_{55}=\frac{1}{4\sqrt{2}}\Big(\bG_{133133} + \bG_{233233}\Big),\\
      \\
         \beta_{12}&=\frac{1}{16} \Big(\bG_{111112} - \bG_{111121} + \bG_{112122} - \bG_{121221} + \bG_{122222} - \bG_{221222}\Big), \\
         \beta_{24}&=\frac{1}{32} \Big(\bG_{111332} - 3\,\bG_{112331} + 2\,\bG_{121331} - 2\,\bG_{122332} + 3\,\bG_{221332} - 
         \,\bG_{222331}\Big),\\
         \beta_{25}&=\frac{1}{16\sqrt{2}}\Big(\bG_{111233} - 3\,\bG_{112133} + 2\,\bG_{121133} - 2\,\bG_{122233} - \bG_{133222} + 
         3\,\bG_{221233}\Big),  \\
         \beta_{34}&=\frac{1}{16\sqrt{2}}\Big(\bG_{111332} + \bG_{112331} - 2 \,\bG_{121331} + 
         2\,\bG_{122332} - \bG_{221332} - \bG_{222331}\Big), \\
         \beta_{35}&=\frac{1}{16} \Big(\bG_{111233} + \bG_{112133} - 2\,\bG_{121133} + 2\,\bG_{122233} - \bG_{133222} - \bG_{221233}\Big), \\
         \beta_{45}&=\frac{1}{4\sqrt{2}}\Big(-\bG_{133332} + \bG_{233331}\Big), \\
         \\
         \gamma_{11}&=\frac{\bG_{333333}}{4}, 
         \qquad
         \gamma_{12}=\frac{1}{8}\Big(\bG_{113333} + \bG_{223333}\Big), 
         \qquad
         \gamma_{13}=\frac{1}{4\sqrt{2}}\Big(\bG_{131333} + \bG_{232333}\Big), 
         \\
         \gamma_{22}&=\frac{1}{32} \Big(3\,\bG_{113113} + 2\,\bG_{113223} + 4\,\bG_{123123} + 3\,\bG_{223223}\Big), \\
         \gamma_{23}&=\frac{1}{16\sqrt{2}}\Big(3\,\bG_{113131} + \bG_{113232} + 2\,\bG_{123132} + 2\,\bG_{123231} + \bG_{131223} + 
         3\,\bG_{223232}\Big), \\
         \gamma_{33}&=\frac{1}{16} \Big(3\,\bG_{131131} + 2\,\bG_{131232} + \bG_{132132} + 2\,\bG_{132231} + \bG_{231231} + 
         3\,\bG_{232232}\Big), \\
         \\
         \zeta_{12}&=\frac{1}{4\sqrt{2}}\Big(\bG_{132333} - \bG_{231333}\Big), \\
         \zeta_{22}&=\frac{1}{16\sqrt{2}}\Big(3\,\bG_{113132} - \bG_{113231} - 2\,\bG_{123131}+ 2\,\bG_{123232} + \bG_{132223} - 
         3\,\bG_{223231}\Big), \\
         \zeta_{31}&=\frac{1}{16} \Big(-\bG_{113132} - \bG_{113231} + 2\,\bG_{123131} - 2\,\bG_{123232} + \bG_{132223} + \bG_{223231}\Big), \\
         \zeta_{32}&=\frac{1}{8} \Big(\bG_{131132} - \bG_{131231} + \bG_{132232} - \bG_{231232}\Big), \\
         \\
         \eta_{11}&=\frac{1}{16} \Big(\bG_{113113} - 2\,\bG_{113223} + 4\,\bG_{123123} + \bG_{223223}\Big), 
         \\
         \eta_{12}&=\frac{1}{16} \Big(\bG_{113131} - \bG_{113232} + 2\,\bG_{123132} + 2\,\bG_{123231} - \bG_{131223} + \bG_{223232}\Big), 
         \\
         \eta_{22}&=\frac{1}{16} \Big(\bG_{131131} - 2\,\bG_{131232} + 3\,\bG_{132132} - 2\,\bG_{132231} + 3\,\bG_{231231} + 
         \bG_{232232}\Big), 
         \\
         \eta_{23}&=\frac{1}{16} \Big(\bG_{131131} - 2\,\bG_{131232} - \bG_{132132} + 6\,\bG_{132231} - \bG_{231231} + \bG_{232232}\Big).       
\end{align*}

\subsection{Indeterminate couple stress model}

We now aim to apply the established invariance theory to a special sub-class of second gradient models: the indeterminate couple stress model (for a very good description of this model we refer to \cite{neff2009subgrid,neff2016some,munch2017modified,ghiba2017variant,neff2016null} and for a determination of the material invariance condition we refer to \cite{munch2018rotational,neff2009subgrid}). The linearization of the indeterminate couple stress model is characterized via the action functional
\begin{equation} 
    \A[u]
    \defi
    \int_\Omega\frac{1}{2}
    \Big(
        \big\langle\C\,\varepsilon,\varepsilon \big\rangle 
        +
        \underbrace{\big\langle\bL\,\nabla\!\curl\!u,\nabla\!\curl\!u \big\rangle}_{\eqcolon\,W_{\curl}(\nabla^2u)}  
    \Big) 
    \dx \,,
\end{equation}
i.e., in the indeterminate couple stress model, the non local term of the deformation energy is defined as a function of the gradient of only the rotational part of the gradient of $u$. Thus, we can immediately remark that
\begin{itemize}
	\item $\nabla\curl u$ is \textbf{not a symmetric} second order tensor and, as a consequence, the curvature elasticity tensor is a fourth order tensor involving only the major symmetry,
	\item considering the orthogonal split of the curvature term in the symmetric and skew symmetric parts
	 $$ 
	      \big\langle\bL\,\nabla\!\curl\!u,\nabla\!\curl\!u \big\rangle
	      =
	      \big\langle\bL_s\,\sym\,\nabla\!\curl u,\sym\,\nabla\!\curl\!u \big\rangle
	      +
	      \big\langle\bL_c\,\skew\,\nabla\!\curl\!u,\skew\,\nabla\!\curl\!u \big\rangle,  
	 $$
	 where
	 $$
	      \bL_s \in\text{Sym}(\Sym,\Sym)\qquad\text{and}\qquad\bL_c\in\text{Sym}(\so,\so),
	 $$
	 we can express the skew-symmetric term as a bilinear form acting on $\R^3$ via the axl operator (see the Appendix \ref{subsec_axl}). Indeed, in this case we have
	$$ 
	    \text{axl}\,\skew\,\nabla\!\curl\!u=\curl\!\curl\!u,
	$$
	and we can introduce a matrix $\widetilde{\bL}_c\in\Sym$ such that
	$$ 
	    \big\langle\bL_c\,\skew\,\nabla\!\curl\!u,\skew\,\nabla\!\curl\!u \big\rangle_{\R^{3\times3}}
	    =
	    \big\langle\widetilde{\bL}_c\,\curl\!\curl\!u,\curl\!\curl\!u \big\rangle_{\R^{3}}
	    \qquad
	    \forall\;\text{admissible}\;u. 
	$$
\end{itemize}

Applying the theoretical machinery developed in Section \ref{Non symmetric theories2}, we can obtain the structure of the elasticity tensor $\bL$ when we ask for the isotropic invariance of the indeterminate couple stress model. Indeed, having that 
$$\bL\in\text{Sym}(\R^{3\times3},\R^{3\times3}), $$
the invariance with respect to the full group $\So$ gives us only 3 independent material constants (see Eqs. \eqref{new parameters}, \eqref{matrix new parameters}). Thus, in this case, we have
\begin{align*}
     \big\langle\bL\,\nabla\!\curl\!u,\nabla\!\curl\!u \big\rangle
     =
     \alpha_1\left\| \sym\,\nabla\!\curl\!u \right\|^2+\frac{\alpha_2}{2}\,(\tr\,\nabla\!\curl\!u)^2
     +\alpha_3 \left\| \skew\,\nabla\!\curl\!u \right\|^2.
\end{align*}
Moreover, remarking that $ \tr\nabla\curl u=\div\curl u=0$,
we can write the deformation energy in the case of isotropy as
\begin{align*}
     \big\langle\bL\,\nabla\!\curl\!u,\nabla\!\curl\!u \big\rangle
     =
     \alpha_1\left\| \text{dev}\,\sym\nabla\!\curl\!u \right\|^2
     +
     \alpha_3 \left\| \skew\nabla\!\curl\!u \right\|^2.
\end{align*}
%
%
%
%
\subsection{Micromorphic type models}\label{Microm}

Our next objective is to apply the well-established theory of invariance to a class of micromorphic-type generalized continua, which have garnered increasing attention in recent years for modeling the mechanical behavior of metamaterials. As we will discuss, the curvature term of the relaxed micromorphic model belongs to a different vector space than that of classical Cauchy continua. To address this discrepancy, we will leverage the findings from Section \ref{Non symmetric theories2}, which will be tailored to suit 2D models. These results have been utilized  in \cite{sarhil2023size} to model the size-effects of metamaterial beams under bending with the aid of the relaxed micromorphic continuum.
We will also compare the symmetry classes of the relaxed micromorphic model with those of the classical micromorphic model.


\subsubsection{3D-relaxed micromorphic model}
We set 
\[
x=(x_1,x_2,x_3)=(\overline{x},x_3),\qquad\textrm{where}\qquad\overline{x}=(x_1,x_2).
\]
The involved kinematic fields are
\begin{align}
          u\colon\Omega\subset\bR^3 &\fr\bR^3, 
          &x &\mapsto u(x)\,=\,\partonb{u_1(x),u_2(x),u_3(x)}\,,
          \nonumber
          \\
          P\colon\Omega\subset\bR^3 &\fr\bR^{3\times 3},
          &x &\mapsto P(x)=
          \begin{pmatrix}
          	P_{11}(x) & P_{12}(x) & P_{13}(x)
          	\\
          	P_{21}(x) & P_{22}(x) & P_{23}(x)
          	\\
          	P_{31}(x) & P_{32}(x) & P_{33}(x)
          \end{pmatrix}
          \,.
          \nonumber
\end{align}
The Curl operator
is defined by
\begin{align}
	\Curl P(x)
	&= 
	\Curl
	\begin{pmatrix}
		P_{11}(x) & P_{12}(x) & P_{13}(x)
		\\
		P_{21}(x) & P_{22}(x) & P_{23}(x)
		\\
		P_{31}(x) & P_{32}(x) & P_{33}(x)
	\end{pmatrix}
	\defi
	\begin{pmatrix}
		\curl\partonb{P_{11}(x), P_{12}(x), P_{13}(x)}
		\\
		\curl\partonb{P_{21}(x), P_{22}(x), P_{23}(x)}\vphantom{\int^{\sum}_{\sum_a}}
		\\
		\curl\partonb{P_{31}(x), P_{32}(x), P_{33}(x)}
	\end{pmatrix}.
\end{align}
The full potential energy density for the linear model consists of the terms
\[
        W(u,\nabla u,P,\Curl  P)
        =
          W_\rre(\sym\nabla u,\sym P)
        +W_\rc(\skew\nabla u,\skew P)
        +W_\rm(\sym P)+W_{\textrm{curv}}(\Curl  P),
\]
where
\begin{align}
	W_\rre(\sym\nabla u,\sym P) &\defi \frac{1}{2}\scalb{\bC_\rre\,\sym(\nabla u-P)}{\sym(\nabla u-P)}_{\bR^{3\times 3}},
	\nonumber 
	\\
	W_\rc(\skew\nabla u,\skew P) &\defi\frac{1}{2}\scalb{\bC_\rc\,\skew(\nabla u-P)}{\skew(\nabla u-P)}_{\bR^{3\times 3}},
	\nonumber
	\\
	W_\rm(\sym P) &\defi\frac{1}{2}\scalb{\bC_\rm\,\sym P}{\sym P}_{\bR^{3\times 3}},
	\\
	W_{\textrm{curv}}(\Curl  P)&\defi\frac{1}{2}\scalb{\bL\,\Curl P}{\Curl  P}_{\bR^{3\times3}},
	\nonumber
\end{align}
and the four involved tensors $\bC_\rre,\bC_\rc,\bC_\rm,\bL$ are
\begin{align}
	\bC_\rre 
	&
	\in\Ela^+(3)\subseteq\otimes^4\bR^3,
	&\bC_\rc 
	&
	\in\textrm{Sym}^+\partonb{\so,\so)},
	\nonumber
	\\
	\bC_\rm 
	&
	\in\Ela^+(3),
	&
	\bL 
	&
	\in\textrm{Sym}^+(\bR^{3\times 3},\bR^{3\times 3}).
	\nonumber
\end{align}

\subsubsection{2D-relaxed micromorphic model, plane strain}

The involved kinematic fields are
\begin{align}
	u\colon\Omega\subset\bR^2 &\fr\bR^2, 
	&(x_1,x_2) &\mapsto u(x_1,x_2)
	\,=
	\,\partonb{u_1(x_1,x_2),u_2(x_1,x_2)}
	\,,
	\nonumber
	\\
	P\colon\Omega\subset\bR^2 &\fr\bR^{2\times 2},
	&(x_1,x_2) &\mapsto P(x_1,x_2)=
	\begin{pmatrix}
		P_{11}(x_{1},x_{2}) & P_{12}(x_{1},x_{2})
		\\
		P_{21}(x_{1},x_{2}) & P_{22}(x_{1},x_{2})
	\end{pmatrix}
	\,.
\end{align}
Given a vector field $v:\Omega\subset\bR^2\fr\bR^2$ we define the 2D-$\curl$ operator as 
\[
\curl_{\,2\text{D}}v(x)\defi v_{2,1}(x)-v_{1,2}(x)\,,
\]
and for a matrix-valued field $P\colon\Omega\subset\bR^2 \fr\bR^{2\times 2}$,
\begin{align}
	\Curl_{\,2\text{D}} P(x)
	&\defi 
	\Curl_{\,2\text{D}}
	\begin{pmatrix}
		P_{11}(x_{1},x_{2}) & P_{12}(x_{1},x_{2})
		\\[1mm]
		P_{21}(x_{1},x_{2}) & P_{22}(x_{1},x_{2})
	\end{pmatrix}
	=
	\begin{pmatrix}
		\curl_{2}\partonb{P_{11}(x_{1},x_{2}),P_{12}(x_{1},x_{2})}
		\\[1mm]
		\curl_{2}\partonb{P_{21}(x_{1},x_{2}),P_{22}(x_{1},x_{2})}
	\end{pmatrix}
	\nonumber
	\\
	&\,=\begin{pmatrix}
		P_{12,1}(x_{1},x_{2})-P_{11,2}(x_{1},x_{2})
		\\[1mm]
		P_{22,1}(x_{1},x_{2})-P_{21,2}(x_{1},x_{2})
	\end{pmatrix}.
\end{align}
The full potential energy density for the linear model consists of the following terms:
\[
W(u,\nabla u,P,\Curl_{\,2\text{D}}\! P)=W_\rre(\sym\nabla u,\sym P)+W_\rc(\skew\nabla u,\skew P)+W_\rm(\sym P)+W_{\textrm{curv}}(\Curl_{\,2\text{D}}\! P),
\]
where
\begin{align}
	W_\rre(\sym\nabla u,\sym P) &\defi \frac{1}{2}\scalb{\bC_\rre\,\sym(\nabla u-P)}{\sym(\nabla u-P)}_{\bR^{2\times 2}},
	\nonumber 
	\\
	W_\rc(\skew\nabla u,\skew P) &\defi\frac{1}{2}\scalb{\bC_\rc\,\skew(\nabla u-P)}{\skew(\nabla u-P)}_{\bR^{2\times 2}},
	\nonumber
	\\
	W_\rm(\sym P) &\defi\frac{1}{2}\scalb{\bC_\rm\,\sym P}{\sym P}_{\bR^{2\times 2}},
	\nonumber
	\\
	W_{\textrm{curv}}(\Curl_{\,2\text{D}}\! P)&\defi\frac{1}{2}\scalb{\bL\,\Curl_{\,2\text{D}}\! P}{\Curl_{\,2\text{D}}\! P}_{\bR^2},
	\nonumber
\end{align}
and the four involved tensors $\bC_\rre,\bC_\rc,\bC_\rm,\bL$ are
\begin{align}
	\bC_\rre &
	\in\Ela^+(2)\subseteq\otimes^4\bR^2,
	&
	\bC_\rc 
	&
	\in\textrm{Sym}\partonb{\mathfrak{so}(2),\mathfrak{so}(2)},
	\nonumber
	\\
	\bC_\rm 
	&
	\in\Ela^+(2),
	&
	\bL 
	&
	\in\textrm{Sym}^+\!(2)\subseteq\bR^{2\times 2},
	\nonumber
\end{align}
where we set $\Ela^+(2)\defi\textrm{Sym}^+\!\partonb{\textrm{Sym}(2),\textrm{Sym}(2)}$. 

\paragraph{2D-relaxed micromorphic model derived from the 3D model:}

Let us give the following definition.

\begin{defn}
	Consider a matrix valued field $P\colon\Omega\fr\bR^{3\times 3}$. We say that $P$ is a \textbf{plane field} if it has the following structure:
	\[
	P(x)=
	\left(
	\begin{array}{ccc}
		P_{11}(x) & P_{12}(x) & P_{13}(x)\\
		P_{21}(x) & P_{22}(x) & P_{23}(x)\\
		P_{31}(x) & P_{32}(x) & P_{33}(x)
	\end{array}\right)
	=
	\left(
	\begin{array}{ccc}
		P_{11}(\overline{x}) & P_{12}(\overline{x}) & 0\\
		P_{21}(\overline{x}) & P_{22}(\overline{x}) & 0\\
		0 & 0 & 0
	\end{array}
	\right).
	\]
	We can identify a plane field with \enquote{its restriction field}
	\[
	\overline{P}:\Omega\fr\bR^{2\times 2},
	\]
	and by an abuse of notation, whenever there is no confusion, we use the same symbol for both.
\end{defn}
If a matrix-valued field $P$ is a plane field we obtain
\begin{align}
	\Curl P(x)
	&= 
	\Curl
	\begin{pmatrix}
		P_{11}(x) & P_{12}(x) & P_{13}(x)
		\\
		P_{21}(x) & P_{22}(x) & P_{23}(x)
		\\
		P_{31}(x) & P_{32}(x) & P_{33}(x)
	\end{pmatrix}
	=
	\begin{pmatrix}
		\curl\partonb{P_{11}(x), P_{12}(x), P_{13}(x)}
		\\
		\curl\partonb{P_{21}(x), P_{22}(x), P_{23}(x)}\vphantom{\int^{\sum}_{\sum_a}}
		\\
		\curl\partonb{P_{31}(x), P_{32}(x), P_{33}(x)}
	\end{pmatrix}
	\nonumber
	\\
	&=
	\begin{pmatrix}
		\partonb{0,0, P_{12,1}(\overline{x})-P_{11,2}(\overline{x})}
		\\
		\partonb{0,0, P_{22,1}(\overline{x})-P_{21,2}(\overline{x})}
		\vphantom{\int^{\sum}_{\sum_a}}
		\\
		\partonb{0,0,0}
	\end{pmatrix}
	=
	\begin{pmatrix}
		0 & 0 & P_{12,1}(\overline{x})-P_{11,2}(\overline{x})
		\\
		0 & 0 & P_{22,1}(\overline{x})-P_{21,2}(\overline{x})
		\vphantom{\int^{\sum}_{\sum_a}}
		\\
		0 & 0 & 0
	\end{pmatrix}
	=
	\left(
	\begin{array}{c c|c}
		0                  & 0 & \multirow{2}{*}{$\Curl_{\,2\text{D}}\overline{P}(\overline{x})$}
		\\ 
		0&0&      
		\\ 
		\hline
		\vphantom{A^{A^{^a}}}0&0&0
	\end{array}
	\right)
	\,.
\end{align}





\subsubsection{Symmetrization of \texorpdfstring{$\bL$}{}}

Introducing the symmetrization identity
\[
\pi:\bR^{2\times2}\fr\textrm{Sym}(2),\qquad \pi_{iajb}=\frac{1}{2}(\delta_{ia}\delta_{jb}+\delta_{ja}\delta_{ib}),
\]
the characters can be computed via 
$
\chi(Q)=\scal{\varphi_Q}{\pi}=\scal{Q^{\otimes 2}}{\pi}_{\bR^{2\times 2\times 2\times 2}}
$
giving\footnote{Keep in mind that \begin{align}
		\tr Q^2=\scaln{\id}{Q^2}_{\bR^2}=\delta_{ij}(Q^2)_{ij}=\delta_{ij}Q_{i\alpha}Q_{\alpha j}=\delta_{ij}Q_{i\alpha}\delta_{\alpha\beta}Q_{\beta j}
		\,.
\end{align}}
\begin{align}
	\chi(Q) &=\frac{1}{2}(\delta_{ia}\delta_{jb}+\delta_{ja}\delta_{ib})Q_{ia}Q_{jb}
	=\frac{1}{2}(\delta_{ia}\delta_{jb}Q_{ia}Q_{jb}+\delta_{ja}\delta_{ib}Q_{ia}Q_{jb})
	\\
	\nonumber
	&=\frac{1}{2}\partonB{\underbrace{\delta_{ia}Q_{ia}}_{\tr Q}
		\underbrace{\delta_{jb}Q_{jb}}_{\tr Q}+\underbrace{\delta_{aj}Q_{jb}\delta_{bi}Q_{ia}}_{\tr Q^2}}
	\nonumber
	=\frac{1}{2}\partonB{(\tr Q)^2+\tr Q^2}.
	\nonumber
\end{align}
The Cayley-Hamilton theorem allows us to express $\chi(Q)$ as a polynomial in $\tr Q$: since
\[
    p(Q)
    =
    Q^2+\underbrace{c_1}_{-\tr Q} Q+\underbrace{\det(Q)}_{=\,\pm1} \id=0,
    \qquad
    \parton{Q\in \O(2)}
\]
we find
$
Q^2=\tr(Q) Q\mp\id
$
and thus
$
\tr Q^2=\tr\partonb{\tr(Q) Q\mp\id}=(\tr Q)^2\mp2.
$
Thus, finally, we have 
\begin{equation}\label{Power2}
	\chi(Q)=(\tr Q)^2\mp1,
\end{equation}
where the sign choice is dictated by 
\[
\begin{cases}
	-1 & \textrm{if }Q\in \SOd,
	\\
	\hphantom{-}1 & \textrm{if }Q\in \O^{-}(2).
\end{cases}
\]
The closed subgroups of $\O(2)$ we account for are $\bZ_2,D_2,D_4$ and $\O(2)$.
\paragraph{\texorpdfstring{$\O(2)$}{}-action:}

Consider
\[
    \O(2)
    =
    \SOd\cup \O_{-}(2)\simeq \SOd\ltimes\underbrace{\graf{\id,R}}_{\simeq\,\bZ_2}
\]
where
\[
    \SOd
    =
    \grafbb{
    	\begin{pmatrix}
    		\cos\theta & -\sin\theta
    		\\
    		\sin\theta & \hphantom{-}\cos\theta
    	\end{pmatrix}\;\Big.\Big|\;\theta\in[0,2\pi)
    }\simeq\fU(1)\simeq\bS^1     
    \quad
    \textrm{and}
    \quad
    \O_{-}(2)
    =
    \grafbb{
    	\begin{pmatrix}
    		\cos\theta & \hphantom{-}\sin\theta
    		\\
    		\sin\theta & -\cos\theta
    	\end{pmatrix}\;\Big.\Big|\;\theta\in[0,2\pi)
    }
\]
and $R=\begin{pmatrix}-1 & 0\\
	\hphantom{-}0 & 1
\end{pmatrix}$ is the reflection across the $x_2$-axis. 
It acts on the space of the symmetric matrices as follows:
\[
    \varphi:\O(2)\times\textrm{Sym}(2)\fr\textrm{Sym}(2),
    \qquad 
    (Q,\bL)\mapsto\varphi(Q,\bL), 
    \quad\textrm{where}
    \quad\partonb{\varphi(Q,\bL)}_{ij}=Q_{ia}Q_{jb}\bL_{ab}.
\]
To be invariant under the accounted action means that 
\[
    \bL\in\Fix^{\,\varphi}_{\O(2)}\textrm{Sym}(2), 
    \qquad\textrm{i.e.}
    \qquad\varphi(Q,\bL)=\bL
    \qquad
    \forall Q\in \O(2).
\]
In this case,
\begin{align}
	\dim \Fix_{\,\O(2)}^{\,\varphi}\textrm{Sym}(2)
	&
	=\int_{\O(2)}\chi(Q)\,d\mu 
	=
	\int_{\SOd\cup \O_{-}(2)}\chi(Q)\,\d\mu
	\\
	\scriptsize{(\textrm{normalisation})}
	&
	=
	\frac{1}{2}
	\parton{\int_{\SOd}\chi(Q)\,\d\mu
		+\int_{\SOd}\underbrace{\chi(RQ)}_{\equiv\,1}\,\d\mu}
	=
	\frac{1}{2}\parton{\frac{1}{2\pi}\int_{0}^{2\pi} \partonB{4\cos^2\theta-1}\,d\theta+1} = 1.
	\nonumber
\end{align}
Let us determine the structure of the admissible constitutive tensors. We have 
\begin{align}
	\sP_{\O(2)}(\bL)
	\defi
	\frac{1}{\modu{\O(2)}}\int_{\O(2)}\varphi(Q,\bL)\,\d\mu
	=\frac{1}{2}(\bL_{11}+\bL_{22})\,\id. 
\end{align}
Thus, in the isotropic planar case, the curvature tensor reduces to $\bL=l\id=l\,\begin{pmatrix}1 & 0\\
	0 & 1
\end{pmatrix},$ with $l\in\bR^+_*$ (since $\bL$ is positive-definite, i.e. $\bL\in\textrm{Sym}^+(2)$, hence $\bL_{11},\bL_{22}>0$, thus $l=1/2(\bL_{11}+\bL_{22})>0$).

\bigskip

N.B. In this case, to be able to integrate over the full group $\O(2)$, which has two connected components, we use the following facts: since $\SOd$ is a closed subgroup of $\O(2)$ and $\SOd$ is compact, $\restn{\Delta_{\O(2)}}_{\SOd}\equiv\Delta_{\SOd}$ (modular functions) and hence, there exists an invariant Radon measure $\nu$ on the quotient $\O(2)/\SOd$ (in this case the normalized counting measure) such that
\begin{align}
	\int_{\O(2)} f(A)\,\d\mu
	&
	=\int_{\O(2)/\SOd} \int_{\SOd} f(AQ)\,\d\mu\,\d\nu
	=\frac{1}{\modu{\O(2)/\SOd}}\sum_{[A]\in \O(2)/\SOd}  \int_{\SOd} f(AQ)\,\d\mu
	\nonumber
	\\
	&=\frac{1}{2}\parton{\int_{\SOd} f(\id Q)\,\d\mu
		+\int_{\SOd} f(R Q)\,\d\mu}.
\end{align}

\paragraph{\texorpdfstring{$D_4$}{}-action: \vcenteredinclude{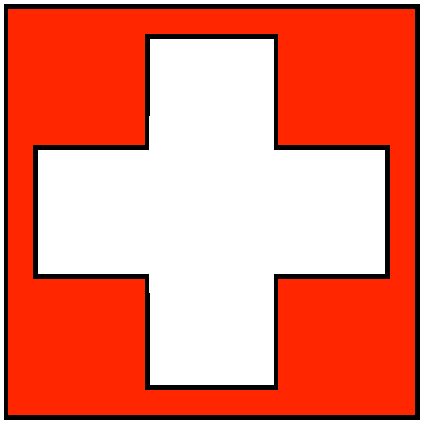}}

   Consider the dihedral group\footnote{
	A dihedral group of order $n$ describes the $2n$ different symmetries of a regular polygon with $n$ sides: $n$ rotational symmetries and $n$ reflection symmetries. The $n$ rotational symmetries are the elements of the cyclic group 
	\[
	     \bZ_n
	     =\graf{\begin{pmatrix}
	     		\cos\frac{2\pi}{k} & -\sin\frac{2\pi}{k}\\
	     		\sin\frac{2\pi}{k} & \cos\frac{2\pi}{k}
	     \end{pmatrix}\stt k\in\{1,\ldots,n\}}
	\]
	and 
	\[
	     D_n\simeq\bZ_n\ltimes\underbrace{\graf{\id,R}}_{\simeq\,\bZ_2},
	\]
	where $R$ is the reflection $\begin{pmatrix}-1 & 0\\
		\hphantom{-}0 & 1
	\end{pmatrix}$.

	} of order four,
 $D_4$ whose cardinality is 8. We have
\begin{align}
	\dim \Fix_{\,D_4}^{\,\varphi}\textrm{Sym}(2)&
	= \frac{1}{\modu{D_4}}\sum_{Q\in D_4}\chi(Q)
	= \frac{1}{\modu{D_4}}\parq{\sum_{Q\in \bZ_{4}}\chi(Q)+\sum_{Q\in R \,\bZ_{4}}\chi(Q)}
	=\frac{1}{8}\parq{\sum_{k\in \graf{1,2,3,4}}\partonB{4\cos^2\frac{k\pi}{2}-1}+4}= 1.
\end{align}
Hence, also in this case, we have the reduction 
$\bL=l\id=l\,\begin{pmatrix}1 & 0\\
	0 & 1
\end{pmatrix},$ with $l\in\bR^+_*$. 

\paragraph{\texorpdfstring{$D_2$}{}-action: \vcenteredinclude{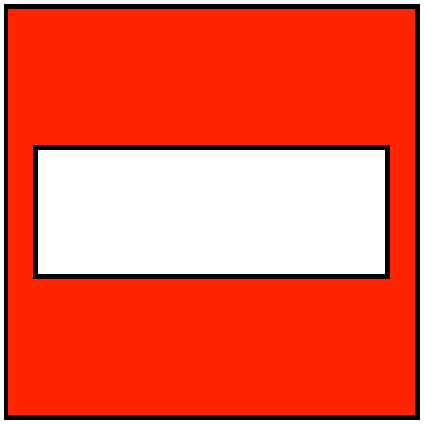}}

Consider the dihedral group $D_2$ whose cardinality is $4$.
We have 
\begin{align}
	\dim \Fix_{\,D_{2}}^{\,\varphi}\textrm{Sym}(2)&
	= \frac{1}{\modu{D_{2}}}\sum_{Q\in D_{2}}\chi(Q)
	= \frac{1}{\modu{D_{2}}}\parq{\sum_{Q\in \bZ_{2}}\chi(Q)+\sum_{Q\in R \,\bZ_{2}}\chi(Q)}
	=
	\frac{1}{4}\parq{2+\sum_{k\in \graf{1,2}}\partonB{4\cos^2 k\pi-1}}
	=
	2.
\end{align}
Therefore, we have the reduction 
$
    \bL
    =
    \begin{pmatrix}
    	l_1 & 0
    	\\
	    0 & l_2
    \end{pmatrix}
$, $l_1,l_2>0$.

\paragraph{\texorpdfstring{$\bZ_{2}$}{}-action: \vcenteredinclude{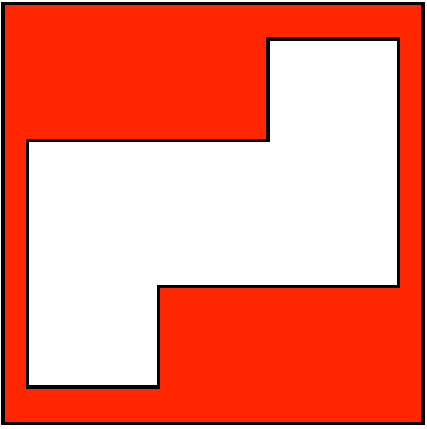}}

Consider the cyclic group $\bZ_2$ whose cardinality is $2$.
We have
\begin{align}
	\dim \Fix_{\,\bZ_{2}}^{\,\varphi}\textrm{Sym}(2)&
	= \frac{1}{\modu{\bZ_{2}}}\sum_{Q\in \bZ_{2}}\chi(Q)
	=\frac{1}{2}\parq{\sum_{k\in \graf{1,2}}\partonB{4\cos^2 k\pi-1}}
	=
	3
	=
	\dim\textrm{Sym}(2)
\end{align}
meaning that $\textrm{Sym}(2)=\Fix_{\,\bZ_{2}}^{\,\varphi}\textrm{Sym}(2)$, i.e., $\sP_{\bZ_2}(\bL)=\bL=\begin{pmatrix} l_1 & l_3\\ l_3 & l_2\end{pmatrix}$ for all $\bL\in\textrm{Sym}(2)$.


\subsubsection{Symmetrization of \texorpdfstring{$\bC_\rre$}{} and \texorpdfstring{$\bC_\rm$}{}}

The symmetrization identity in this case is 
\[
     \Pi:\bR^{2\times 2\times 2\times 2}\fr\Ela(2),
     \qquad
     \Pi
     =\sym(\pi\otimes\pi),
\]
giving component-wise
\begin{align}
	\Pi_{iajblcdk} & =\frac{1}{8}(\delta_{ia}\delta_{jb}\delta_{kc}\delta_{ld}+\delta_{ia}\delta_{jb}\delta_{lc}\delta_{kd}+\delta_{ja}\delta_{ib}\delta_{lc}\delta_{kd}+\delta_{ja}\delta_{ib}\delta_{kc}\delta_{ld}
	\\
	& \quad+\delta_{ka}\delta_{lb}\delta_{ic}\delta_{jd}+\delta_{ka}\delta_{lb}\delta_{jc}\delta_{id}+\delta_{la}\delta_{kb}\delta_{jc}\delta_{id}+\delta_{la}\delta_{kb}\delta_{ic}\delta_{jd}). \nonumber
\end{align}
The characters can be computed via 
$
\chi(Q)=\scal{\varphi_Q}{\pi}=\scal{Q^{\otimes 4}}{\Pi}_{\otimes^8\bR^{2}}
$
giving
\begin{align}
	\chi(Q) &=\frac{1}{8}\left[(\tr Q)^4+2(\tr Q)^{2}\tr Q^2+2\,\tr Q^4+3(\tr Q^2)^{2}\right].
\end{align}
The Cayley-Hamilton theorem allows us to express $\chi(Q)$ as a polynomial in $\tr Q$.\footnote{Avoiding explicit powers of $Q$ reduces the computational cost of evaluating $\chi(Q)$.} From eq. \eqref{Power2} we already know that $Q^2=\tr(Q) Q\pm\id$ and that $\tr Q^2=\tr\partonb{\!\tr(Q) Q\mp\id}=(\tr Q)^2\mp2$. Hence we need to find the expression of $\tr Q^4$ as a functions of the powers of $\tr Q$:
\begin{align*}
	Q^3 &=QQ^2=Q\partonb{\!\tr(Q) Q\mp\id}=\tr(Q) Q^2\mp Q, 
	\\
	\Longrightarrow\quad \tr Q^3 
	&=\tr\parton{\tr(Q) Q^2\mp Q} 
	=\tr Q \tr Q^2\mp \tr Q
	=\tr Q \parton{(\tr Q)^2\mp2}\mp \tr Q
	=\tr Q \parton{(\tr Q)^2\mp3},
	\\
	\\
	Q^4 &=QQ^3=Q\partonb{\!\tr(Q) Q^2\mp Q}=\tr(Q) Q^3\mp Q^2,
	\\
	\Longrightarrow\quad \tr Q^4 
	&=\tr\parton{\tr(Q) Q^3\mp Q^2} 
	=\tr Q \tr Q^3\mp \tr Q^2
	=(\tr Q)^2 \parton{(\tr Q)^2\mp3}\mp \parton{(\tr Q)^2\mp2}
	=(\tr Q)^4\mp 4(\tr Q)^2+2.
\end{align*}
Hence we obtain
\begin{align*}
	\chi(Q) & =\frac{1}{8}\left[(\tr Q)^{4}+2(\tr Q)^{2}\tr Q^{2}+2\,\tr Q^{4}+3(\tr Q^{2})^{2}\right]
	\\
	& =\frac{1}{8}\biggl[(\tr Q)^{4}+2(\tr Q)^{2}\parton{(\tr Q)^{2}\mp2}+2\parton{(\tr Q)^{4}\mp4(\tr Q)^{2}+2}+3\parton{(\tr Q)^{2}\mp2}^{2}\biggr]
	\\
	& =\frac{1}{8}\biggl[(\tr Q)^{4}+2(\tr Q)^{4}\mp4(\tr Q)^{2}+2(\tr Q)^{4}\mp8(\tr Q)^{2}+4+3\parton{(\tr Q)^{4}+4\mp4(\tr Q)^{2}}\biggr]
	\\
	& =\frac{1}{8}\biggl[(\tr Q)^{4}+2(\tr Q)^{4}\mp4(\tr Q)^{2}+2(\tr Q)^{4}\mp8(\tr Q)^{2}+4+3(\tr Q)^{4}+12\mp12(\tr Q)^{2}\biggr]
	\\
	& =\frac{1}{8}\biggl[8(\tr Q)^{4}\mp24(\tr Q)^{2}+16\biggr]
	=(\tr Q)^{4}\mp3(\tr Q)^{2}+2,
\end{align*}
i.e.
\[
\chi(Q)=\begin{cases}
	(\tr Q)^{4}-3(\tr Q)^{2}+2 & \textrm{if }Q\in SO(2),
	\\
	\\
	(\tr Q)^{4}+3(\tr Q)^{2}+2 & \textrm{if }Q\in O^{-}(2).
\end{cases}
\]
\paragraph{\texorpdfstring{$D_4$}{}-action:}
Consider the dihedral group $D_4$ whose cardinality is 8. Then 
\begin{align*}
		\dim \Fix_{\,C_{4v}}^{\,\varphi}\Ela(2)
	&
	= \frac{1}{\modu{C_{4v}}}\sum_{Q\in C_{4v}}\chi(Q)
	= \frac{1}{\modu{C_{4v}}}\parq{\sum_{Q\in C_{4}}\chi(Q)+\sum_{Q\in R C_{4}}\chi(Q)}
	\\
	&=\frac{1}{8}\sum_{k\in \graf{1,2,3,4}}\parqB{2\partonb{ \cos (\pi  k)+\cos (2 \pi  k)+1}+2}= 
	3.
\end{align*}
The structure of the symmetrized tensors is obtained via 
\begin{align}
	\sP_{D_4}(\bC)&
	= \frac{1}{\modu{D_4}}\sum_{Q\in D_4}\varphi(Q,\bC)
	= \frac{1}{\modu{D_4}}\sum_{Q\in \bZ_{4}}\partonB{\varphi(Q,\bC)+\varphi(QR,\bC)},
\end{align}
giving 

\begin{equation}
	\tcbhighmath[drop fuzzy shadow]{
		\begin{pmatrix}
			\widetilde \bC_{11} & \widetilde \bC_{12} & 0
			\\
			\widetilde \bC_{12} & \widetilde \bC_{11} & 0
			\\
			0 & 0& \widetilde \bC_{22}
		\end{pmatrix}
	}
\end{equation}
after considering a suitable Voigt isomorphism.

\subsection{2D-micromorphic model}

The major difference with the relaxed micromorphic model is the curvature term. Indeed, in the classical Eringen-Mindlin micromorphic model, we have that 
\begin{align}
	W_{\textrm{curv}}(\nabla P)&\defi\frac{1}{2}\scalb{\gL\,\nabla P}{\nabla P}_{\bR^{2\times 2\times 2}},
\end{align}
where 
\[
    \gL\in\text{Sym}^+(\otimes^3\bR^2,\otimes^3\bR^2)\subseteq \text{Sym}(\otimes^3\bR^2,\otimes^3\bR^2)
\]
with the latter being a vector space of dimension $36$. Moreover, the classical micromorphic model would allow for additional mixed terms as 
$
     \scal{\sym(\nabla\,u-P)}{\sym P}.
$
We set 
\[
    \High(2)
    \defi 
    \text{Sym}(\otimes^3\bR^2,\otimes^3\bR^2)
    \qquad
    \textrm{and}
    \qquad
    \High^+\!(2)
    \defi 
    \text{Sym}^+\!(\otimes^3\bR^2,\otimes^3\bR^2).
\]
%
The only interesting term is the curvature term. In this case we remind that 
\[
        P\in\bR^{2\times 2},
        \qquad
        \nabla P\in\otimes^3\bR^2,
        \qquad
        \gL\in\High^+\!(2)
        \defi
        \textrm{Sym}^+\!\parton{\otimes^3\bR^2,\otimes^3\bR^2}.
\]
The resultant action of $\fO(2)$ is
\begin{equation}
	\varphi: \fO(2)\times\High(2)\fr\High(2),
	\qquad
	(Q,\gL)
	\mapsto
	\varphi(Q,\gL),
	\quad
	\text{where}
	\quad
	\partonb{\varphi(Q,\gL)}_{ijklmn}
	=
	Q_{ia}Q_{jb}Q_{kc}Q_{ld}Q_{me}Q_{nf}\gL_{abcdef},
	\nonumber
\end{equation}
and the symmetrization identity is
\begin{equation}
	\Pi^{\,\High(2)}: 
	\underbrace{\parton{\otimes^3\bR^2}\otimes\parton{\otimes^3\bR^2}}_{=\,\otimes^6\bR^2} \twoheadrightarrow
	\underbrace{\textrm{Sym}\parton{\otimes^3\bR^2,\otimes^3\bR^2}}_{=:\,\High(2)},
	\qquad
	\Pi_{aibjchdkemfn}^{\,\High(2)} 
	=\frac{1}{2}\partonb{
		\delta_{ai}\delta_{bj}\delta_{ch}\delta_{dk}\delta_{em}\delta_{fn}
		+
		\delta_{ak}\delta_{bm}\delta_{cn}\delta_{di}\delta_{ej}\delta_{fh}
	}.
	\nonumber
\end{equation}
The characters can be computed via 
$
     \chi(Q)
     =
     \textrm{tr}\,(\varphi_Q\circ\Pi^{\High(2)})
     =
     \scal{\varphi_Q\circ\Pi^{\High(2)}}{\otimes^6\id_2}_{\otimes^{1\!2}\bR^{2}},
$
giving
\begin{align}
	\chi(Q) &=\partonb{
		\delta_{ai}\delta_{bj}\delta_{ch}\delta_{dk}\delta_{em}\delta_{fn}
		+
		\delta_{ak}\delta_{bm}\delta_{cn}\delta_{di}\delta_{ej}\delta_{fh}
	}Q_{ia}Q_{jb}Q_{hc}Q_{kd}Q_{me}Q_{nf}
	\nonumber
	\\
	&
	=
	\frac{1}{2}\Big(
	\delta_{ai}\delta_{bj}\delta_{ch}\delta_{dk}\delta_{em}\delta_{fn}Q_{ia}Q_{jb}Q_{hc}Q_{kd}Q_{me}Q_{nf}
	+
	\delta_{ak}\delta_{bm}\delta_{cn}\delta_{di}\delta_{ej}\delta_{fh}
	Q_{ia}Q_{jb}Q_{hc}Q_{kd}Q_{me}Q_{nf}
	\Big)
	\nonumber
	\\
	&
	=
	\frac{1}{2}\partonB{
		\underbrace{\delta_{ia}Q_{ia}}_{\tr Q}
		\underbrace{\delta_{jb}Q_{jb}}_{\tr Q}
		\underbrace{\delta_{hc}Q_{hc}}_{\tr Q}
		\underbrace{\delta_{kd}Q_{kd}}_{\tr Q}
		\underbrace{\delta_{me}Q_{me}}_{\tr Q}
		\underbrace{\delta_{nf}Q_{nf}}_{\tr Q}
	%
	+
		\underbrace{\delta_{ak}Q_{kd}\delta_{di}Q_{ia}}_{\tr Q^2}
		\underbrace{\delta_{bm}Q_{me}\delta_{ej}Q_{jb}}_{\tr Q^2}
		\underbrace{\delta_{cn}Q_{nf}\delta_{fh}Q_{hc}}_{\tr Q^2}
	}
	\nonumber
	\\
	&
	=
	\frac{1}{2}\partonB{(\tr Q)^6+(\tr Q^2)^3}.
\end{align}
Via the obtained formula $\tr Q^2=(\tr Q)^2\mp2$ we find
\begin{align}
	(\tr Q^2)^3 &=((\tr Q)^2\mp2)^3=(\tr Q)^6\mp6(\tr Q)^4+12(\tr Q)^2\mp8
\end{align}
and thus
\begin{align}
	\chi(Q) &=\frac{1}{2}\partonB{(\tr Q)^6+(\tr Q)^6\mp6(\tr Q)^4+12(\tr Q)^2\mp8}=(\tr Q)^6\mp3(\tr Q)^4+6(\tr Q)^2\mp4.
\end{align}
\paragraph{\texorpdfstring{$D_4$}{}-action:}
Concerning the number of independent components, we have
\begin{align*}
		\dim \Fix_{\,D_4}^{\,\varphi}\!\High(2)
	&
	= \frac{1}{\modu{D_4}}\sum_{Q\in D_4}\chi(Q)
	= \frac{1}{\modu{D_4}}\parq{\sum_{Q\in \bZ_{4}}\chi(Q)+\sum_{Q\in R \bZ_{4}}\chi(Q)}
	\\
	&
	=\frac{1}{8}\sum_{k\in \graf{1,2,3,4}}\parqB{2 \partonb{6 + 9 \cos(k \pi) + 3 \cos(2 k \pi) + \cos(3 k \pi)}+2}= 
	10.
\end{align*}
The symmetrized tensor $\sP_{D_4}(\gL)$ has the structure
\[
	\tcbhighmath[drop fuzzy shadow]{
	\left(
	\def\arraystretch{1.5}
	\begin{array}{c|c}
		   \begin{array}{cccc}
		   	      \widetilde \gL_{11} & \widetilde \gL_{12} & \widetilde \gL_{13} & \widetilde \gL_{14} 
		   	      \\
		   	       & \widetilde \gL_{22} & \widetilde \gL_{23} & \widetilde \gL_{24}
		   	      \\
		   	       \text{sym}&  & \widetilde \gL_{33} & \widetilde \gL_{34} 
		   	      \\
		   	       &  & & \widetilde \gL_{44}
		   \end{array}
		   &
		   0
		   \\
		   \hline
		   0
		   &
		    \begin{array}{cccc}
		    	\widetilde \gL_{11} & \widetilde \gL_{12} & \widetilde \gL_{13} & \widetilde \gL_{14} 
		    	\\
		    	& \widetilde \gL_{22} & \widetilde \gL_{23} & \widetilde \gL_{24}
		    	\\
		    	\text{sym}&  & \widetilde \gL_{33} & \widetilde \gL_{34} 
		    	\\
		    	&  & & \widetilde \gL_{44}
		    \end{array}
	\end{array}
	\right)
}
\]
after considering a suitable Voigt isomorphism. 
%
%
By directly comparing the two curvature terms for the $D_4$ case, one for the relaxed micromorphic model and the one for the classical micromorphic model, we can observe that accounting for $\Curl P$ as a curvature term significantly reduces the number of elastic moduli. This simplification is advantageous when attempting to characterize them. \cor{In this scenario, we also provide the explicit expression for the energy density:
\[
W_{\textrm{Eringen}}(\nabla u,P,\nabla P)
=
\underbrace{\frac{1}{2}\scal{\bCe \, \sym(\nabla u-P)}{\sym(\nabla u-P)}}_{W_{\textrm{e}}(\nabla u,P)}
+
\underbrace{\frac{1}{2}\scal{\bCm \, \sym P}{\sym P}}_{W_{\textrm{micro}}(P)}
+
\underbrace{\frac{1}{2} \mu\,L_c^2
	\scal{\gL\,\nabla P}{\nabla P}}_{W_{\text{curv}}(\nabla P)}
\]
where
\begin{align*}
	W_{\text{curv}}(\nabla P)
	&=
	\frac{1}{2} \mu\,L_c^2
	\scal{\gL\,\nabla P}{\nabla P}_{\bR^2\otimes\bR^2\otimes\bR^2}
	\\[2mm]
	&=
	\frac{1}{2} \mu\,L_c^2\bigg[
	(P_{11,1}^2+P_{22,2}^2) \, \widetilde \gL_{11}
	+
	2 (P_{11,1} \,P_{12,2}
	+
	P_{21,1} \, P_{22,2}) \, \widetilde\gL_{12}
	+
	2 P_{11,1} \, P_{22,1} \, \widetilde\gL_{13}
	\\[2mm]
	&
	\quad
	+
	2 P_{11,2} \, P_{22,2} \, \widetilde\gL_{13}
	+
	2 P_{11,1} \, P_{21,2} \, \widetilde\gL_{14}
	+
	2 \,P_{12,1} \, P_{22,2} \, \widetilde\gL_{14}
	+
	\partonb{P_{12,2}^2+P_{21,1}^2} \, \widetilde\gL_{22}
	\\[2mm]
	&
	\quad
	+
	2 P_{11,2} \, P_{21,1} \, \widetilde\gL_{23}
	+
	2 P_{12,2} \, P_{22,1} \, \widetilde\gL_{23}
	+
	2 \, P_{12,1} \, P_{21,1} \, \widetilde\gL_{24}
	+
	2 \, P_{12,2} \, P_{21,2} \, \widetilde\gL_{24}
	\\[2mm]
	&
	\quad
	+
	P_{11,2}^2 \, \widetilde\gL_{33}
	+
	P_{22,1}^2 \, \widetilde\gL_{33}
	+
	2 \, P_{11,2}\,  P_{12,1} \, \widetilde\gL_{34}
	+
	2 \, P_{21,2} \, P_{22,1} \, \widetilde\gL_{34}
	+
	\partonb{P_{12,1}^2+P_{21,2}^2} \, \widetilde\gL_{44}\bigg].
\end{align*}
}

\paragraph{\texorpdfstring{$D_2$}{}-action:}
Concerning the number of independent components, we have
\begin{align*}
	\dim \Fix_{\,D_2}^{\,\varphi}\!\High(2)
	&
	= 
	\frac{1}{\modu{D_2}}\sum_{Q\in D_2}\chi(Q)
	= 
	\frac{1}{\modu{D_2}}\parq{\sum_{Q\in \bZ_{2}}\chi(Q)+\sum_{Q\in R \bZ_{2}}\chi(Q)}
	=
	20.
\end{align*}
The symmetrized tensor $\sP_{D_2}(\gL)$ has the structure
\[
\tcbhighmath[drop fuzzy shadow]{
	\left(
	\def\arraystretch{1.5}
	\begin{array}{cccc|cccc}
		\widetilde \gL_{11} & \widetilde \gL_{12} & \widetilde \gL_{13} & \widetilde \gL_{14} & 
		0 & 0 & 0 & 0 \\
		& \widetilde \gL_{22} & \widetilde \gL_{23} & \widetilde \gL_{24} & 
		0 & 0 & 0 & 0 \\
		\text{sym} &  & \widetilde \gL_{33} & \widetilde \gL_{34} & 
		0 & 0 & 0 & 0 \\
		& & & \widetilde \gL_{44} & 
		0 & 0 & 0 & 0 \\
		\hline
		0 & 0 & 0 & 0 & \widetilde \gL_{55} & \widetilde \gL_{56} & \widetilde \gL_{57} & \widetilde \gL_{58} \\
		0 & 0 & 0 & 0 & & \widetilde \gL_{66} & \widetilde \gL_{67} & \widetilde \gL_{68} \\
		0 & 0 & 0 & 0 & \text{sym} & & \widetilde \gL_{77} & \widetilde \gL_{78} \\
		0 & 0 & 0 & 0 &  & &  & \widetilde \gL_{88} \\
	\end{array}
	\right)
}
\]
after introducing a suitable Voigt isomorphism.
\paragraph{\texorpdfstring{$\bZ_2$}{}-action:}
Concerning the number of independent components, we have
\begin{align*}
	\dim \Fix_{\,\bZ_2}^{\,\varphi}\!\High(2)
	&
	= 
	\frac{1}{\modu{\bZ_2}}\sum_{Q\in \bZ_2}\chi(Q)
	=
	36.
\end{align*}
The symmetrized tensor $\sP_{\bZ_2}(\gL)$ has the structure
\[
\tcbhighmath[drop fuzzy shadow]{
	\left(
	\def\arraystretch{1.5}
	\begin{array}{c|c}
		\begin{array}{cccc}
			\widetilde \gL_{11} & \widetilde \gL_{12} & \widetilde \gL_{13} & \widetilde \gL_{14} 
			\\
			 & \widetilde \gL_{22} & \widetilde \gL_{23} & \widetilde \gL_{24}
			\\
			\text{sym} &  & \widetilde \gL_{33} & \widetilde \gL_{34} 
			\\
			 &  &  & \widetilde \gL_{44}
		\end{array}
		&
		\begin{array}{cccc}
			\widetilde \gL_{15} & \widetilde \gL_{16} & \widetilde \gL_{17} & \widetilde \gL_{18} 
			\\
			\widetilde \gL_{25} & \widetilde \gL_{26} & \widetilde \gL_{27} & \widetilde \gL_{28}
			\\
			\widetilde \gL_{35} & \widetilde \gL_{36} & \widetilde \gL_{37} & \widetilde \gL_{38} 
			\\
			\widetilde \gL_{45} & \widetilde \gL_{46} & \widetilde \gL_{47} & \widetilde \gL_{48}
		\end{array}
		\\
		\hline
		\text{sym}
		&
		\begin{array}{cccc}
			\widetilde \gL_{55} & \widetilde \gL_{56} & \widetilde \gL_{57} & \widetilde \gL_{58} 
			\\
			& \widetilde \gL_{66} & \widetilde \gL_{67} & \widetilde \gL_{68}
			\\
			\text{sym} & & \widetilde \gL_{77} & \widetilde \gL_{78} 
			\\
			 &  &  & \widetilde \gL_{88}
		\end{array}
	\end{array}
	\right)
}
\]
after introducing a suitable Voigt isomorphism.

\section*{Conclusions}

We applied the fundamental theoretical framework developed in the work by Danescu \cite{danescu1997number} to determine the number of independent components and the representations of tensor classes involved in generalized continuum models. This approach offers several advantages, primarily its simplicity and clarity in implementation.

It is worth noting that in recent years, various research directions \cite{olive2017minimal,desmoratspace,auffray2017handbook,kolev2018characterization,abramian2020recovering,desmorat2023computation} have emerged with the goal of establishing a priori the permissible symmetry classes for a given category of tensors, building upon and extending the findings presented in previous works \cite{forte1996symmetry}.

\paragraph*{Acknowledgements.}
We would like to express our sincere gratitude to the reviewers for their valuable feedback and insightful comments on our manuscript. In particular, we appreciate the clarification provided on Remark 6, which significantly enhanced the precision and depth of our discussion.

\section*{Appendix}

\appendix

\section{Haar integration }

We want to show how it is possible to built up the Haar measure on the Lie group $\So$ directly using the Cardan's angles parametrization. To facilitate a better understanding of how the Haar measure operates, we first provide some classical examples. The general procedure, which we will present in Appendix~\ref{section:local_chart}, will extend beyond the initial examples.

\paragraph*{Haar measure on the multiplicative group of positive reals:}
\label{exa: r}Consider the multiplicative group $\left(\R^{+}_*,\cdot\right)$.
The Lebesgue measure $\lambda$ is not invariant with
respect to the multiplication of real numbers but, instead of $\lambda$,
we can define the measure 
\begin{equation}
     \lambda_{*}(\mathsf{A})
     \defi
     \underset{x\in\mathsf{A}}{\int}\frac{1}{x}\,\d\lambda(x),
     \qquad
     \textrm{for all }\mathsf{A}\in\sB(\R^{+}),
     \label{eq: r}
\end{equation}
where $\mathscr{B}(\R^{+})$ denotes the $\sigma-$algebra of Borel sets. If $\mathsf{A}=\left[a,b\right]$ with $a,b>0$, we can explicitly
calculate the introduced measure: 
\[
    \lambda_{*}([a,b])
    \defi
    \int_{a}^{b}\frac{1}{x}\,\d\lambda(x)
    =
    \log b-\log a.
\]
To verify that $\lambda_{*}$ is indeed a Haar measure, we can simply show the invariance
with respect to the left multiplication: using a simple change of variable ($y=x_{0}\,x\;\rightsquigarrow\;d\lambda(y)=x_{0}\,\d\lambda(x)$), we find
\begin{align*}
    \lambda_{*}(x_{0}\,\mathsf{A}) 
    & 
    =
    \underset{y\in x_{0}\,\mathsf{A}}{\int}\frac{1}{y}\,d\lambda(y)
    =
    \underset{x\in\mathsf{A}}{\int}\frac{x_{0}\,\d\lambda(x)}{x_{0}\,x}=\lambda_{*}(A),
\end{align*}
and, in the specific example in which $\mathsf{A}=\left[a,b\right]$
with $a,b>0$,
\begin{align*}
\lambda_{*}\!\left(\left[x_{0}\,a,x_{0}\,b\right]\right) & =\log\left(x_{0}\,b\right)-\log\left(x_{0}\,a\right)=\log x_{0}+\log b-\log x_{0}-\log a=\log b-\log a=\lambda_{*}\!\left(\left[a,b\right]\right).
\end{align*}

\begin{figure}[H]
	\begin{centering}
		\includegraphics[scale=0.35]{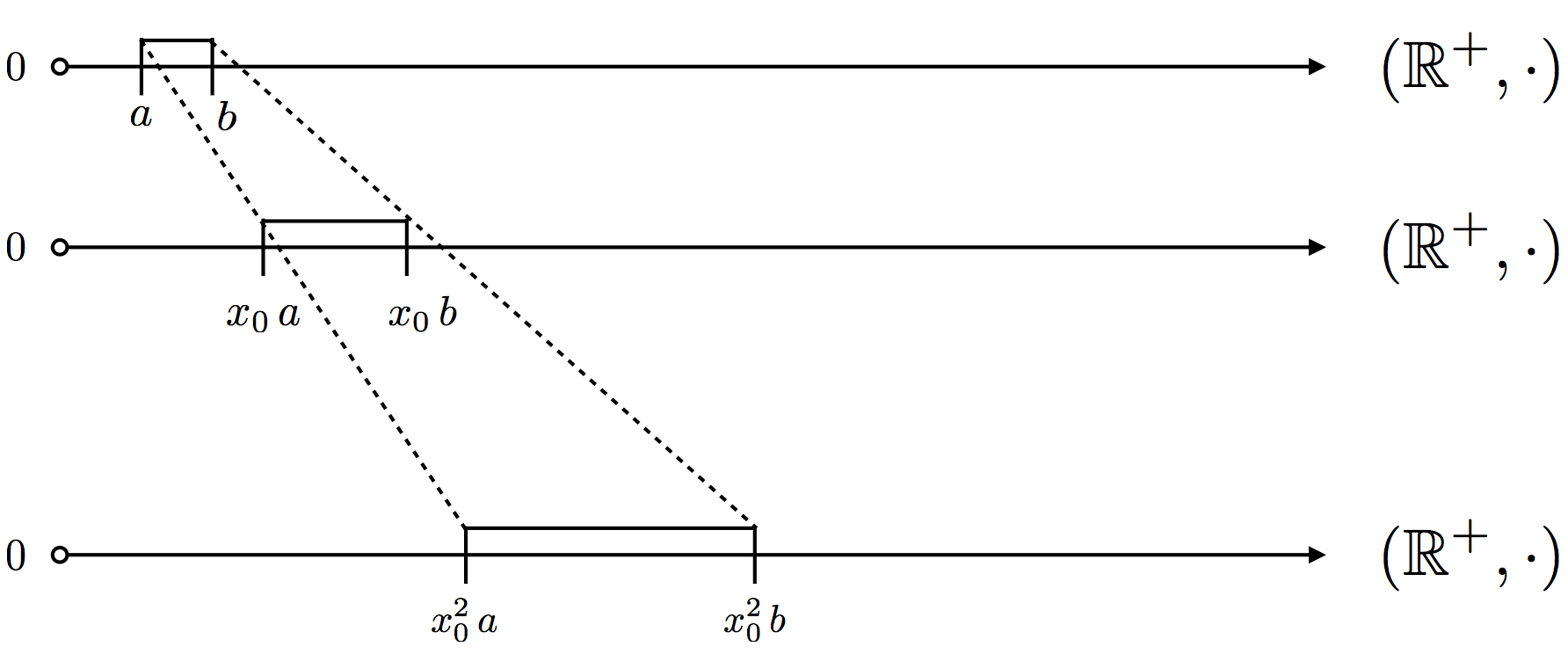}
		\begin{tikzpicture}
              \node[](1)at(-1, 2) {$\lambda_*\left( \left[x_0^2\,a,x_0^2\,b \right] \right)=\lambda_*\left(\left[x_0\,a,x_0\,b \right]\right) =\lambda_*\left(\left[a,b \right]\right) $};
              \node[](2)at(-1, 1.4) {$\lambda\left( \left[x_0\,a,x_0\,b \right] \right)=x_0\,\lambda\left(\left[a,b \right]\right) \neq\lambda\left(\left[a,b \right]\right) $};
              \node[](3)at(-1, 0.8) {$\lambda\left( \left[x_0^2\,a,x_0^2\,b \right] \right)=x_0^2\,\lambda\left(\left[a,b \right]\right) \neq\lambda\left(\left[a,b \right]\right) $};
		\end{tikzpicture}	
		\par\end{centering}
		\caption{Difference between Lebesgue and Haar measures on $\left(\protect\R^{+},\cdot\right)$}
\end{figure}

\paragraph*{Haar measure on the multiplicative group of the invertible matrices:}
\label{exa:rr}Now, consider the multiplicative group $\left(\mathrm{GL}^{+}\!\left(n\right),\cdot\right)$,
where $\mathrm{GL}^{+}\!\left(n\right)$ is the open subset of $\R^{n\times n}$
constituted by all matrices with positive determinant. The Lebesgue
measure $\lambda{}^{n\times n}\left(M\right)$ on $\R^{n\times n}$
is not invariant with respect to the multiplication between matrices,
but, similarly to $\R^+_*$,
we can define the measure
\begin{equation}\label{haar matrices}
     \lambda_{*}^{n\times n}(\mathsf{A})
     \defi
     \underset{M\in\mathsf{A}}{\int}\frac{1}{\det^{n}M}\,\d\lambda{}^{n\times n}(M)
     \qquad
     \textrm{for}\;\mathsf{A}\;\text{measurable}.
\end{equation}
Thanks to the change of variables formula, we can again
verify that the introduced measure (\ref{haar matrices}) is a Haar measure:
\begin{align*}
    \lambda_{*}^{n\times n}(M_{0}\,\mathsf{A}) 
    & 
    =
    \underset{N\in M_{0}\mathsf{A}}{\int}\frac{1}{\left(\det N\right)^{n}}\,\d\lambda{}^{n\times n}(N)
    \quad
    \left(\textrm{Change of variables: }N
    =
    M_{0}\,M\;\rightsquigarrow\;\d\lambda{}^{n\times n}(N)
    =
    \left({\textstyle \det}M_{0}\right)^{n}\d\lambda{}^{n\times n}(M)\right)
    \\
    & 
    =
    \underset{M\in\mathsf{A}}{\int}\frac{\left(\det M_{0}\right)^{n}\,\d\lambda{}^{n\times n}\left(M\right)}{\left(\det M_{0}\right)^{n}\,(\det M)^{n}}
    =
    \lambda_{*}^{n\times n}(\mathsf{A}).
\end{align*}
In order to best illustrate the meaning of the Haar measure, let us
consider the simpler situation in which $n=2$, $\mathsf{A}$ is a
four-dimensional rectangular set $[a,b]\times[a,b]\times[a,b]\times[a,b]$
and $\left\{ \mathsf{C}_{i}\right\} _{i=1}^{k}$ is a partition in
small 4-dimensional cubes of $\mathsf{A}$. Then
\begin{align*}
     \lambda_{*}^{2\times 2}\left(M_{0}\,\mathsf{A}\right) & =\underset{N\in M_{0}\mathsf{A}}{\int}\frac{1}{\left(\det N\right)^{2}}\,\d\lambda{}^{2\times2}(N)
     \\
     \left(M_{i}\textrm{ is the central point of }\mathsf{C}_{i}\right) 
     & 
     \approx
     \sum_{i=1}^{k}\frac{1}{(\det N_{i})^{2}}\,\lambda^{2\times2}(M_{0}\fC_{i})
     \overset{N_{i}=M_{0}M_{i}}{=}\sum_{i=1}^{k}\frac{1}{(\det M_{0})^{2}(\det M_{i})^{2}}\,(\det M_{0})^2\,\lambda^{2\times2}(\fC_{i})
     \\
     & 
     =
     \sum_{i=1}^{k}\frac{1}{\left(\det M_{i}\right)^{2}}\,\lambda{}^{2\times2}\left(\mathsf{C}_{i}\right)\approx\underset{M\in\mathsf{A}}{\int}\frac{1}{\left(\det M\right)^{2}}\,\d\lambda{}^{2\times2}(M)
     =
     \lambda_{*}^{2\times 2}(\mathsf{A}).
\end{align*}

\paragraph{Haar measure on $\text{SO}(2)$:}

Recall that
$$ \text{SO}(2)=\left\lbrace R_\vartheta\defi \begin{pmatrix}\cos\vartheta & -\sin\vartheta\\
\sin\vartheta & \cos\vartheta
\end{pmatrix} \Big. \Big|\; \vartheta\in\left[0,2\pi \right) \right\rbrace.  $$
This Lie group is compact and thus admits an unique left-invariant normalized Haar measure (see \cite{deitmar2014principles}). In order to explicitly obtain this measure, let us remark that every real-valued integrable function $\widehat{f}$ on $\text{SO}(2)$ can be identified with a $2\pi$-periodic function $f_{\text{per}}\colon\R\fr\R$ by setting
\begin{equation*}
    f_{\text{per}}(\vartheta)
    =
    \widehat{f}(R_\vartheta)
\end{equation*}
on $\left[0,2\pi \right)$ and extending $f_{\text{per}}$ to $\R$ by periodicity.
In this way, we can define the Haar measure $\mu$ on  $\text{SO}(2)$ by
\begin{equation}\label{Haar circle}
     \int_ {\text{SO}(2)}  \widehat{f}(R_\vartheta)\,\d\mu
     \defi
     \frac{1}{2\pi}\int_0^{2\pi} f_{\text{per}}(\vartheta)\,\d\vartheta,
\end{equation}
where $\vartheta$ is the Lebesgue measure on $\R$. In order to verify that the stated definition gives a left invariant measure, we have to show that
$$ \int_ {\text{SO}(2)} \left( \widehat{f}\circ L_{R_{\vartheta_0}}\right) (R_\vartheta)\,\d\mu=\int_ {\text{SO}(2)}  \widehat{f}(R_\vartheta)\,\d\mu \qquad\forall\vartheta_0\in\R. $$
Recalling that
$$ R_{\vartheta_0}\, R_\vartheta=\begin{pmatrix}\cos\vartheta_0 & -\sin\vartheta_0\\
\sin\vartheta_0 & \cos\vartheta_0
\end{pmatrix}\,\begin{pmatrix}\cos\vartheta & -\sin\vartheta\\
\sin\vartheta & \cos\vartheta
\end{pmatrix}=\begin{pmatrix}\cos(\vartheta_0+\vartheta) & -\sin(\vartheta_0+\vartheta)\\
\sin(\vartheta_0+\vartheta) & \cos(\vartheta_0+\vartheta)
\end{pmatrix}=R_{\vartheta_0+\vartheta}, $$
we find
\begin{align}
   \int_ {\text{SO}(2)} \left( \widehat{f}\circ L_{R_{\vartheta_0}}\right) (R_\vartheta)\,\d\mu
   &=
   \int_ {\text{SO}(2)} \widehat{f} (R_{\vartheta_0+\vartheta})\,\d\mu=\frac{1}{2\pi}\int_0^{2\pi} f_{\text{per}}(\vartheta_0+\vartheta)\,\d\vartheta
   \nonumber
   =
   \frac{1}{2\pi}\int_0^{2\pi} f_{\text{per}}(\vartheta)\,\d\vartheta=\int_ {\text{SO}(2)}  \widehat{f}(R_\vartheta)\,\d\mu
\end{align}
for all $\vartheta_0\in\R$.

\paragraph{Haar measure on $\text{O}(2)$:}
Setting
$$ \text{O}_{-}(2)\defi\left\lbrace \begin{pmatrix}\cos\vartheta & \sin\vartheta\\
\sin\vartheta & -\cos\vartheta
\end{pmatrix} \Big. \Big|\; \vartheta\in\left[0,2\pi \right) \right\rbrace \subseteq\R^{2\times2},  $$
we observe that
$$ \text{O}(2)=\text{SO}(2)\cup\text{O}_{-}(2). $$
The group $\text{O}(2)$ is compact as well, but unlike $\text{SO}(2)$ it has two disconnected components. For this reason, the idea for defining the Haar measure consists in adapting the reasoning for $\text{SO}(2)$ but considering that, in this case, the \enquote{size} of the involved group is doubled. We set
\begin{align}\label{Haar O(2)}
     \int_ {\text{O}(2)}  \widehat{f}(R_\vartheta)\,\d\mu
     &=\int_ {\text{SO}(2)}  \widehat{f}(R^+_\vartheta)\,\d\mu+\int_ {\text{O}_-(2)}  \widehat{f}(R^-_\vartheta)\,\d\mu 
     =
     \frac{1}{4\pi}\left(\int_0^{2\pi} f^+_{\text{per}}(\vartheta)\,\d\vartheta+\int_0^{2\pi} f^-_{\text{per}}(\vartheta)\,\d\vartheta \right),
\end{align} 
where 
$$  f^+_{\text{per}}(\vartheta)=\widehat{f}(R_\vartheta)\;\text{for}\;R_\vartheta\in\text{SO}(2)\quad\text{and}\quad f^-_{\text{per}}(\vartheta)=\widehat{f}(R_\vartheta)\;\text{for}\;R_\vartheta\in\text{O}_-(2)
$$
and the factor $1/4\pi$ is taken to normalize the Haar measure.
\subsection{Haar measure induced by a local chart}
\label{section:local_chart}

Let $\sG$ be a Lie group and $(\sU,\varphi)$ a local chart with $\sU$ open subset of $\R^n$ such that $0\in\sU$ and $\varphi(0)=e$. In this Appendix, following \cite[Exercice 1.8 pag.32]{golubitsky2012singularities}, we show how it is possible to define a Haar measure on $\varphi(\sU)$ starting from the Lebesgue measure on $\sU$. 

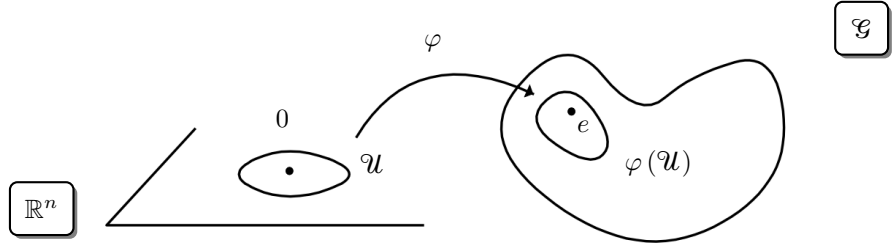
\begin{figure}[H]
	\begin{centering}
		\begin{tikzpicture}
		\tikzset{
			nodeoformula/.style={rectangle,rounded corners=0.1cm,drop shadow={shadow xshift=0.5mm, shadow yshift=-0.5mm,opacity=1},draw=black, top color=white, bottom color=white, thick, inner sep=2mm, minimum size=2em, text centered},
			nodepoint/.style={circle,draw=gray,fill=gray,inner sep=0.8mm}
		}
		\node[anchor=south west,inner sep=0] at (0,0) {\includegraphics[scale=0.55]{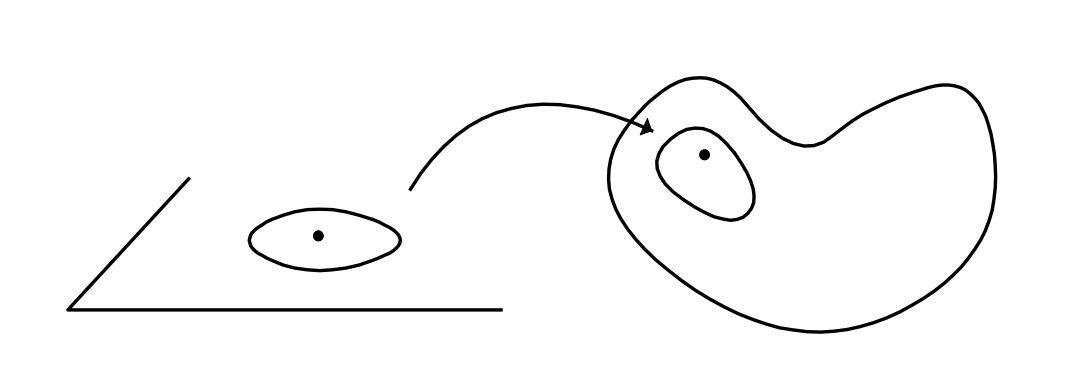}};
		\node[nodeoformula](1)at(-0.2, 1) {$\mathbb{R}^n$};
		\node[nodeoformula](2)at(10.7, 3.4) {$\sG$};
		\node[style=none](3)at(3, 2.2) {$0$};
		\node[style=none](4)at(7, 2.1) {$e$};
		\node[style=none](5)at(4.2, 1.6) {$\sU$};
		\node[style=none](5)at(8, 1.6) {$\varphi\,(\sU)$};
		\node[style=none](6)at(5, 3.2) {$\varphi$};
		
		\end{tikzpicture}
		\par\end{centering}
	\caption{\label{fig:Left-invariance-of-a}Local chart for a Lie group.}
\end{figure} 
\noindent In order to define the Haar integral we need to introduce some auxiliary function. For every $h\in\varphi(\sU)$ we can find an open neighbourhood of the origin $\sV_h$ on which the function\footnote{Such an open neighbourhood $\sV_h$ of the origin always exists. Indeed, consider an open neighbourhood $A_h$ of $h\in\varphi(\sU)$ in $\varphi(\sU)$. Then, $L_{h^{-1}}(A_h)$ is an open neighbourhood of the identity $e$ and thus the intersection $\varphi(\sU)\cap L_{h^{-1}}(A_h)$  is an open neighbourhood of $e$ contained in $\varphi(\sU)$. Thus, considering $\varphi^{-1}(\varphi(\sU)\cap L_{h^{-1}}(A_h))$, we obtain  an open neighbourhood of the origin in $\R^n$ on which the function $\psi_h$ is well defined.}(see Figures \ref{Vh} and \ref{psih})
$$ \psi_h\colon\sV_h\subseteq\sU\fr\sU\qquad\psi_h\defi\varphi^{-1}\circ L_h\circ\varphi$$
is well defined.
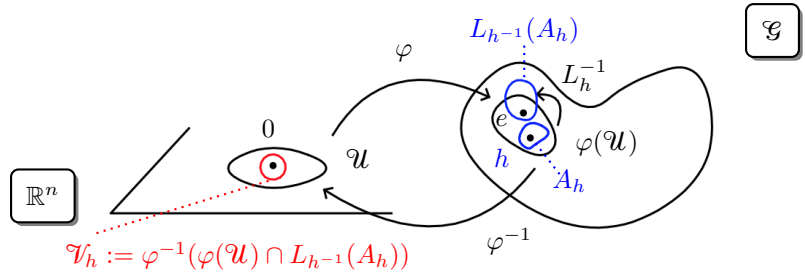
\begin{figure}[H]
	\begin{centering}
		\begin{tikzpicture}
		\tikzset{
			nodeoformula/.style={rectangle,rounded corners=0.1cm,drop shadow={shadow xshift=0.5mm, shadow yshift=-0.5mm,opacity=1},draw=black, top color=white, bottom color=white, thick, inner sep=2mm, minimum size=2em, text centered},
			nodepoint/.style={circle,draw=gray,fill=gray,inner sep=0.8mm}
		}
		\node[anchor=south west,inner sep=0] at (0,0) {\includegraphics[scale=0.55]{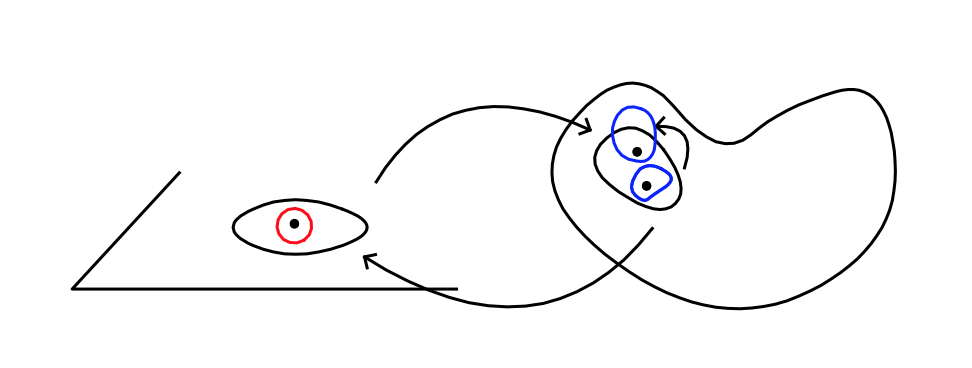}};
		\node[nodeoformula](1)at(-0.2, 1) {$\mathbb{R}^n$};
		\node[nodeoformula](2)at(9.5, 3.2) {$\sG$};
		\node[style=none](3)at(2.8, 1.9) {$0$};
		\node[style=none](4)at(5.9, 2) {$e$};
		\node[style=none](4)at(5.9, 1.5) {\textcolor{blue}{$h$}};
		\node[style=none](5)at(4, 1.5) {$\sU$};
		\node[style=none](5)at(2.4, 0.2) {$\textcolor{red}{\sV_h\defi\varphi^{-1}(\varphi(\sU)\cap L_{h^{-1}}(A_h))}$};
		\node[style=none](5)at(7.3, 1.7) {$\varphi(\sU)$};
		\node[style=none](4)at(6.8, 1.2) {\textcolor{blue}{$A_h$}};
		\node[style=none](4)at(6.2, 3.2) {\textcolor{blue}{$L_{h^{-1}}(A_h)$}};
		\node[style=none](5)at(7, 2.6) {$L_h^{-1}$};
		\node[style=none](6)at(4.6,2.9) {$\varphi$};
		\node[style=none](6)at(6, 0.4) {$\varphi^{-1}$};
		\draw[thick,blue,dotted] (6.35, 1.7) -- (6.7, 1.25);
		\draw[thick,blue,dotted] (6.2, 3) -- (6.2, 2.55);
		\draw[thick,red,dotted] (2.8, 1.2) -- (0.5, 0.5);
		\end{tikzpicture}
		\par\end{centering}
	\caption{\label{Vh}Construction of the neighbourhood $\sV_h$ (i.e. the domain of the functions $\psi_h$ for $h\in\varphi(\sU)$) starting from an open neighbourhood $A_h$ of the considered element $h$ of $\sG$.}
\end{figure}
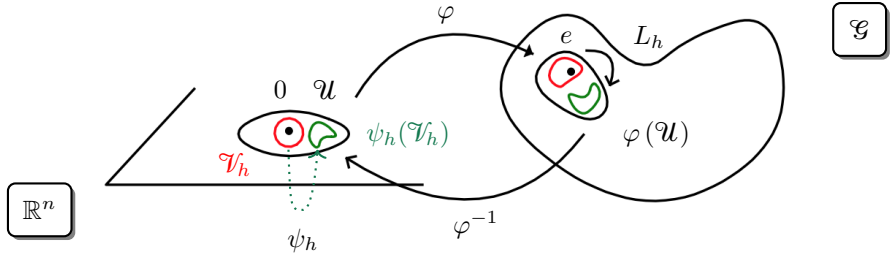
\begin{figure}[H]
	\begin{centering}
		\begin{tikzpicture}
		\tikzset{
			nodeoformula/.style={rectangle,rounded corners=0.1cm,drop shadow={shadow xshift=0.5mm, shadow yshift=-0.5mm,opacity=1},draw=black, top color=white, bottom color=white, thick, inner sep=2mm, minimum size=2em, text centered},
			nodepoint/.style={circle,draw=gray,fill=gray,inner sep=0.8mm}
		}
		\node[anchor=south west,inner sep=0] at (0,0) {\includegraphics[scale=0.55]{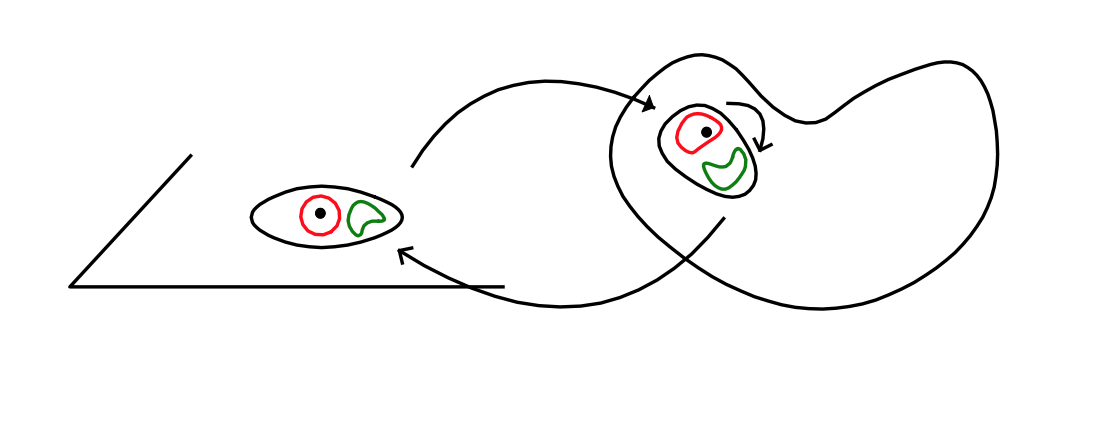}};
		\node[nodeoformula](1)at(-0.2, 1) {$\mathbb{R}^n$};
		\node[nodeoformula](2)at(10.7, 3.4) {$\sG$};
		\node[style=none](3)at(3, 2.6) {$0$};
		\node[style=none](4)at(6.8, 3.3) {$e$};
		\node[style=none](5)at(3.6, 2.6) {$\sU$};
		\node[style=none](5)at(3.3, 0.6) {$\psi_h$};
		\node[style=none](5)at(2.4, 1.6) {$\textcolor{red}{\sV_h}$};
		\node[style=none](5)at(4.7, 2) {$\textcolor{tsforestgreen}{\psi_h(\sV_h)}$};
		\node[style=none](5)at(8, 2) {$\varphi\,(\sU)$};
		\node[style=none](5)at(7.9, 3.3) {$L_h$};
		\node[style=none](6)at(5.2, 3.6) {$\varphi$};
		\node[style=none](6)at(5.6, 0.8) {$\varphi^{-1}$};
		\draw[->,tsforestgreen,thick,shorten >=2mm,shorten <=2mm,dotted] plot [smooth,tension=1.5] coordinates{(3.1,2) (3.3,1) (3.5,2)};
		\end{tikzpicture}
		\par\end{centering}
	\caption{\label{psih}Construction of functions $\psi_h$ for $h\in\varphi(\sU)$.}
\end{figure} 
Setting
$$ 
     J_h
     \defi
     \det[ \nabla\,\psi_h(0)]
     \qquad
     \forall h\in\varphi(\sU), 
$$
for every function $f$ such that $\text{supp}\,f\subseteq\varphi(\sU)$ we define the desired integral in the following way: 
$$
     \int_{\text{supp}\,f} f(g)\,\d\mu
     \defi
     \int_{\varphi^{-1}(\text{supp}\,f)}f(\varphi(x))
     \,
     \frac{1}{\left| \det\left[ \nabla\,\psi_{\varphi(x)}(0) \right] \right| }\,\dx.
$$

\begin{rem}
	The underlying concept behind this definition of the Haar integral is to adjust the Lebesgue measure on a point-by-point basis, resulting in a measure on $\varphi(\sU)$ that remains invariant under left multiplications. This pointwise correction at the position $x$ is represented by the term $1/\left| J_{\varphi(x)} \right|=1/\left| \det\left[ \nabla\,\psi_{\varphi(x)}(0) \right] \right|$, which is always finite due to the fact that $\psi_{\varphi(x)}$ is a diffeomorphism, ensuring that $J_{\varphi(x)} =\det\left[ \nabla\,\psi_{\varphi(x)}(0) \right]\neq0$.
\end{rem}

We want to show, with a direct calculation, that this definition is invariant with respect to left multiplications. Let us assume that we have a function $f$ and an element $h\in\varphi(\sU)$ such that $L_{h^{-1}}(\text{supp}(f))\subseteq\varphi(\sU)$. We have to show that
$$ \int_{\text{supp}\,f} f(g)\,\d\mu=\int_{\text{supp}\,f\circ L_h} (f\circ L_h)(g)\,\d\mu, $$
i.e., denoting with $x$ the points in $\varphi^{-1}(\text{supp}\,f)$ and with $y$ the points in $\varphi^{-1}(\text{supp}\,f\circ L_h)$, that
\begin{equation}\label{change of coordinates}
     \int_{\varphi^{-1}(\text{supp}\,f)}f(\varphi(x))\,\frac{1}{\left| J_{\varphi(x)} \right| }\,\dx
     =
     \int_{\varphi^{-1}(\text{supp}\,f\circ L_h)}(f\circ L_h)(\varphi(y))\,\frac{1}{\left| J_{\varphi(y)} \right| }\,\textrm{dy}.
\end{equation}
First of all, we can express $y$ as a function of $x$. Indeed we have that
\begin{equation}\label{xy}
y=\psi_{h^{-1}}(x).
\end{equation}
In this manner, we can effect a change of coordinates in the second integral of equation \eqref{change of coordinates} with the hope that this transformation yields results akin to those in the first integral. Consequently, by virtue of the relation described in \eqref{xy}, we attain the following outcome:
\begin{itemize}
	\item change of domain of integration: starting from the relation
	\begin{equation}\label{domains2}
	\text{supp}\,(f\circ L_h)=L_{h^{-1}}(\text{supp}\,f),
	\end{equation}
	and considering the change of variable $y=\psi_{h^{-1}}(x)$, we obtain
	\begin{equation}\label{domains}
	\psi_{h^{-1}}(\varphi^{-1}(\text{supp}\,f))=\varphi^{-1}\circ L_{h^{-1}}\circ\varphi\,(\varphi^{-1}(\text{supp}\,f))=(\varphi^{-1}\circ L_{h^{-1}})(\text{supp}\,f)=\varphi^{-1}(\text{supp}\,(f\circ L_h)),
	\end{equation}
	\item  relation between differentials
	\begin{equation}\label{differential}
	     \textrm{dy}
	     =
	     \left|\det\left[  \nabla\,\psi_{h^{-1}}(x)\right]  \right|\,\dx,
	\end{equation}
	\item for the integrating function we have
	\begin{align} \label{functions}
	     f\circ L_h(\varphi(y))&=f\circ L_h(\varphi(\psi_{h^{-1}}(x)))=f\circ L_h(\varphi(\varphi^{-1}\circ L_{h^{-1}}\circ\varphi(x)))
	     = f\circ L_h(L_{h^{-1}}\circ\varphi(x)))=f(\varphi(x)),
	\end{align}
	\item and finally for the correction term
	\begin{align*}
	    J_{\varphi(y)}
	    &
	    =
	    \det\left[ \text{D}\,\psi_{\varphi(y)} (0)\right]= \det\left[ \text{D}\,\psi_{\varphi(\psi_{h^{-1}}(x))} (0)\right]
	    =
	    \det\left[ \text{D}\,\psi_{\varphi(\varphi^{-1}\circ L_{h^{-1}}\circ\varphi(x))} (0)\right]
	   =
	   \det\left[ \text{D}\,\psi_{L_{h^{-1}}(\varphi(x))} (0)\right]. 
	\end{align*}
	Remarking that
	\begin{align*}
	\psi_{L_{h^{-1}}(\varphi(x))} (0)&=(\varphi^{-1}\circ L_{L_{h^{-1}}(\varphi(x))}\circ\varphi)(0)=(\varphi^{-1}\circ L_{h^{-1}}\circ L_{\varphi(x)}\circ\varphi)(0)\\
	&=(\varphi^{-1}\circ L_{h^{-1}}\circ\varphi\circ\varphi^{-1}\circ L_{\varphi(x)}\circ\varphi)(0)=(\psi_{h^{-1}}\circ\psi_{\varphi(x)})(0)\\
	&=\psi_{h^{-1}}(\psi_{\varphi(x)}(0)),
	\end{align*}
	by the chain rule we have
	\begin{align*}
	\det\left[ \text{D}\,\psi_{(L_{h^{-1}}(\varphi(x)))} (0)\right]&=\det\left[ \text{D}\,(\psi_{h^{-1}}\circ\psi_{\varphi(x)})(0)\right]=
	\det\left[ \text{D}\,\psi_{h^{-1}}(\psi_{\varphi(x)}(0))\, \text{D}\,\psi_{\varphi(x)}(0)\right]\\
	&= \det\left[ \text{D}\,\psi_{h^{-1}}(\psi_{\varphi(x)}(0))\right]\det\left[ \text{D}\,\psi_{\varphi(x)}(0)\right].
	\end{align*}
	Moreover, we have also that
	$$ \psi_{\varphi(x)}(0)=(\varphi^{-1}\circ L_{\varphi(x)}\circ\varphi)(0)=(\varphi^{-1}\circ L_{\varphi(x)})(e)=\varphi^{-1}(\varphi(x))=x,$$
	thanks to which we finally find
	\begin{equation}\label{correction factors}
	\det\left[ \text{D}\,\psi_{\varphi(y)} (0)\right]=\det\left[ \text{D}\,\psi_{(L_{h^{-1}}(\varphi(x)))} (0)\right]=  \det\left[ \text{D}\,\psi_{h^{-1}}(x)\right]\det\left[ \text{D}\,\psi_{\varphi(x)}(0)\right].
	\end{equation}
\end{itemize}
Thus, inserting \eqref{domains}, \eqref{differential}, \eqref{functions} and \eqref{correction factors} into the second integral of \eqref{change of coordinates}, we obtain the desired identity. Indeed
\begin{align*}
     \int_{\varphi^{-1}(\text{supp}\,f\circ L_h)}
     &
     (f\circ L_h)(\varphi(y))
     \,
     \frac{1}{\left| \det\left[ \nabla\,\psi_{\varphi(y)}(0) \right] \right| }
     \,\textrm{dy}
     \\
     &
     =
     \int_{\varphi^{-1}(\text{supp}\,f)}f(\varphi(x))
     \,
     \underbrace{\frac{1}{\left| \det\left[ \nabla\,\psi_{h^{-1}}(x)\right] \right| }\,\frac{1}{\left| \det\left[ \nabla\,\psi_{\varphi(x)}(0)\right] \right| }}_{\frac{1}{\vphantom{\int^{A}}| \det[ \nabla\,\psi_{\varphi(y)}(0) ] | }}
     \,
     \underbrace{\left|\det\left[  \nabla\,\psi_{h^{-1}}(x)\right]  \right|\,\dx}_{\textrm{dy}}.
\end{align*}

This technique proves especially advantageous when it is possible to cover a Lie group using a single chart, one that covers it entirely except for a subset of zero measure.
 In such instances, no compatibility condition arises between the support of a given function and the left translation by any arbitrary element within the group.


In the subsequent paragraph, it will be demonstrated that in the case of 
$\So$, we precisely find ourselves in the aforementioned scenario: the local chart that covers 
$\So$ but leaves out a subset of zero measure is defined by the Cardan angles.

\subsubsection{Cardan angles}


In the literature 
numerous local charts have been devised for the compact group\footnote{It is known that to cover $\So$ we need at least four local charts (see for example \cite{grafarend2011minimal}).} $\So$. Among these, the Euler angles chart stands as one of the most prominent. However, for the specific objectives under consideration, the Euler angles chart is not the suitable choice. This is primarily due to the fact that it does not encompass the origin, as $0$ lies outside its domain, and the identity element $\id$ is excluded from its range. Unfortunately, this incompatibility with our current approach presents a significant hurdle\footnote{ The Euler angles parametrization is provided by considering the mapping \cite[pag. 497]{marsden1995introduction}
	
	\[
	\mathscr{E}:\left(0,2\pi\right)\times\left(0,\pi\right)\times\left(0,2\pi\right)\fr\So,\qquad\mathscr{E}\left(\phi,\theta,\psi\right):=R_{\psi}R_{\theta}R_{\phi},
	\]
	where
	\begin{align*}
	R_{\phi} & =\begin{pmatrix}\cos\phi & -\sin\phi & 0\\
	\sin\phi & \cos\phi & 0\\
	0 & 0 & 1
	\end{pmatrix}, & R_{\theta} & =\begin{pmatrix}1 & 0 & 0\\
	0 & \cos\theta & -\sin\theta\\
	0 & -\sin\theta & \cos\theta
	\end{pmatrix}, & R_{\psi} & =\begin{pmatrix}\cos\psi & -\sin\psi & 0\\
	\sin\psi & \cos\psi & 0\\
	0 & 0 & 1
	\end{pmatrix}.
	\end{align*}
	The three real numbers $\left(\phi,\theta,\psi\right)$ are called
	\textbf{Euler angles}.}:
	We are unable to directly apply our argument to derive the Haar measure representation within  this chart.
	However, in lieu of considering this parametrization of $\So$, we may opt for the one provided by the Cardan angles.

Consider the bijective map (see \cite{diebel2006representing} for more details)

\begin{equation}
    \sC:\left[-\pi,\pi\right)\times\left[-\frac{\pi}{2},\frac{\pi}{2}\right)\times\left[-\pi,\pi\right)\fr\So,
    \qquad
    \sC\left(\phi,\theta,\psi\right)
    \defi 
    R_{\phi}\,R_{\theta}\,R_{\psi},
\end{equation}
where
\begin{align*}
R_{\phi} & =\begin{pmatrix}1 & 0 & 0\\
0 & \cos\phi & \sin\phi\\
0 & -\sin\phi& \cos\phi
\end{pmatrix}, & R_{\theta} & =\begin{pmatrix}\cos\theta & 0 & -\sin\theta\\
0 & 1 & 0\\
\sin\theta & 0 & \cos\theta
\end{pmatrix}, & R_{\psi} & =\begin{pmatrix}\cos\psi & -\sin\psi & 0\\
\sin\psi & \cos\psi & 0\\
0 & 0 & 1
\end{pmatrix}.
\end{align*}

The three real numbers $\left(\phi,\theta,\psi\right)$ are called
\textbf{Cardan angles}. The map $\mathscr{C}$ is not the desired local chart
because its domain of definition is not an open subset of $\R^{3}$.
Its restriction to the open subset 
\[
\Delta\defi\left(-\pi,\pi\right)\times\left(-\frac{\pi}{2},\frac{\pi}{2}\right)\times\left(-\pi,\pi\right)\,,
\]
however, is a chart (indeed $\left. \mathscr{C}\right|_\Delta $ is injective and its Jacobian has always maximal rank) which covers $\So$ except for a zero measure subset (the
image of union of the three surfaces $\Sigma=\left(\left\{ -\pi\right\} \times\left[-\frac{\pi}{2},\frac{\pi}{2}\right)\times\left[-\pi,\pi\right)\right)\cup\left(\left[-\pi,\pi\right)\times\left\{ -\frac{\pi}{2}\right\} \times\left[-\pi,\pi\right)\right)\cup\left(\left[-\pi,\pi\right)\times\left[-\frac{\pi}{2},\frac{\pi}{2}\right)\times\left\{ -\pi\right\} \right)$).

This local chart possesses all the required properties to build the adapted Haar measure. In order to compute $J_{\mathscr{C}(\phi,\theta,\psi)}$, we need an explicit expression for the inverse map $\mathscr{C}^{-1}$: denoting by
\begin{equation}
    \text{atan2}(y,x)
    =
    \begin{cases}	   
		\arctan\frac{y}{x}&\text{if}\,x>0
		\\
		\arctan\frac{y}{x}+\pi&\text{if}\,x<0\,\text{and}\,y\geq0
		\\
		\arctan\frac{y}{x}-\pi&\text{if}\,x<0\,\text{and}\,y<0
		\\
		+\frac{\pi}{2}&\text{if}\,x=0\,\text{and}\,y>0
		\\
		-\frac{\pi}{2}&\text{if}\,x=0\,\text{and}\,y<0
		\\
		\text{not defined}&\text{if}\,x=0\,\text{and}\,y=0,
	\end{cases}
\end{equation}
we have
$$\mathscr{C}^{-1}(Q)=\begin{pmatrix}\textrm{atan2}\left(Q_{23},Q_{33}\right)\\
-\arcsin\left(Q_{13}\right)\\
\textrm{atan2}\left(Q_{12},Q_{11}\right)
\end{pmatrix} \qquad\forall Q\in\mathscr{C}(\Delta). $$
In this way, the map 
$$\psi_{\,\mathscr{C}(\phi,\theta,\psi)}:\R^3\supseteq\sV\fr\R^3,\qquad(\alpha,\beta,\gamma)\mapsto \psi_{\,\mathscr{C}(\phi,\theta,\psi)}(\alpha,\beta,\gamma)\defi(\mathscr{C}^{-1}\circ L_{\mathscr{C}(\phi,\theta,\psi)}\circ\mathscr{C}) (\alpha,\beta,\gamma)$$
is given by 
$$ 
       \mathscr{C}^{-1}\Big[ \underbrace{\mathscr{C}(\phi,\theta,\psi) \,\mathscr{C} (\alpha,\beta,\gamma)}_{\begin{matrix}\text{{\scriptsize matrix product of}}
       	\\
	   \text{{\scriptsize these two matrices}}
	   \end{matrix}} \Big]
       =
       \begin{pmatrix}
       	       \textrm{atan2}\left((\mathscr{C}(\phi,\theta,\psi) \,\mathscr{C} (\alpha,\beta,\gamma))_{23},(\mathscr{C}(\phi,\theta,\psi) \,\mathscr{C} (\alpha,\beta,\gamma))_{33}\right)
       	       \\[2mm]
               -\arcsin\left((\mathscr{C}(\phi,\theta,\psi) \,\mathscr{C} (\alpha,\beta,\gamma))_{13}\right)
               \\[2mm]
               \textrm{atan2}\left((\mathscr{C}(\phi,\theta,\psi) \,\mathscr{C} (\alpha,\beta,\gamma))_{12},(\mathscr{C}(\phi,\theta,\psi) \,\mathscr{C} (\alpha,\beta,\gamma))_{11}\right)
       \end{pmatrix} .  
$$
Making the explicit calculation we find that 
$$
      \frac{1}{\left| \det\left[ \text{D}\,\psi_{\,\mathscr{C}(\phi,\theta,\psi)}(0,0,0)\right]  \right| }
      =
      \left|\cos\theta \right|.  
$$
In this way 
\begin{align*}
     \int_{\text{SO}(3)} f(Q)\,\d\mu
     &=
     \frac{1}{m(\Delta)}\int_\Delta f(\sC(\phi,\theta,\psi))
     \,
     \frac{1}{\left| \det\left[ \text{D}\,\psi_{\,\sC(\phi,\theta,\psi)}(0,0,0)\right]  \right| }
     \, \d\phi\,\d\theta\,\d\psi
     \\
     &
     =
     \frac{1}{8\,\pi^2}\int_{-\pi}^\pi\int_{-\frac{\pi}{2}}^{\frac{\pi}{2}}\int_{-\pi}^\pi f(R_{\phi}R_{\theta}R_{\psi})\,\left| \cos\theta \right| \, \d\phi\,\d\theta\,\d\psi,
\end{align*}
and considering the linear change of variables
$$ 
    \phi'=\phi+\pi,\qquad\theta'=\theta+\frac{\pi}{2},\qquad\psi'=\psi+\pi
$$
we obtain the classical expression for the integration over $\So$
$$ 
    \int_{\text{SO}(3)} f(Q)\,\d\mu
    =
    \frac{1}{8\,\pi^2}\int_{0}^{2\pi}\hspace{-3mm}\int_{0}^{\pi}\hspace{-2mm}\int_{0}^{2\pi} \hspace{-2mm}f(R_{\phi'}R_{\theta'}R_{\psi'})\,\sin\theta'\, \d\phi'\,\d\theta'\,\d\psi'. 
$$

\subsection{Geometry of the space of elasticity tensors, Harmonic decompositions ans stratifications}\label{Geometry of the space of elasticity}

In this appendix, our main objective is to explore the geometric structure of  $\Ela(3)$. 
Additionally, we aim to demonstrate how the action of $\SO$, introduced in eq.\eqref{eq:main action}, induces a decomposition of $\Ela(3)$ into disjoint subspaces, referred to as \textbf{strata}. The elements of a stratum are all the elements of the vector space whose isotropy groups are conjugate. This provides us with a way to \enquote{a priori} determine the permissible symmetry classes for the tensors under consideration, which correspond to the closed subgroups of $\SO$ with non-empty corresponding strata.

\subsubsection{Harmonic tensors}

In this Paragraph, we want to introduce the building blocks to generalize the Cartan-decomposition of $\bR^{3\times3}$ to $\Ela(3)$. First of all, let us introduce the subspace of totally symmetric tensors
\[
    \Tots(3)
    \defi
    \graf{
    	\bC\in\Ela(3)\stt \bC_{ijkl}
    	=
    	\bC_{\sigma(i)\sigma(j)\sigma(k)\sigma(l)
    	}
    	\quad\forall\sigma\in\gS(4)
    },
\]
where $\gS(4)$ is the permutation group of 4 elements. The associated projection operator is denoted by $P_{\Tots(3)}$. Setting $\bC^s\defi P_{\Tots(3)}(\bC)$, we have
\[
    \bC^s_{ijkl}
    =
    \frac{1}{3}
    \partonb{\bC_{ijkl}+\bC_{ikjl}+\bC_{ilkj}}.
\]

\vspace{-5mm}

The complementary space is the space $\gS^{(2,2)}$ of tensors whose Young tableau is {\footnotesize \young(12,34)}, indicating that the tensor is antisymmetric under the exchange of pairs of indices at positions $(1,3)$ and $(2,4)$. The relative projection is denoted by $P_{\gS^{(2,2)}}$ and setting  $\bC^a\defi P_{\gS^{(2,2)}}(\bC)$ we get 
\[
    \bC^a_{ijkl}
    =
    \frac{1}{3}
    \partonb{2 \, \bC_{ijkl}-\bC_{ikjl}-\bC_{ilkj}}.
\]
Therefore
\[
    \Ela(3)
    \simeq
    \Tots(3)
    \oplus
    \gS^{(2,2)}.
\]
In the context of $\Ela(3)$, $\Tots(3)$ and $\gS^{(2,2)}$ respectively fulfill the roles of $\Sym$ and $\so$ for $\bR^{3\times3}$. Now, we aim to further decompose $\Tots(3)$ to mimic the decomposition $\Sym\simeq\Dev\Sym\oplus\langle\id\rangle$ where $\Dev\Sym\defi\mathfrak{sl}(3)\cap\Sym$. To achieve this, we need to generalize the trace operator. The simplest approach is to consider the scalar product with the symmetrization identity $\Pi^{\Sym}$ introduced in Eq.\eqref{Symmetrization identity}:
\[
    \Tr(\bC)=
    \scalb{\bC}{\Pi^{\Sym}}_{\otimes^4\bR^3}.
\] 
For an element $\bD\in\Tots(3)$, due to the symmetry of the indices, this is equivalent to considering the iterated trace
\[
    \Tr(\bD)=\tr(\tr(\bD)).
\]
We can hence introduce the space of the $4^{\text{th}}$ order Harmonic tensors as
\[
    \Harg_4(3)
    \defi
    \graf{\bC\in\Tots(3)\stt\Tr(\bC)=0}
    \qquad
    \text{and}
    \qquad
    \Tots(3)
    \simeq
    \Harg_4(3)
    \oplus
    \Harg_2(3)
    \oplus
    \langle\id\rangle,
\]
where, to have a consistent notation, we have put $\Harg_2(3)\defi\Dev\Sym$. It
is well known, the validity of the following isomorphisms of vector spaces
\[
     \Harg_4(3)
     \simeq
     \Har_4(3)
     \simeq
     \mathscr{H\!a\!r}_4(3)
\]
where $\Har_4(3)$ is the vector space of the traceless homogeneous polynomials over $\bR^3$ and $\mathscr{H\!a\!r}_4(3)$ is the space of the spherical harmonics of degree $4$ over $\bR^3$.

\subsubsection{Irreducible representations of $\SO$ and harmonic decomposition of $\Ela(3)$}

A linear action of a group over a vector space can be reinterpreted as a linear representation of the group (and, with an abuse of notation, we denote both using the same symbol). We briefly remind that a representation of a group $\cG$ on a vector space $\bV$ is a group homomorphism $ \varphi\in\Hom(\cG,\textrm{GL}(\bV))$. A representation is said to be irreducible if there are no invariant subspaces. Here, a subspace $\bW$ is said to be $(\cG,\varphi)-$invariant if $\varphi(g)(\bW)\subseteq\bW$ for every $g\in\cG$. When the considered group is compact, every linear representation can be decomposed as a (finite) direct sum of irreducible linear representations, unique up to $\cG-$isomorphisms (for example, \cite[Thm2.5, pag. 35]{golubitsky2012singularities}). 
The considered case of the representation associated with the action $(\Ela(3),\varphi,\SO)$ satisfies all these requirements. Therefore, it is crucial for us to determine the decomposition of $\varphi$ in irreducible factors, which is known as the harmonic decomposition of $\Ela(3)$. Firstly, we need to determine all the irreducible linear representations of $\SO$. This is a classical result (see, for example, \cite[Thm7.3+7.5, pag. 111]{golubitsky2012singularities}) summarized as follows

\begin{thm}
	The representations 
	\[
	    \widetilde \rho_n:\SO
	    \xrightarrow{\phantom{aaa}}
	    \textrm{GL}\big(\mathscr{H\!a\!r}_n(3)\big),
	    \qquad
	    Q
	    \mapsto
	    \widetilde \rho_n(Q),
	    \quad
	    \widetilde \rho_n(Q)(\psi(x))
	    =
	    \psi(Q^{-1} \, x)
	\]
	
	are irreducible for every $n\in\bN$. Moreover,every irreducible representation of $\SO$ is isomorphic to some $\widetilde \rho_n$.
\end{thm}

Once the building blocks for linear representations of $\SO$ have been established, we obtain the desired decomposition ( see for example \cite[Section 3.2]{forte1996symmetry})

\begin{thm}\label{Harmonic decomposition}
	 The Harmonic decomposition of the space $\Ela(3)$ is as follows:
	\[
	    \Ela(3)
	    \simeq
	    \Harg_4(3)
	    \oplus
	    \Harg_2(3)
	    \oplus
	    \Harg_2(3)
	    \oplus
	    \langle\id\rangle
	    \oplus
	    \langle\id\rangle.
	\]
	Moreover, all the introduced subspaces are $\SO-$invariant w.r.t.\ the action of $\varphi$ and the isomorphism is equivariant.
\end{thm}

\subsubsection{Stratification and isotropy classes}

We begin by introducing the concept of space stratification resulting from the action of a topological group $\cG$ over a topological space $X$. Let us denote by $\Clo(\cG)$ the set of all the closed subgroups of $\cG$. The natural action of $\cG$ over $\Clo(\cG)$ is the conjugation:
\[
    \co:\cG\times\Clo(\cG)\longrightarrow\Clo(\cG),
    \qquad
    (g,H)
    \longmapsto
    \co(g,H)
    \defi
    g^{-1}\,H\,g.
\]
The quotient space is denoted by $[\Clo(\cG)]_{\co}$ and an element $[H]$ is a conjugacy class of closed subgroups of $\cG$.
\begin{defn}
	Let us consider the $\cG-$space $(X,\varphi,\cG)$ and consider $[H]\in[\Clo(\cG)]_{\co}$. The \textbf{stratum} $\Sigma_{[H]}\subset X$ associated to $[H]$ through the action $\varphi$ is the set
	\[
	    \Sigma_{[H]}
	    \defi
	    \graf{
	    	x\in X\stt \cG_x\in[H]
	    }
	\]
	where $\cG_x\in\Clo(\cG)$ (said the \textbf{isotropy group of $x$}) is the closed subgroup of $\cG$ of the elements of $\cG$ leaving $x$ fixed, i.e.,
	\[
	    \cG_x
	    \defi
	    \graf{g\in\cG\stt\varphi(g,x)=x}.
	\]
	An \textbf{isotropy classe} of $(X,\varphi,\cG)$ is an element $[H]\in[\Clo(\cG)]_{\co}$ such that $\Sigma_{[H]}\neq\varnothing$. The set of all the isotropy classes, denoted by
	\[
	   \Isot^\varphi_\cG(X)
	   \defi
	   \graf{[H]\in[\Clo(\cG)]_{\co}\stt \Sigma_{[H]}\neq\varnothing},
	\]
	is said an \textbf{isotropy type} of $X$. The partition of $X$ in the disjoint subsets $\{\Sigma_{[H]}\}_{[H]\in\Isot^\varphi_\cG(X)}$ is said \textbf{isotropic stratification} of $X$. 
\end{defn}

The set $\Isot^\varphi_\cG(X)$ inherits a partial ordering $\preccurlyeq$ from the partial order given by the inclusion of subgroups over $\Clo(\cG)$ which is defined as follows: we say that $[H_1]\preccurlyeq[H_2]$ if $H_1$ is conjugate to a subgroup of $H_2$. The isotropic stratification also inherits a contravariant partial order: $[H_1]\preccurlyeq[H_2]\;\Leftrightarrow\;\Sigma_{[H_2]}\preccurlyeq\Sigma_{[H_1]}$.
When a Lie group acts over a finite dimension vector space, then the induced stratification has an interesting simple structure (see for example \cite{abud1983geometry}). 

\begin{thm}
	Let us consider a finite linear representation $(\varphi,V)$ of a compact Lie group $\cG$. Then,
	\begin{enumerate}
		\item Each stratum is a smooth submanifold of $V$
		\item the topological boundary of a stratum contains the strata of inferior dimension,
		\item There exists a maximal stratum which is an open and dense subset of $V$,
		\item There is only a finite number of strata and
		\[
		V
		=
		\bigsqcup_{\mathclap{[H]\,\in\,\Isot^\varphi_\cG(\bV)}} \; \Sigma_{[H]}.
		\] 
	\end{enumerate}
\end{thm}

The problem we are tackling fits the hypothesis of the mentioned theorem. Indeed, our objective, translated into this new language, is to find the isotropy type determined by the action $\varphi$ introduce in \eqref{eq:main action} of the Lie group $\SO$ on the finite-dimensional vector space $\Ela(3)$.

Thanks to theorem \ref{Harmonic decomposition}, because the action $\varphi$ is equivariant w.r.t.\ the Harmonic decomposition of $\Ela(3)$, if we are able to find the isotropy classes of each single factor of the Harmonic decomposition and combine them in a suitable way, we can expect to obtain $\Isot^\varphi_\cG(\Ela(3))$ from  the knowledge of the isotropy type of the factors. This is possible and in literature two methods have been proposed, one in \cite{forte1996symmetry} and another in \cite{azzi2023clips,auffray2017handbook,olive2013symmetry,olive2014symmetry}. The advantage of the methods proposed in \cite{azzi2023clips,auffray2017handbook,olive2013symmetry,olive2014symmetry}, based on the clips operation, is related to the fact that it allows the development of an algorithmic procedure that can be easily generalized to other classes of tensors. Given the isotropy classes, $[H_1]$ and $[H_2]$, their clips is defined as 
\[
    [H_1]
    \circledcirc
    [H_2]
    \defi
    \bigl\{[H_1\cap g\,H_2\,g^{-1}]\bigr\}_{g\in\cG}.
\]
Considering two linear representation $\varphi_1\in\Hom(\cG,\bV_1)$ and  $\varphi_2\in\Hom(\cG,\bV_2)$ of the compact Lie groups $\cG$, it is possible to define the clips between $\Isot^{\varphi_1}_\cG(\bV_1)$ and $\Isot^{\varphi_2}_\cG(\bV_2)$ as
\[
    \Isot^{\varphi_1}_\cG(\bV_1)
    \circledcirc
    \Isot^{\varphi_2}_\cG(\bV_2)
    \defi
    \bigcup_{
    	\substack{
    		[H_1]\in\Isot^{\varphi_1}_\cG(\bV_1) 
    		\\[1mm] 
    		[H_2]\in\Isot^{\varphi_2}_\cG(\bV_2)
    	}
    }
    [H_1]
    \circledcirc
    [H_2].
\] 
Then, it is proved that 
\[
    \Isot^{\varphi_1\oplus\varphi_2}_\cG(\bV_1\oplus\bV_2)
    =
    \Isot^{\varphi_1}_\cG(\bV_1)
    \circledcirc
    \Isot^{\varphi_2}_\cG(\bV_2),
\]
and hence, in general, if $\varphi\in\Hom(\cG,\textrm{GL}(\bV))$ decomposes as the direct sum of irreducible representations, $\varphi\simeq\bigoplus_i\varphi_i^{\oplus\alpha_i}\in\bigoplus_i\Hom(\cG,\textrm{GL}(\bV_i))^{\oplus\alpha_i}$, then
\[
    \Isot^{\varphi}_\cG(\bV)
    =
    \circledcirc_i\Big(\circledcirc_{\alpha_i}\Isot^{\varphi_i}_\cG(\bV_i)\Big).
\]
In this way, computing the clips between the isotropy classes of the irreducible representations involved in the harmonic decomposition of $\Ela(3)$ (see for example \cite{ihrig1984pattern,olive2019effective}), it is possible to finally state that the non-trivial strata of $\Ela(3)$ are those corresponding to the following conjugacy classes of closed subgroups of $\SO$:
\[
    \Isot^\varphi_\cG(\Ela(3))
    =
    \bigl\{[\id],[Z_2],[D_2],[D_3],[D_4],[\mathcal O],[O(2)],[\SO]\bigr\},
\]
for a total of 8 distinct conjugacy classes.

\section{Useful isomorphisms between vector spaces}
\subsection{Voigt isomorphism}

In order to work with vectors and matrices, we define certain isomorphisms
between finite-dimensional vector spaces. Let $\left\{ e_{1},e_{2},e_{3}\right\} $ be the canonical basis\footnote{$e_{1}=\left(1,0,0\right)^{T},e_{2}=\left(0,1,0\right)^{T},e_{3}=\left(0,0,1\right)^{T}$}
of $\R^{3}$
and let $\left\{ \epsilon_{i}\right\} _{i=1}^{6}$ be the canonical
basis of $\R^{6}$. We can consider the isomorphism $\MM:\Sym\fr\R^{6}$
defined as follows:
\begin{align*}
\MM\left(e_{1}\otimes e_{1}\right)=\epsilon_{1}\,, &  & \MM\left(e_{2}\otimes e_{2}\right)=\epsilon_{2}\,, &  & \MM\left(e_{3}\otimes e_{3}\right)=\epsilon_{3}\,,\\
\MM\left(e_{2}\otimes e_{3}+e_{3}\otimes e_{2}\right)=2\,\epsilon_{4}\,, &  & \MM\left(e_{3}\otimes e_{1}+e_{1}\otimes e_{3}\right)=2\,\epsilon_{5}\,, &  & \MM\left(e_{1}\otimes e_{2}+e_{2}\otimes e_{1}\right)=2\,\epsilon_{6}\,.
\end{align*}
Seen as a tensor $\MM\in\R^{6}\otimes\R^{3\times3}$, its action
on a matrix $X$ is $\left(\MM\,X\right)_{\alpha}=\mathfrak{M}_{\alpha ij}X_{ij},$
where
\[
\mathfrak{M}_{\alpha ij}=\widetilde{\delta}_{\alpha1}\delta_{i1}\delta_{j1}+\widetilde{\delta}_{\alpha2}\delta_{i2}\delta_{j2}+\widetilde{\delta}_{\alpha3}\delta_{i3}\delta_{j3}+\widetilde{\delta}_{\alpha4}\left(\delta_{i2}\delta_{j3}+\delta_{i3}\delta_{j2}\right)+\widetilde{\delta}_{\alpha5}\left(\delta_{i1}\delta_{j3}+\delta_{i3}\delta_{j1}\right)+\widetilde{\delta}_{\alpha6}\left(\delta_{i1}\delta_{j2}+\delta_{i2}\delta_{j1}\right),
\]
and $\delta_{ij}$ is the usual Kronecker delta in $\R^{3}$ while
$\widetilde{\delta}_{\alpha\beta}$ is the Kronecker delta in $\R^{6}$.
Fixing the index $\alpha$ the components of $\MM$ can be represented
using the following matrices: 
\begin{align*}
     \mathfrak{M}_{1ij}
     =
     \begin{pmatrix}
     	1 & 0 & 0\\
     	0 & 0 & 0\\
     	0 & 0 & 0
     \end{pmatrix}, 
     &  & 
     \mathfrak{M}_{2ij}
     =
     \begin{pmatrix}
     	0 & 0 & 0\\
     	0 & 1 & 0\\
     	0 & 0 & 0
     \end{pmatrix}, 
     &  & 
     \mathfrak{M}_{3ij}
     =
     \begin{pmatrix}
     	0 & 0 & 0\\
     	0 & 0 & 0\\
     	0 & 0 & 1
     \end{pmatrix},
     \\[5mm]
     \mathfrak{M}_{4ij}
     =
     \begin{pmatrix}
     	0 & 0 & 0\\
     	0 & 0 & 1\\
     	0 & 1 & 0
     \end{pmatrix}, 
     &  & 
     \mathfrak{M}_{5ij}
     =
     \begin{pmatrix}
     	0 & 0 & 1\\
     	0 & 0 & 0\\
     	1 & 0 & 0
     \end{pmatrix}, 
     &  & 
     {\displaystyle \mathfrak{M}_{6ij}
     =
     \begin{pmatrix}
     	0 & 1 & 0\\
     		1 & 0 & 0\\
     		0 & 0 & 0
     \end{pmatrix}}.
\end{align*}
In this way, given 

$$X=\begin{pmatrix}X_{11} & X_{12} & X_{13}\\
X_{12} & X_{22} & X_{23}\\
X_{13} & X_{23} & X_{33}
\end{pmatrix}\in\textrm{Sym}\left(3\right)$$

\noindent we have\footnote{N.B. The Voigt isomorphism is not an isometry between the two involved spaces. If we want to consider an isometry, we should change the transformation in the following way:
$$
X\mapsto\left(X_{11},X_{22},X_{33},\sqrt{2}\,X_{23},\sqrt{2}\,X_{13},\sqrt{2}\,X_{12}\right)^{T}.	
$$
This transformation is known in the literature as the \textbf{Mandel isomorphism}.
}

\begin{equation}
     \MM\,X:=\left(X_{11},X_{22},X_{33},2\,X_{23},2\,X_{13},2\,X_{12}\right)^{T}\in\bR^6.
     \label{eq:Voigt1-1-1}
\end{equation}
The inverse map\footnote{I.e.\ the map verifying $\MM\circ\MM^{-1}=\id_{\bR^{3\times3}}$ or component-wise $\MM_{\alpha ij}\circ\MM^{-1}_{kl\beta}=\delta_{ij}\delta_{lk}$ and $\MM^{-1}\circ\MM=\id_{\bR^{6}}$ or component-wise $\mathfrak{M}_{ij\alpha}^{-1}\,\mathfrak{M}_{\beta ij}=\widetilde{\delta}_{\alpha\beta}$.} $\MM^{-1}:\R^{6}\fr\Sym$
is given in coordinates by
\[
\mathfrak{M}_{ij\alpha}^{-1}=\widetilde{\delta}_{\alpha1}\delta_{i1}\delta_{j1}+\widetilde{\delta}_{\alpha2}\delta_{i2}\delta_{j2}+\widetilde{\delta}_{\alpha3}\delta_{i3}\delta_{j3}+\frac{1}{2}\left(\widetilde{\delta}_{\alpha4}\left(\delta_{i2}\delta_{j3}+\delta_{i3}\delta_{j2}\right)+\widetilde{\delta}_{\alpha5}\left(\delta_{i1}\delta_{j3}+\delta_{i3}\delta_{j1}\right)\right)+\frac{1}{2}\widetilde{\delta}_{\alpha6}\left(\delta_{i1}\delta_{j2}+\delta_{i2}\delta_{j1}\right),
\]
and it acts by $\left(\MM^{-1}\, x\right)_{ij}=\mathfrak{M}_{ij\alpha}^{-1}\,x_{\alpha}.$
From the definition of $\MM$, we can define
\begin{equation}\label{M bar}
    \underline{\MM}:
    \Ela(3)
    \to
    \textrm{Sym}\partonn{\R^{6},\R^{6}},
\end{equation}
which associates to an elasticity tensor $\mathbb{C}$ the second-order tensor
$\widetilde{\mathbb{C}}:=\underline{\MM}\left(\mathbb{C}\right)$
defined by
\[
       \big\langle \widetilde{\mathbb{C}}\, a,b\big\rangle _{\R^{6}}
       =
       \left\langle \mathbb{C}\:(\MM^{-1}\, a),\MM^{-1}\, b\right\rangle _{\R^{3\times3}}\qquad\forall a,b\in\R^{6}.
\]
Component-wise, the tensor $\widetilde{\mathbb{C}}$ is given by
\[
\widetilde{\mathbb{C}}_{\alpha\beta}=\mathbb{C}_{ijkl}\,\mathfrak{M}_{ij\alpha}^{-1}\,\mathfrak{M}_{kl\beta}^{-1}.
\]
%
%
%
%
\subsection{axl isomorphism}\label{subsec_axl}
A direct isomorphism can be established between $\so$ (which is a vector space of dimension 3) and $\R^3$ in the following way:
$$ \axl:\so\fr\R^3,\qquad\axl\,\begin{pmatrix}
0 & -a_3 & a_2\\
a_3 & 0 & -a_1\\
-a_2 & a_1 & 0
\end{pmatrix}\defi(a_1,a_2,a_3), $$
which component-wise gives 
$$ 
    ( \axl A )_k
    =
    -\frac{1}{2}\epsilon_{kij}\,A_{ij},  
$$
where $\epsilon_{kij}$ is the Levi-Civita tensor. For any $v\in\bR^3$, the identity
$
    A\, v
    =
    \axl A\times v 
$
holds.
Its inverse operator is
$$ \anti:\R^3\fr\so,\qquad (\anti\,a)_{ij}=-\epsilon_{ijk}\,a_k\,.$$
Following the same line as done for the isomorphism $\underline{\MM}$, we introduce the isomorphism
$$\underline{\text{Axl}}:\text{Sym}(\so,\so)\fr\Sym,\qquad\cc\mapsto\widetilde{\mathbb{C}}_c$$
defined by
$$ \big\langle \widetilde{\mathbb{C}}_c\, a,b \big\rangle_{\R^3}=\left\langle  \cc\,\anti\,a,\anti\,b\right\rangle   $$
Component-wise, the tensor $\widetilde{\mathbb{C}}_c$ is given by
$$ (\widetilde{\mathbb{C}}_c)_{mn}=(\cc)_{ijhk}\,\epsilon_{ijm}\epsilon_{hkn}. $$

\subsection{Computation of characters}

In order to obtain the most convenient expression of $\chi\left(Q\right)$
for the numerical computation (that which involves the less number
of powers) we express, thanks to the Cayley-Hamilton Theorem the trace
of the powers of a given matrix $Q$ in term of the powers of the
trace of $Q$. For a matrix $Q\in\R^{3\times3}$ the Hamilton-Cayley
theorem states that the identity
\begin{equation}
Q^{3}-\textrm{tr}\left(Q\right)Q^{2}+\frac{1}{2}\left(\left(\textrm{tr}\,Q\right)^{2}-\textrm{tr}\,Q^{2}\right)Q-\det\left(Q\right)\id=0\label{H-C theorem}
\end{equation}
holds. From (\ref{H-C theorem}) we easily derive that
\[
Q^{3}=\textrm{tr}\left(Q\right)Q^{2}-\frac{1}{2}\left(\left(\textrm{tr}\,Q\right)^{2}-\textrm{tr}\,Q^{2}\right)Q+\id\quad\Longleftrightarrow\quad Q^{2}=\textrm{tr}\left(Q\right)Q-\frac{1}{2}\left(\left(\textrm{tr}\,Q\right)^{2}-\textrm{tr}\,Q^{2}\right)\id+Q^{T}
\]
for $Q\in\SO$, from which we obtain
\begin{equation}
\textrm{tr}\,Q^{2}=\left(\textrm{tr}\,Q\right)^{2}-\frac{3}{2}\left(\left(\textrm{tr}\,Q\right)^{2}-\textrm{tr}\,Q^{2}\right)+\textrm{tr}\,Q\quad\Longleftrightarrow\quad\textrm{tr}\,Q^{2}=\left(\textrm{tr}\,Q\right)^{2}-2\,\textrm{tr}\,Q.\label{2power}
\end{equation}
Having the expression of $\textrm{tr}\,Q^{2}$ as a polynomial of
the powers of $\textrm{tr}\,Q$, we can calculate the same relation
also for $\textrm{tr}\,Q^{3}:$

\begin{align}
     \textrm{tr}\,Q^{3} 
     & =\textrm{tr}\left(\textrm{tr}\left(Q\right)Q^{2}-\frac{1}{2}\left(\left(\textrm{tr}\,Q\right)^{2}-\textrm{tr}\,Q^{2}\right)Q+\id\right)
     =\textrm{tr}\,Q\,\textrm{tr}\,Q^{2}-\frac{1}{2}\left(\textrm{tr}\,Q\right)^{2}\textrm{tr}\,Q+\frac{1}{2}\textrm{tr}\,Q^{2}\,\textrm{tr}\,Q+3
     \nonumber \\
     &
     =-\frac{1}{2}\left(\textrm{tr}\,Q\right)^{3}+\frac{3}{2}\textrm{tr}\,Q^{2}\,\textrm{tr}\,Q+3
     =-\frac{1}{2}\left(\textrm{tr}\,Q\right)^{3}+\frac{3}{2}\left(\left(\textrm{tr}\,Q\right)^{2}-2\,\textrm{tr}\,Q\right)\textrm{tr}\,Q+3
     \nonumber \\
     &
     =-\frac{1}{2}\left(\textrm{tr}\,Q\right)^{3}+\frac{3}{2}\left(\textrm{tr}\,Q\right)^{3}-3\left(\textrm{tr}\,Q\right)^{2}+3
     =\left(\textrm{tr}\,Q\right)^{3}-3\left(\textrm{tr}\,Q\right)^{2}+3.\label{3power}
\end{align}
Now, observing that 
\[
Q^{4}=QQ^{3}=\textrm{tr}\left(Q\right)Q^{3}-\frac{1}{2}\left(\left(\textrm{tr}\,Q\right)^{2}-\textrm{tr}\,Q^{2}\right)Q^{2}+Q,
\]
we find
\begin{align}
\textrm{tr}\,Q^{4} & =\textrm{tr}\left(\textrm{tr}\left(Q\right)Q^{3}-\frac{1}{2}\left(\left(\textrm{tr}\,Q\right)^{2}-\textrm{tr}\,Q^{2}\right)Q^{2}+Q\right)\nonumber \\
& =\textrm{tr}\,Q\,\textrm{tr}\,Q^{3}-\frac{1}{2}\left(\left(\textrm{tr}\,Q\right)^{2}-\textrm{tr}\,Q^{2}\right)\textrm{tr}\,Q^{2}+\textrm{tr}\,Q\nonumber \\
& =\textrm{tr}\,Q\left(\left(\textrm{tr}\,Q\right)^{3}-3\left(\textrm{tr}\,Q\right)^{2}+3\right)-\frac{1}{2}\left(\left(\textrm{tr}\,Q\right)^{2}-\left(\left(\textrm{tr}\,Q\right)^{2}-2\,\textrm{tr}\,Q\right)\right)\left(\left(\textrm{tr}\,Q\right)^{2}-2\,\textrm{tr}\,Q\right)+\textrm{tr}\,Q\nonumber \\
& =\left(\textrm{tr}\,Q\right)^{4}-3\left(\textrm{tr}\,Q\right)^{3}+3\textrm{tr}\,Q-\textrm{tr}\,Q\left(\left(\textrm{tr}\,Q\right)^{2}-2\,\textrm{tr}\,Q\right)+\textrm{tr}\,Q\nonumber \\
& =\left(\textrm{tr}\,Q\right)^{4}-3\left(\textrm{tr}\,Q\right)^{3}+3\textrm{tr}\,Q-\left(\textrm{tr}\,Q\right)^{3}+2\left(\textrm{tr}\,Q\right)^{2}+\textrm{tr}\,Q\nonumber \\
& =\left(\textrm{tr}\,Q\right)^{4}-4\left(\textrm{tr}\,Q\right)^{3}+2\left(\textrm{tr}\,Q\right)^{2}+4\,\textrm{tr}\,Q.\label{4power}
\end{align}
Thus, thanks to relations (\ref{2power}),(\ref{3power}) and (\ref{4power}),
we can compute the expression of the characters for all the considered
elasticity frameworks as function of the powers of $\textrm{tr}\,Q$.

\subsubsection{Classical elasticity }

In classical elasticity framework we found

\[
\chi(Q)=\frac{1}{8}\left[(\textrm{tr}\,Q)^{4}+2(\textrm{tr}\,Q)^{2}\textrm{tr}\,Q^{2}+2\,\textrm{tr}\,Q^{4}+3(\textrm{tr}\,Q^{2})^{2}\right],
\]
thus
\begin{align*}
\chi(Q) & =\frac{1}{8}\left[(\textrm{tr}\,Q)^{4}+2(\textrm{tr}\,Q)^{2}\textrm{tr}\,Q^{2}+2\,\textrm{tr}\,Q^{4}+3(\textrm{tr}\,Q^{2})^{2}\right]\\
& =\frac{1}{8}\left[(\textrm{tr}\,Q)^{4}+2(\textrm{tr}\,Q)^{2}\left(\left(\textrm{tr}\,Q\right)^{2}-2\,\textrm{tr}\,Q\right)+2\left(\left(\textrm{tr}\,Q\right)^{4}-4\left(\textrm{tr}\,Q\right)^{3}+2\left(\textrm{tr}\,Q\right)^{2}+4\,\textrm{tr}\,Q\right)+3(\left(\textrm{tr}\,Q\right)^{2}-2\,\textrm{tr}\,Q)^{2}\right]\\
& =\frac{1}{8}\left[(\textrm{tr}\,Q)^{4}+2\left(\textrm{tr}\,Q\right)^{4}-4\left(\textrm{tr}\,Q\right)^{3}+2\left(\textrm{tr}\,Q\right)^{4}-8\left(\textrm{tr}\,Q\right)^{3}+4\left(\textrm{tr}\,Q\right)^{2}+8\,\textrm{tr}\,Q+3\left(\textrm{tr}\,Q\right)^{4}+12\left(\textrm{tr}\,Q\right)^{2}-12\left(\textrm{tr}\,Q\right)^{3}\right]\\
& =\frac{1}{8}\left[8(\textrm{tr}\,Q)^{4}-24\left(\textrm{tr}\,Q\right)^{3}+16\left(\textrm{tr}\,Q\right)^{2}+8\,\textrm{tr}\,Q\right]
=
(\textrm{tr}\,Q)^{4}-3\left(\textrm{tr}\,Q\right)^{3}+2\left(\textrm{tr}\,Q\right)^{2}+\textrm{tr}\,Q.
\end{align*}

\subsubsection{Second gradient elasticity}

For the second gradient model we found
\paragraph{Cases $\left(V_{\,1},\protect\G,\varphi\right)$ and $\left(\underline{V}_{\,1},\protect\G,\underline{\varphi}\right)$:}
\begin{align*}
\left\langle \Pi_{\,V_{1}},\varphi_{Q}\right\rangle  & =\left(\Pi_{\,V_{1}}\right)_{iajbkcldme}Q_{ia}Q_{jb}Q_{kc}Q_{ld}Q_{me}\\
& =\frac{1}{4}(\delta_{ia}\delta_{jb}\delta_{kc}\delta_{ld}\delta_{me}+\delta_{ia}\delta_{kb}\delta_{jc}\delta_{ld}\delta_{me}+\delta_{ia}\delta_{jb}\delta_{kc}\delta_{md}\delta_{le}+\delta_{ia}\delta_{kb}\delta_{jc}\delta_{md}\delta_{le})Q_{ia}Q_{jb}Q_{kc}Q_{ld}Q_{me}\\
& =\frac{1}{4}(\delta_{ia}\delta_{jb}\delta_{kc}\delta_{ld}\delta_{me}Q_{ia}Q_{jb}Q_{kc}Q_{ld}Q_{me}+\delta_{ia}\delta_{kb}\delta_{jc}\delta_{ld}\delta_{me}Q_{ia}Q_{jb}Q_{kc}Q_{ld}Q_{me}\\
& \quad+\delta_{ia}\delta_{jb}\delta_{kc}\delta_{md}\delta_{le}Q_{ia}Q_{jb}Q_{kc}Q_{ld}Q_{me}+\delta_{ia}\delta_{kb}\delta_{jc}\delta_{md}\delta_{le}Q_{ia}Q_{jb}Q_{kc}Q_{ld}Q_{me}\\
& =\frac{1}{4}(\underbrace{\delta_{ia}Q_{ia}}_{\textrm{tr}Q}\underbrace{\delta_{jb}Q_{jb}}_{\textrm{tr}Q}\underbrace{\delta_{kc}Q_{kc}}_{\textrm{tr}Q}\underbrace{\delta_{ld}Q_{ld}}_{\textrm{tr}Q}\underbrace{\delta_{me}Q_{me}}_{\textrm{tr}Q}+\underbrace{\delta_{ia}Q_{ia}}_{\textrm{tr}Q}\underbrace{Q_{jb}\delta_{bk}Q_{kc}\delta_{cj}}_{\textrm{tr}Q^{2}}\underbrace{\delta_{ld}Q_{ld}}_{\textrm{tr}Q}\underbrace{\delta_{me}Q_{me}}_{\textrm{tr}Q}\\
& \quad+\underbrace{\delta_{ia}Q_{ia}}_{\textrm{tr}Q}\underbrace{\delta_{jb}Q_{jb}}_{\textrm{tr}Q}\underbrace{\delta_{kc}Q_{kc}}_{\textrm{tr}Q}\underbrace{\delta_{el}Q_{ld}\delta_{dm}Q_{me}}_{\textrm{tr}Q^{2}}+\underbrace{Q_{jb}\delta_{bk}Q_{kc}\delta_{cj}}\underbrace{\delta_{ia}Q_{ia}}_{\textrm{tr}Q}\underbrace{\delta_{dm}Q_{me}\delta_{el}Q_{ld}}_{\textrm{tr}Q^{2}}\\
& =\frac{1}{4}((\textrm{tr}\,Q)^{5}+(\textrm{tr}\,Q)^{3}\textrm{tr}\,Q^{2}+(\textrm{tr}\,Q)^{3}\textrm{tr}\,Q^{2}+\textrm{tr}\,Q\left(\textrm{tr}\,Q^{2}\right)^{2}
=
\frac{1}{4}((\textrm{tr}\,Q)^{5}+2\,(\textrm{tr}\,Q)^{3}\textrm{tr}\,Q^{2}+\textrm{tr}\,Q\left(\textrm{tr}\,Q^{2}\right)^{2},
\end{align*}
and
\begin{align*}
\left\langle \Pi\,_{\underline{V}_{\,1}},\varphi_{Q}\right\rangle  & =\left(\Pi\,_{\underline{V}_{\,1}}\right)_{iajbkcldme}Q_{ia}Q_{jb}Q_{kc}Q_{ld}Q_{me}\\
& =\frac{1}{4}(\delta_{ia}\delta_{jb}\delta_{kc}\delta_{ld}\delta_{me}+\delta_{ja}\delta_{ib}\delta_{kc}\delta_{ld}\delta_{me}+\delta_{ia}\delta_{jb}\delta_{kc}\delta_{md}\delta_{le}+\delta_{ja}\delta_{ib}\delta_{kc}\delta_{md}\delta_{le})Q_{ia}Q_{jb}Q_{kc}Q_{ld}Q_{me}\\
& =\frac{1}{4}(\delta_{ia}\delta_{jb}\delta_{kc}\delta_{ld}\delta_{me}Q_{ia}Q_{jb}Q_{kc}Q_{ld}Q_{me}+\delta_{ja}\delta_{ib}\delta_{kc}\delta_{ld}\delta_{me}Q_{ia}Q_{jb}Q_{kc}Q_{ld}Q_{me}\\
& \quad+\delta_{ia}\delta_{jb}\delta_{kc}\delta_{md}\delta_{le}Q_{ia}Q_{jb}Q_{kc}Q_{ld}Q_{me}+\delta_{ja}\delta_{ib}\delta_{kc}\delta_{md}\delta_{le}Q_{ia}Q_{jb}Q_{kc}Q_{ld}Q_{me}\\
& =\frac{1}{4}(\underbrace{\delta_{ia}Q_{ia}}_{\textrm{tr}Q}\underbrace{\delta_{jb}Q_{jb}}_{\textrm{tr}Q}\underbrace{\delta_{kc}Q_{kc}}_{\textrm{tr}Q}\underbrace{\delta_{ld}Q_{ld}}_{\textrm{tr}Q}\underbrace{\delta_{me}Q_{me}}_{\textrm{tr}Q}+\underbrace{\delta_{aj}Q_{jb}\delta_{bi}Q_{ia}}_{\textrm{tr}Q^{2}}\underbrace{\delta_{kc}Q_{kc}}_{\textrm{tr}Q}\underbrace{\delta_{ld}Q_{ld}}_{\textrm{tr}Q}\underbrace{\delta_{me}Q_{me}}_{\textrm{tr}Q}\\
& \quad+\underbrace{\delta_{ia}Q_{ia}}_{\textrm{tr}Q}\underbrace{\delta_{jb}Q_{jb}}_{\textrm{tr}Q}\underbrace{\delta_{kc}Q_{kc}}_{\textrm{tr}Q}\underbrace{\delta_{el}Q_{ld}\delta_{dm}Q_{me}}_{\textrm{tr}Q^{2}}+\underbrace{\delta_{aj}Q_{jb}\delta_{bi}Q_{ia}}_{\textrm{tr}Q^{2}}\underbrace{\delta_{kc}Q_{kc}}_{\textrm{tr}Q}\underbrace{\delta_{dm}Q_{me}\delta_{el}Q_{ld}}_{\textrm{tr}Q^{2}}\\
& =\frac{1}{4}((\textrm{tr}\,Q)^{5}+(\textrm{tr}\,Q)^{3}\textrm{tr}\,Q^{2}+(\textrm{tr}\,Q)^{3}\textrm{tr}\,Q^{2}+\textrm{tr}\,Q\left(\textrm{tr}\,Q^{2}\right)^{2}
=
\frac{1}{4}((\textrm{tr}\,Q)^{5}+2\,(\textrm{tr}\,Q)^{3}\textrm{tr}\,Q^{2}+\textrm{tr}\,Q\left(\textrm{tr}\,Q^{2}\right)^{2}.
\end{align*}

\paragraph{Cases $\left(V_{2},\protect\G,\varphi_{2}\right)$ and $\left(\underline{V}_{\,2},\protect\G,\underline{\varphi}_{2}\right)$:}
\begin{align*}
\left\langle \Pi_{\,V_{2}},\varphi_{Q}\right\rangle  & =\left(\Pi_{\,V_{2}}\right)_{iajbkcldmenf}Q_{ia}Q_{jb}Q_{kc}Q_{ld}Q_{me}Q_{nf}\\
& =\frac{1}{8}(\delta_{ia}\delta_{jb}\delta_{kc}\delta_{ld}\delta_{me}\delta_{nf}+\delta_{ia}\delta_{kb}\delta_{jc}\delta_{ld}\delta_{me}\delta_{nf}+\delta_{ia}\delta_{jb}\delta_{kc}\delta_{ld}\delta_{ne}\delta_{mf}+\delta_{ia}\delta_{kb}\delta_{jc}\delta_{ld}\delta_{ne}\delta_{mf}\\
& \quad+\delta_{la}\delta_{mb}\delta_{nc}\delta_{id}\delta_{je}\delta_{kf}+\delta_{la}\delta_{nb}\delta_{mc}\delta_{id}\delta_{je}\delta_{kf}+\delta_{la}\delta_{mb}\delta_{nc}\delta_{id}\delta_{ke}\delta_{jf}+\delta_{la}\delta_{nb}\delta_{mc}\delta_{id}\delta_{ke}\delta_{jf})Q_{ia}Q_{jb}Q_{kc}Q_{ld}Q_{me}Q_{nf},\\
& =\frac{1}{8}(\delta_{ia}\delta_{jb}\delta_{kc}\delta_{ld}\delta_{me}\delta_{nf}Q_{ia}Q_{jb}Q_{kc}Q_{ld}Q_{me}Q_{nf}+\delta_{ia}\delta_{kb}\delta_{jc}\delta_{ld}\delta_{me}\delta_{nf}Q_{ia}Q_{jb}Q_{kc}Q_{ld}Q_{me}Q_{nf}\\
& \quad+\delta_{ia}\delta_{jb}\delta_{kc}\delta_{ld}\delta_{ne}\delta_{mf}Q_{ia}Q_{jb}Q_{kc}Q_{ld}Q_{me}Q_{nf}+\delta_{ia}\delta_{kb}\delta_{jc}\delta_{ld}\delta_{ne}\delta_{mf}Q_{ia}Q_{jb}Q_{kc}Q_{ld}Q_{me}Q_{nf}\\
& \quad+\delta_{la}\delta_{mb}\delta_{nc}\delta_{id}\delta_{je}\delta_{kf}Q_{ia}Q_{jb}Q_{kc}Q_{ld}Q_{me}Q_{nf}+\delta_{la}\delta_{nb}\delta_{mc}\delta_{id}\delta_{je}\delta_{kf}Q_{ia}Q_{jb}Q_{kc}Q_{ld}Q_{me}Q_{nf}\\
& \quad+\delta_{la}\delta_{mb}\delta_{nc}\delta_{id}\delta_{ke}\delta_{jf}Q_{ia}Q_{jb}Q_{kc}Q_{ld}Q_{me}Q_{nf}+\delta_{la}\delta_{nb}\delta_{mc}\delta_{id}\delta_{ke}\delta_{jf}Q_{ia}Q_{jb}Q_{kc}Q_{ld}Q_{me}Q_{nf})\\
& =\frac{1}{8}(\underbrace{\delta_{ia}Q_{ia}}_{\textrm{tr}Q}\underbrace{\delta_{jb}Q_{jb}}_{\textrm{tr}Q}\underbrace{\delta_{kc}Q_{kc}}_{\textrm{tr}Q}\underbrace{\delta_{ld}Q_{ld}}_{\textrm{tr}Q}\underbrace{\delta_{me}Q_{me}}_{\textrm{tr}Q}\underbrace{\delta_{nf}Q_{nf}}_{\textrm{tr}Q}+\underbrace{\delta_{ia}Q_{ia}}_{\textrm{tr}Q}\underbrace{Q_{jb}\delta_{bk}Q_{kc}\delta_{cj}}_{\textrm{tr}Q^{2}}\underbrace{\delta_{ld}Q_{ld}}_{\textrm{tr}Q}\underbrace{\delta_{me}Q_{me}}_{\textrm{tr}Q}\underbrace{\delta_{nf}Q_{nf}}_{\textrm{tr}Q}\\
& \quad+\underbrace{\delta_{ia}Q_{ia}}_{\textrm{tr}Q}\underbrace{\delta_{jb}Q_{jb}}_{\textrm{tr}Q}\underbrace{\delta_{kc}Q_{kc}}_{\textrm{tr}Q}\underbrace{\delta_{ld}Q_{ld}}_{\textrm{tr}Q}\underbrace{Q_{me}\delta_{en}Q_{nf}\delta_{fm}}_{\textrm{tr}Q^{2}}+\underbrace{\delta_{ia}Q_{ia}}_{\textrm{tr}Q}\underbrace{Q_{jb}\delta_{bk}Q_{kc}\delta_{cj}}_{\textrm{tr}Q^{2}}\underbrace{\delta_{ld}Q_{ld}}_{\textrm{tr}Q}\underbrace{Q_{me}\delta_{en}Q_{nf}\delta_{fm}}_{\textrm{tr}Q^{2}}\\
& \quad+\underbrace{\delta_{al}Q_{ld}\delta_{di}Q_{ia}}_{\textrm{tr}Q^{2}}\underbrace{\delta_{bm}Q_{me}\delta_{ej}Q_{jb}}_{\textrm{tr}Q^{2}}\underbrace{\delta_{cn}Q_{nf}\delta_{fk}Q_{kc}}_{\textrm{tr}Q^{2}}+\underbrace{\delta_{bn}Q_{nf}\delta_{fk}Q_{kc}\delta_{cm}Q_{me}\delta_{ej}Q_{jb}}_{\textrm{tr}Q^{4}}\underbrace{\delta_{al}Q_{ld}\delta_{di}Q_{ia}}_{\textrm{tr}Q^{2}}\\
& \quad+\underbrace{\delta_{bm}Q_{me}\delta_{ek}Q_{kc}\delta_{cn}Q_{nf}\delta_{fj}Q_{jb}}_{\textrm{tr}Q^{4}}\underbrace{\delta_{al}Q_{ld}\delta_{di}Q_{ia}}_{\textrm{tr}Q^{2}}+\underbrace{\delta_{al}Q_{ld}\delta_{di}Q_{ia}}_{\textrm{tr}Q^{2}}\underbrace{\delta_{bn}Q_{nf}\delta_{fj}Q_{jb}}_{\textrm{tr}Q^{2}}\underbrace{\delta_{cm}Q_{me}\delta_{ek}Q_{kc}}_{\textrm{tr}Q^{2}})\\
& =\frac{1}{8}((\textrm{tr}\,Q)^{6}+(\textrm{tr}\,Q)^{4}\textrm{tr}\,Q^{2}+(\textrm{tr}\,Q)^{4}\textrm{tr}\,Q^{2}+(\textrm{tr}\,Q)^{2}\left(\textrm{tr}\,Q^{2}\right)^{2}
+
(\textrm{tr}\,Q^{2})^{3}+\textrm{tr}\,Q^{4}\textrm{tr}\,Q^{2}+\textrm{tr}\,Q^{4}\textrm{tr}\,Q^{2}+\left(\textrm{tr}\,Q^{2}\right)^{3})\\
& =\frac{1}{8}\left((\textrm{tr}\,Q)^{6}+2(\textrm{tr}\,Q)^{4}\textrm{tr}\,Q^{2}+(\textrm{tr}\,Q)^{2}\left(\textrm{tr}\,Q^{2}\right)^{2}+2\,\textrm{tr}\,Q^{4}\textrm{tr}\,Q^{2}+2\left(\textrm{tr}\,Q^{2}\right)^{3}\right)
\end{align*}
and
\begin{align*}
\left\langle \Pi_{\,\underline{V}_{\,2}},\varphi_{Q}\right\rangle  & =\left(\Pi_{\,\underline{V}_{\,2}}\right)_{iajbkcldmenf}Q_{ia}Q_{jb}Q_{kc}Q_{ld}Q_{me}Q_{nf}\\
& =\frac{1}{8}(\delta_{ia}\delta_{jb}\delta_{kc}\delta_{ld}\delta_{me}\delta_{nf}+\delta_{ja}\delta_{ib}\delta_{kc}\delta_{ld}\delta_{me}\delta_{nf}+\delta_{ia}\delta_{jb}\delta_{kc}\delta_{md}\delta_{le}\delta_{nf}+\delta_{ja}\delta_{ib}\delta_{kc}\delta_{md}\delta_{le}\delta_{nf}\\
& \quad+\delta_{la}\delta_{mb}\delta_{nc}\delta_{id}\delta_{je}\delta_{kf}+\delta_{ma}\delta_{lb}\delta_{nc}\delta_{id}\delta_{je}\delta_{kf}+\delta_{la}\delta_{mb}\delta_{nc}\delta_{jd}\delta_{ie}\delta_{kf}+\delta_{ma}\delta_{lb}\delta_{nc}\delta_{jd}\delta_{ie}\delta_{kf})Q_{ia}Q_{jb}Q_{kc}Q_{ld}Q_{me}Q_{nf},\\
& =\frac{1}{8}(\delta_{ia}\delta_{jb}\delta_{kc}\delta_{ld}\delta_{me}\delta_{nf}Q_{ia}Q_{jb}Q_{kc}Q_{ld}Q_{me}Q_{nf}+\delta_{ja}\delta_{ib}\delta_{kc}\delta_{ld}\delta_{me}\delta_{nf}Q_{ia}Q_{jb}Q_{kc}Q_{ld}Q_{me}Q_{nf}\\
& \quad+\delta_{ia}\delta_{jb}\delta_{kc}\delta_{md}\delta_{le}\delta_{nf}Q_{ia}Q_{jb}Q_{kc}Q_{ld}Q_{me}Q_{nf}+\delta_{ja}\delta_{ib}\delta_{kc}\delta_{md}\delta_{le}\delta_{nf}Q_{ia}Q_{jb}Q_{kc}Q_{ld}Q_{me}Q_{nf}\\
& \quad+\delta_{la}\delta_{mb}\delta_{nc}\delta_{id}\delta_{je}\delta_{kf}Q_{ia}Q_{jb}Q_{kc}Q_{ld}Q_{me}Q_{nf}+\delta_{ma}\delta_{lb}\delta_{nc}\delta_{id}\delta_{je}\delta_{kf}Q_{ia}Q_{jb}Q_{kc}Q_{ld}Q_{me}Q_{nf}\\
& \quad+\delta_{la}\delta_{mb}\delta_{nc}\delta_{jd}\delta_{ie}\delta_{kf}Q_{ia}Q_{jb}Q_{kc}Q_{ld}Q_{me}Q_{nf}+\delta_{ma}\delta_{lb}\delta_{nc}\delta_{jd}\delta_{ie}\delta_{kf}\delta_{jd}Q_{ia}Q_{jb}Q_{kc}Q_{ld}Q_{me}Q_{nf})\\
& =\frac{1}{8}(\underbrace{\delta_{ia}Q_{ia}}_{\textrm{tr}Q}\underbrace{\delta_{jb}Q_{jb}}_{\textrm{tr}Q}\underbrace{\delta_{kc}Q_{kc}}_{\textrm{tr}Q}\underbrace{\delta_{ld}Q_{ld}}_{\textrm{tr}Q}\underbrace{\delta_{me}Q_{me}}_{\textrm{tr}Q}\underbrace{\delta_{nf}Q_{nf}}_{\textrm{tr}Q}+\underbrace{\delta_{aj}Q_{jb}\delta_{bi}Q_{ia}}_{\textrm{tr}Q^{2}}\underbrace{\delta_{kc}Q_{kc}}_{\textrm{tr}Q}\underbrace{\delta_{ld}Q_{ld}}_{\textrm{tr}Q}\underbrace{\delta_{me}Q_{me}}_{\textrm{tr}Q}\underbrace{\delta_{nf}Q_{nf}}_{\textrm{tr}Q}\\
& \quad+\underbrace{\delta_{ia}Q_{ia}}_{\textrm{tr}Q}\underbrace{\delta_{jb}Q_{jb}}_{\textrm{tr}Q}\underbrace{\delta_{kc}Q_{kc}}_{\textrm{tr}Q}\underbrace{\delta_{el}Q_{ld}\delta_{dm}Q_{me}}_{\textrm{tr}Q^{2}}\underbrace{\delta_{nf}Q_{nf}}_{\textrm{tr}Q}+\underbrace{\delta_{aj}Q_{jb}\delta_{bi}Q_{ia}}_{\textrm{tr}Q^{2}}\underbrace{\delta_{kc}Q_{kc}}_{\textrm{tr}Q}\underbrace{\delta_{dm}Q_{me}\delta_{el}Q_{ld}}_{\textrm{tr}Q^{2}}\underbrace{\delta_{nf}Q_{nf}}_{\textrm{tr}Q}\\
& \quad+\underbrace{\delta_{al}Q_{ld}\delta_{di}Q_{ia}}_{\textrm{tr}Q^{2}}\underbrace{\delta_{bm}Q_{me}\delta_{ej}Q_{jb}}_{\textrm{tr}Q^{2}}\underbrace{\delta_{cn}Q_{nf}\delta_{fk}Q_{kc}}_{\textrm{tr}Q^{2}}+\underbrace{\delta_{am}Q_{me}\delta_{ej}Q_{jb}\delta_{bl}Q_{ld}\delta_{di}Q_{ia}}_{\textrm{tr}Q^{4}}\underbrace{\delta_{cn}Q_{nf}\delta_{fk}Q_{kc}}_{\textrm{tr}Q^{2}}\\
& \quad+\underbrace{\delta_{al}Q_{ld}\delta_{dj}Q_{jb}\delta_{bm}Q_{me}\delta_{ei}Q_{ia}}_{\textrm{tr}Q^{4}}\underbrace{\delta_{cn}Q_{nf}\delta_{fk}Q_{kc}}_{\textrm{tr}Q^{2}}+\underbrace{\delta_{am}Q_{me}\delta_{ei}Q_{ia}}_{\textrm{tr}Q^{2}}\underbrace{\delta_{bl}Q_{ld}\delta_{dj}Q_{jb}}_{\textrm{tr}Q^{2}}\underbrace{\delta_{cn}Q_{nf}\delta_{fk}Q_{kc}}_{\textrm{tr}Q^{2}})\\
& =\frac{1}{8}((\textrm{tr}\,Q)^{6}+(\textrm{tr}\,Q)^{4}\textrm{tr}\,Q^{2}+(\textrm{tr}\,Q)^{4}\textrm{tr}\,Q^{2}+(\textrm{tr}\,Q)^{2}\left(\textrm{tr}\,Q^{2}\right)^{2}\\
& \quad+(\textrm{tr}\,Q^{2})^{3}+\textrm{tr}\,Q^{4}\textrm{tr}\,Q^{2}+\textrm{tr}\,Q^{4}\textrm{tr}\,Q^{2}+\left(\textrm{tr}\,Q^{2}\right)^{3})\\
& =\frac{1}{8}\left((\textrm{tr}\,Q)^{6}+2(\textrm{tr}\,Q)^{4}\textrm{tr}\,Q^{2}+(\textrm{tr}\,Q)^{2}\left(\textrm{tr}\,Q^{2}\right)^{2}+2\,\textrm{tr}\,Q^{4}\textrm{tr}\,Q^{2}+2\left(\textrm{tr}\,Q^{2}\right)^{3}\right).
\end{align*}
Thus, applying the Cayley-Hamilton Theorem gives
\begin{align*}
\chi(Q) & =\frac{1}{8}\left((\textrm{tr}\,Q)^{6}+2(\textrm{tr}\,Q)^{4}\textrm{tr}\,Q^{2}+(\textrm{tr}\,Q)^{2}\left(\textrm{tr}\,Q^{2}\right)^{2}+(\textrm{tr}\,Q^{2})^{3}+2\,\textrm{tr}\,Q^{4}\textrm{tr}\,Q^{2}+2\left(\textrm{tr}\,Q^{2}\right)^{3}\right)\\
& =\frac{1}{8}\Big((\textrm{tr}\,Q)^{6}+2(\textrm{tr}\,Q)^{4}\left(\left(\textrm{tr}\,Q\right)^{2}-2\,\textrm{tr}\,Q\right)+(\textrm{tr}\,Q)^{2}\left(\left(\textrm{tr}\,Q\right)^{2}-2\,\textrm{tr}\,Q\right)^{2}\\
& \quad+2\left(\left(\textrm{tr}\,Q\right)^{4}-4\left(\textrm{tr}\,Q\right)^{3}+2\left(\textrm{tr}\,Q\right)^{2}+4\,\textrm{tr}\,Q\right)\left(\left(\textrm{tr}\,Q\right)^{2}-2\,\textrm{tr}\,Q\right)+2\left(\left(\textrm{tr}\,Q\right)^{2}-2\,\textrm{tr}\,Q\right)^{3}\Big)\\
& =\frac{1}{8}\Big((\textrm{tr}\,Q)^{6}+2(\textrm{tr}\,Q)^{6}-4(\textrm{tr}\,Q)^{5}+\left(\textrm{tr}\,Q\right)^{6}+4\left(\textrm{tr}\,Q\right)^{4}-4\left(\textrm{tr}\,Q\right)^{5}\\
& \quad+2\left(\textrm{tr}\,Q\right)^{6}-8\left(\textrm{tr}\,Q\right)^{5}+4\left(\textrm{tr}\,Q\right)^{4}+8\left(\textrm{tr}\,Q\right)^{3}-4\left(\textrm{tr}\,Q\right)^{5}+16\left(\textrm{tr}\,Q\right)^{4}-8\left(\textrm{tr}\,Q\right)^{3}-16\left(\textrm{tr}\,Q\right)^{2}\\
& \quad+2\left(\textrm{tr}\,Q\right)^{6}-16\left(\textrm{tr}\,Q\right)^{3}-12\left(\textrm{tr}\,Q\right)^{5}+24\left(\textrm{tr}\,Q\right)^{4}\Big)\\
& =\frac{1}{8}\Big(8(\textrm{tr}\,Q)^{6}-32\left(\textrm{tr}\,Q\right)^{5}+48\left(\textrm{tr}\,Q\right)^{4}-16\left(\textrm{tr}\,Q\right)^{3}-16\left(\textrm{tr}\,Q\right)^{2}\Big)\\
& =(\textrm{tr}\,Q)^{6}-4\left(\textrm{tr}\,Q\right)^{5}+6\left(\textrm{tr}\,Q\right)^{4}-2\left(\textrm{tr}\,Q\right)^{3}-2\left(\textrm{tr}\,Q\right)^{2}.
\end{align*}

\subsubsection{Non symmetric theories}\label{Non symmetric theories}

In this case we have 
\begin{align*}
     \overline{\Pi}	
     &
     = \frac{1}{2}\left(\delta_{ia}\delta_{jb}\delta_{kc}\delta_{ld}+\delta_{ka}\delta_{lb}\delta_{ic}\delta_{jd}\right)Q_{ia}Q_{jb}Q_{kc}Q_{ld}  
     =
     \frac{1}{2}\left(\delta_{ia}\delta_{jb}\delta_{kc}\delta_{ld}Q_{ia}Q_{jb}Q_{kc}Q_{ld}+\delta_{ka}\delta_{lb}\delta_{ic}\delta_{jd}Q_{ia}Q_{jb}Q_{kc}Q_{ld}\right)
     \\
     &
     =
     \frac{1}{2}\big(\underbrace{\delta_{ia}Q_{ia}}_{\text{tr}Q}\underbrace{\delta_{jb}Q_{jb}}_{\text{tr}Q}\underbrace{\delta_{kc}Q_{kc}}_{\text{tr}Q}\underbrace{\delta_{ld}Q_{ld}}_{\text{tr}Q}+\underbrace{Q_{ia}\delta_{ak}Q_{kc}\delta_{ci}}_{\text{tr}Q^2}\underbrace{Q_{jb}\delta_{bl}Q_{ld}\delta_{dj}}_{\text{tr}Q^2}\big)
     =
     \frac{1}{2}\left((\textrm{tr}\,Q)^{4}+\left(\textrm{tr}\,Q^{2}\right)^{2}\right),
\end{align*}
and so the Cayley-Hamilton Theorem yields
\begin{align*}
     \chi(Q) 
     & 
     =
     \frac{1}{2}\left((\textrm{tr}\,Q)^{4}+\left(\textrm{tr}\,Q^{2}\right)^{2}\right)
     =
     \frac{1}{2}\left((\textrm{tr}\,Q)^{4}+\left(\left(\textrm{tr}\,Q\right)^{2}-2\,\textrm{tr}\,Q\right)^{2}\right)
     \\
     & 
     =
     \frac{1}{2}\left((\textrm{tr}\,Q)^{4}+\left(\textrm{tr}\,Q\right)^{4}-4\left(\textrm{tr}\,Q\right)^{3}+4\left(\textrm{tr}\,Q\right)^{2}\right)
     =(\textrm{tr}\,Q)^{4}-2\left(\textrm{tr}\,Q\right)^{3}+2\left(\textrm{tr}\,Q\right)^{2}.
\end{align*}

{\footnotesize{}\bibliographystyle{plain}
	\bibliography{library-2}
}{\footnotesize \par}

%
%
\end{document}